\documentclass[graybox]{svmult}
\usepackage[round,sort,comma,numbers]{natbib}
\usepackage{slashed} 
\usepackage{journal-names}

\usepackage{float}
\usepackage[caption = false]{subfig}

\usepackage[export]{adjustbox}

\usepackage{mathptmx}       
\usepackage{helvet}         
\usepackage{courier}        
\usepackage{type1cm}        
\usepackage{makeidx}         
\usepackage{graphicx}        
\usepackage{multicol}        
\usepackage[bottom]{footmisc}

\def\gta{\mathrel{\raise.3ex\hbox{$>$}\mkern-14mu
             \lower0.6ex\hbox{$\sim$}}}
\def\lta{\mathrel{\raise.3ex\hbox{$<$}\mkern-14mu
             \lower0.6ex\hbox{$\sim$}}}
\def\simless{\mathbin{\lower 3pt\hbox
     {$\rlap{\raise 5pt\hbox{$\char'074$}}\mathchar"7218$}}}   
\def\Msun{{\rm M}_\odot}                                       

\begin{document}
\title*{High-frequency variability in neutron-star low-mass X-ray binaries}
\author{Mariano M\'endez and Tomaso M. Belloni}
\institute{Mariano M\'endez  \at Kapteyn Astronomical Institute, University of Groningen,
P.O. Box 800, 9700 AV Groningen, The Netherlands
\email{mariano@astro.rug.nl}
\and
Tomaso M. Belloni  \at INAF - Osservatorio Astronomico di Brera,
Via E. Bianchi 46, I-23807, Merate, Italy
\email{tomaso.belloni@inaf.it}
}

\maketitle

\abstract{Binary systems with a neutron-star primary accreting from a companion star display variability in the X-ray band on time scales ranging from years to milliseconds. With frequencies of up to $\sim$$1300$ Hz, the kilohertz quasi-periodic oscillations (kHz QPOs) represent the fastest variability observed from any astronomical object. The sub-millisecond time scale of this variability implies that the kHz QPOs are produced in the accretion flow very close to the surface of the neutron star, providing a unique view of the dynamics of matter under the influence of some of the strongest gravitational fields in the Universe. This offers the possibility to probe some of the most extreme predictions of General Relativity, such as dragging of inertial frames and periastron precession at rates that are sixteen orders of magnitude faster than those observed in the solar system and, ultimately, the existence of a minimum distance at which a stable orbit around a compact object is possible. Here we review the last twenty years of research on kHz QPOs, and we discuss the prospects for future developments in this field.}


\section{Introduction}
\label{sec:intro}

Fast time variability from accreting X-ray binaries has become in the past decades a very important tool for our understanding of the process of accretion onto compact objects. As emission properties change on time scales well below a second, it is impossible to ignore variability while concentrating solely on spectral analysis. It was the Rossi X-ray Timing Explorer (RXTE) satellite with its large-area PCA instrument that allowed us to probe very fast time scales, below 10 milliseconds. In this regime, we are exploring the accreting flow very close to the compact object, whether it is a black hole or a neutron star. So deep into the potential well the effects of General Relativity in the strong-field regime can be observable and timing analysis is a very direct way to explore them. 

In neutron-star binaries, the phenomenology is particularly rich and complex, with the presence of Quasi-Periodic Oscillations (QPOs) at frequencies of hundreds of Hz, and even faster than 1 kHz. The frequencies of these oscillations are linked to fundamental frequencies in a gravitational field, allowing us to probe General Relativity in extreme gravitational fields. Most of what we learned comes from the RXTE satellite, which ended in 2012, although new information is provided by current missions such as Astrosat and NICER. New, much more sensitive, instruments are being planned, like the Chinese-European satellite eXTP, and when they become operative we expect a real explosion of new results. This will take several years; here we review the current state of research for neutron-star binaries, concentrating on high-frequency oscillations.

\section{History}

The history of aperiodic variability from X-Ray Binaries began in the early days of X-ray astronomy. After several detections of spurious pulse periods from Cygnus X-1 were reported, the idea that the observed variability was the result of an incoherent process (shot noise) was put forward \citep{Terrell-1972}. Further observations with more advanced instrumentation led to the production of the first statistically significant Power Density Spectra (PDS) from which the strong noise from this black hole candidate was defined and found to be more complex than a simple shot noise \citep[see e.g.][]{Nolan-1981}.

For Neutron-Star Low-Mass X-ray Binaries (NS LMXB), the first important result was the discovery of Quasi-Periodic Oscillations (QPO) in GX~5--1 in 1985 with the EXOSAT satellite \cite{vdk-1985}, obtained serendipitously when looking for X-ray pulsations (see the chapter by Patruno \& Watts, this volume). From this first detection, QPOs were soon observed in many other sources of the same class.
The detection of a signal with a very defined frequency provided the first precise time scale measurement of the accretion flow. The first interpretation of QPOs was in terms of the spin frequency of the neutron star. The values of their frequencies (typically $\sim$$1$ Hz) and the fact that they were not coherent or constant in time excluded that they could be a direct observation of the spin. However, models that interpreted them as a beat between the rotation of the neutron star and the orbital motion at the inner radii of the accretion flow were proposed and were able to explain the observations \citep[see e.g.][]{Alpar-1985, Lamb-1985}. For this to be the case, the presence of a non-negligible magnetic field is necessary. Additional types of QPOs with different properties were also found, the origin of which was even more difficult to explain.

A few years later, similar QPOs started being observed from black-hole binaries (BHB), thanks to the new all-sky monitors that were able to discover transient systems (since most of the BHB are transient). Their frequency was lower ($1-10$ Hz), but they appeared to be rather similar to those in NS LMXB \citep[see e.g.][]{Miyamoto-1991}. In addition, the broad-band noise component connected to QPOs was found, at least in some source states, to be extremely similar between the two classes. Since black holes do not have a solid surface nor a magnetic field, the NS models that depend on either could not be applied to black holes.

The launch of RXTE at the end of 1995 opened the way to the detection of high-frequency ($>$100 Hz) features in the PDS of accreting binaries. For NS systems, new quasi-periodic peaks at frequencies of hundreds of Hertz, called kilohertz QPOs (kHz QPOs), were discovered first in the brightest source, Sco~X--1 \citep{vdk-1996}, then soon in many other NS LMXBs. The model involving a beat with the neutron star spin was adapted to interpret these high frequencies, as the peaks often appear in pairs with roughly the same separation \citep[e.g.][]{Strohmayer-1996a, Ford-1997}, but new data presented problems for the model, which had to be abandoned. RXTE also allowed to bring the BHB QPOs into a phenomenological scheme that appears to be connected to that of NS LMXB low-frequency QPOs \citep{Casella-2005}. The kHz phenomenon appears to be very common in bright NS LMXBs. RXTE also discovered high-frequency QPOs (HFQPOs) from BHBs, but they are extremely rare to the extent that, excluding one peculiar source that had many detections \citep{Morgan-1997,Belloni-2013}, only a handful of them were found in the sixteen years of operation of the satellite \citep{Belloni-2012}. RXTE also led to the discovery of other fast-timing phenomena from NS LMXBs that have completely changed our knowledge of these systems: burst oscillations, accreting millisecond pulsars and intermittent pulsars, all of which are dealt with in other chapters of this book.

After the end of the RXTE mission it has become much more difficult to detect fast-timing aperiodic phenomena, because missions like XMM-Newton, Chandra or Swift are not optimised for timing studies and do not yield the high count rates that are needed. In the recent years, the launch of the Indian satellite Astrosat, which contains an instrument similar to the main one on board RXTE \citep{Agrawal-2006} and of the NICER experiment on board the International  Space Station \citep{Gendreau-2012} have opened a new window onto these phenomena, while future missions like eXTP are being studied.


\section{Basic frequencies close to a neutron star} 
\label{sec:basic}

The accretion flow around a neutron star is a very complex physical system. In order to study the time variability of the emitted flux, it is important to consider the expected characteristic time scales that might be observed, leaving aside the issue of the mechanism that will give rise to flux variability.
\begin{itemize}

\item Neutron stars in LMXBs are expected to be rapidly rotating, based on evolutionary scenarios \citep{Tauris-2010}. An obvious characteristic time to consider is the rotational period of the central object, which would manifest itself in the form of a coherent signal. Rotational frequencies higher than 100 Hz are known for 26 systems, with the fastest being currently 620 Hz (see chapter by Patruno \& Watts and \cite{Watts-2012}).

\item A particle orbiting a compact object defines an obvious time scale, that of the period of its orbit (hereafter dynamical time scale $t_K$). In the vicinity of a neutron star, the space time is affected by the presence of the compact object and an expression from General Relativity has to be used.

\item The accretion flow around a compact object is made of different components whose physical nature and emission properties are very varied, more for a neutron star
than for a black hole \citep[see e.g.][and references therein]{Lin-2007}. Depending on the model and on the source state, we have: (a) the surface of the neutron star, onto which the accreting matter is deposited; (b) a boundary layer between the star and the accretion flow, where the speed of the material in the disc needs to drop rather quickly to adjust to the slower rotation speed of the neutron-star surface; (c) a geometrically thin accretion disc; (d) a Comptonising medium whose spatial location is not yet firmly established; (d) a relativistic jet where matter is ejected from the system at a speed close to that of light. In addition, although the magnetic field of the neutron star in a LMXB is expected to be low, of the order of $10^8$G, nevertheless the presence of a magnetosphere has influence onto the accretion flow.
\end{itemize}

Close the surface of the neutron star surface, matter orbits with a speed close to half the speed of light, $c$. A number of fundamental time scales can be identified. The light-crossing time, $t_{LC}$, is shorter than a millisecond and is potentially detectable in time delays between signals. For sub-Keplerian flows, the free-fall time scale $t_{ff}$ can become important. In an optically thick and geometrically thin disc, in addition to the shortest characteristic time scale corresponding to the {\it dynamical} timescale $t_K$ (see above), other important time scales are the {\it viscous} time scale, $t_{disc}$, on which matter diffuses through the disc due to viscosity, the {\it vertical} time scale, $t_z$, on which vertical deviations from the hydrostatic equilibrium are damped, and the {\it thermal} time scale, $t_{th}$. on which deviations from thermal equilibrium are damped \citep[see][]{fkr}. In the innermost regions of an accretion disc around a neutron star, $t_K$ and $t_z$ are of the order of milliseconds, $t_{th}$ is higher by a factor of a few and $t_{visc}$ is much higher. Moreover, all these time scales increase moving away from the neutron star, and have similar functional dependences on the orbital radius.

With the exception of the neutron star spin, all other timescales apply also to the case of black holes. Since the inner orbits of the accretion flow are comparable in radius between the two objects, similar frequencies are expected, 
although the mass difference will yield faster time scales for neutron stars.
Notice that General Relativity predicts the presence of an innermost stable orbit around a compact object, which naturally imposes a lower limit on all these time scales.


\section{Timing phenomenology: QPOs 101}
\label{sec:qpo101}

In this section we provide a brief introduction to the study of variability using Fourier power density spectra, we explain the concept of variability components in the Fourier power spectrum of accreting LMXBs, and we discuss the properties of one of those components, the high-frequency quasi-periodic oscillations in neutron-star LMXBs, the so-called kHz QPOs. In subsequent sections we expand on some of the properties of the kHz QPOs in more detail. Because this is meant to be a very general introduction to the topics discussed later in this chapter, and to improve the readability, in this section we try to keep the references to the minimum necessary. We give the appropriate references when we discuss the topics introduced here in more detail in the rest of the chapter.

A useful way to characterise the variability of a source is to use the Fourier power density spectrum (PDS) of the source light curve. The PDS gives the square of the amplitude, called power, of the variability in the light curve at each frequency over a range of frequencies \citep[see][for a full explanation]{vdk-1989}. The great advantage of using the Fourier PDS instead of studying the light curves directly is that, while in a light curve one is bound to study the variability over a single, broad range of time scales, from the longest time scale equal to the length of the observation to the shortest time scale equal to the time resolution of the light curve (more precisely, twice the time resolution), in the PDS one can isolate a certain range of frequencies (or equivalently time scales) to study those separately. For instance, it would be very difficult (to say the least) to study a weak, short-period, quasi-periodic signal (e.g., a truly periodic signal with a period that changes randomly during the observation time) in a light curve when that signal is superimposed to another signal that changes stochastically over a long time scale. The reason for this complication is that the two signals would be mixed up in the light curve; one would only be able to study the amplitude of the variability over the total range of time scales combined, and hence only see the combined effect of the two processes. On a PDS, however, one can isolate certain time scales to study the phenomena independently. Perhaps the best example is the case of a strictly periodic signal, e.g., from a pulsar; even if the pulsations appear on top of a very noisy light curve, the signal of the pulsar can be easily identified in the PDS. This is so because the amplitude of the variability of the pulsar signal is spread over all time bins in the light curve, but the power is concentrated in a few frequency bins (ideally one) in a PDS. The same applies to signals that are not strictly periodic; the advantage in these cases is, again, that in the PDS one can isolate, and study separately, the properties of different variability components that are present in the light curve, but span only a limited range of frequencies, whereas this is impossible using the light curve directly. Also because of this, one final advantage of using the PDS is that, at each frequency, one can easily separate and subtract the part of the variability in the light curve due to the Poisson nature of the signal. The power per unit frequency of a constant, Poisson dominated, signal is also a constant that, when the units of the PDS are chosen conveniently \citep{Leahy-1983}, is equal to $2$. In the remainder of this chapter we will use the PDS to characterise the variability components observed in accreting X-ray sources. 

Without entering into too much details, a PDS gives the power per unit frequency of a signal as a function of frequency. The units of the power can be chosen arbitrarily, but the important point we want to make here is that this power is per unit frequency (therefore the word density in the name power density spectrum; as is common in the literature, here we use loosely the word power to refer to power density). The total power in a light curve over a certain range of time scales is the integral of the PDS with respect to frequency over the corresponding range of frequencies; this quantity is no longer a density (per unit frequency) and, because of Parseval's theorem, this integral is equal to the total variance in the light curve in that particular frequency range. By choosing the appropriate PDS normalisation, this quantity can represent the fractional root-mean square variability, also known as fractional rms, in the light curve over a range of frequencies \citep[see][for details]{vdk-1989}.
 
After producing the PDS of a light curve, the power can be fitted as a function of frequency with (a combination of) all kinds of mathematical functions, and use the parameters of those functions to characterise the properties of the components that those functions represent. Ideally those functions would have some underlying theoretical meaning but, even if they do not, one can still deduce interesting properties of the processes that produce that variability, and eventually about the sources themselves, from the parameters of those functions. For instance, a mathematical function that is commonly used to fit the PDS is a Lorentzian, or Cauchy, function:

\begin{equation}
\displaystyle P(\nu) = \frac{\Delta}{2\pi} \frac{N}{(\nu -\nu_0)^2 + (\frac{\Delta}{2})^2}.
\label{eq:lor}
\end{equation}

This function has three parameters: The centroid frequency, $\nu_0$, measures the frequency at which this variability component peaks in the PDS. When we fit a Lorentzian to a QPO, the centroid frequency of the Lorentzian provides information about the dynamics of the process that produces the QPO, e.g. an orbital frequency in the disc, or the frequency of a standing wave in the accretion disc. The next parameter is the full-width at half-maximum (FWHM), $\Delta$, which measures the range of frequencies over which the power of this component contributes significantly to the variability. Instead of the width, some authors use the quality factor, $Q$, (sometimes also called the coherence, but we will reserve the name coherence for another property of the QPO signal), defined as the ratio of the centroid frequency and the FWHM of the QPO, $Q=\nu_0/\Delta$, to characterise the width of the Lorentzian. A narrow Lorentzian would then have a high quality factor. The width or, equivalently, the quality factor, can provide information about the lifetime of the process that produces the QPO, or how much the frequency of the QPO changes over the time interval that was used to produce the PDS. On the other hand, an initially very narrow QPO could be broadened if the oscillations are damped in an intervening medium between the source and the observer, e.g., an X-ray corona very close to the accreting object. Finally, the normalisation, $N$, equal to the integral of the Lorentzian from $-\infty$ to $\infty$, measures the total power of that variability component. As mentioned earlier, the integral of the power density over a certain frequency range gives the power contributed by, in this case, the Lorentzian component that represents the QPO and, because of Parseval's theorem, this is the part of the variance in the light curve that is produced by the QPO. The amplitude of the QPO is the square root of $N$, and is usually expressed as the rms variability of the signal that produces the QPO divided by the average intensity of the source (and normally given in percent), the so-called rms fractional amplitude, or rms amplitude for short \citep{Belloni-1990, Miyamoto-1991}. When it is not normalised by the average intensity, this amplitude is called the absolute rms variability \citep{Uttley-2001}. Both the fractional and the absolute rms amplitudes provide a measure of the variability of the light curve of the source over the range of frequencies (or, equivalently, times scales) where the QPO dominates the power spectrum. The rms amplitude as a function of energy provides information about the radiative process that produces the QPO. 

A narrow component, with a $Q$ factor larger than 2, is usually called a QPO. The definition is a bit vague (should a component with a $Q$ factor just a bit smaller or bigger than 2 be also called a QPO?), but it has been useful, and hence it sticked. Components that have $Q<2$ are usually called bumps and, if the central frequency of this component is at $\nu_0$$=$$0$, they are called zero-centred Lorentzians. In general, all components that produce power over a broad frequency range are called broad-band noise components. (Notice that, in this case, the word noise refers to variability from the source.) Sometimes a broad-band noise component can be fitted by a combination of several, relatively broad and weak, Lorentzians. Since a Lorentzian is the Fourier transform of a sine (or cosine) function whose amplitude drops exponentially with time, a so-called shot, there have been many attempts to understand the variability in these sources in terms of a combination of shot noise components, with different periods, amplitudes and decay times, that add up together to produce the observed light curve. (The decay time of a shot in the light curve is inversely proportional to the FWHM of the Lorentzian in the Fourier PDS.) In recent years, however, it has been shown that the variability in these sources is inconsistent with additive shots, but it is rather a multiplicative process \citep{Uttley-2005}. This raises the question of whether the bumps and broad-band noise components in the PDS of these sources are in reality a combination of several narrow Lorentzians. This standpoint is attractive because a broad-band noise component is complex, and it is difficult to assign a characteristic frequency (time scale) to it, whereas relatively narrow QPOs give good frequencies which are easier to extract and follow over time, and can be treated in a model-independent way. This approach has been tried in a few cases, and interesting correlations among the properties of those (sometimes weak) Lorentzians have emerged. We will mention some of those in the coming sections. 

The Rossi RXTE mission yielded thousands of high-sensitivity observations of X-ray binaries and revolutionised our knowledge of these objects. RXTE observations covered a large range of states of dozens of X-ray binaries, unveiling details of the variability of these objects that helped us understand them more deeply. One of the discoveries of RXTE was the existence of very-high frequency quasi-periodic variability components, up to $\sim 1200$ Hz, in several NS LMXBs. These variability components are what we call the kHz QPOs. Other variability components were also studied with RXTE, including low-frequency QPOs and broad-band noise components. We will mention some of those in passing when necessary, but here we will concentrate mainly on the properties of the kHz QPOs. 

At a very basic level, NS LMXBs can be subdivided into three classes: (i) persistent sources at high luminosity, historically called ``Z'' sources, that can reach luminosities close to the Eddington limit for a neutron star, (ii) persistent and transient sources that can become rather bright but, with top luminosities of 0.1-0.2 Eddington, do not reach the same high luminosities as the Z sources, historically called ``atoll" sources, and (iii) faint sources which, even when transient, remain at low luminosities, below 0.01 Eddington. 

\begin{figure}
\centering
\subfloat[]{\label{fig:lowlum}
\includegraphics[width=0.34\textwidth, angle=90]{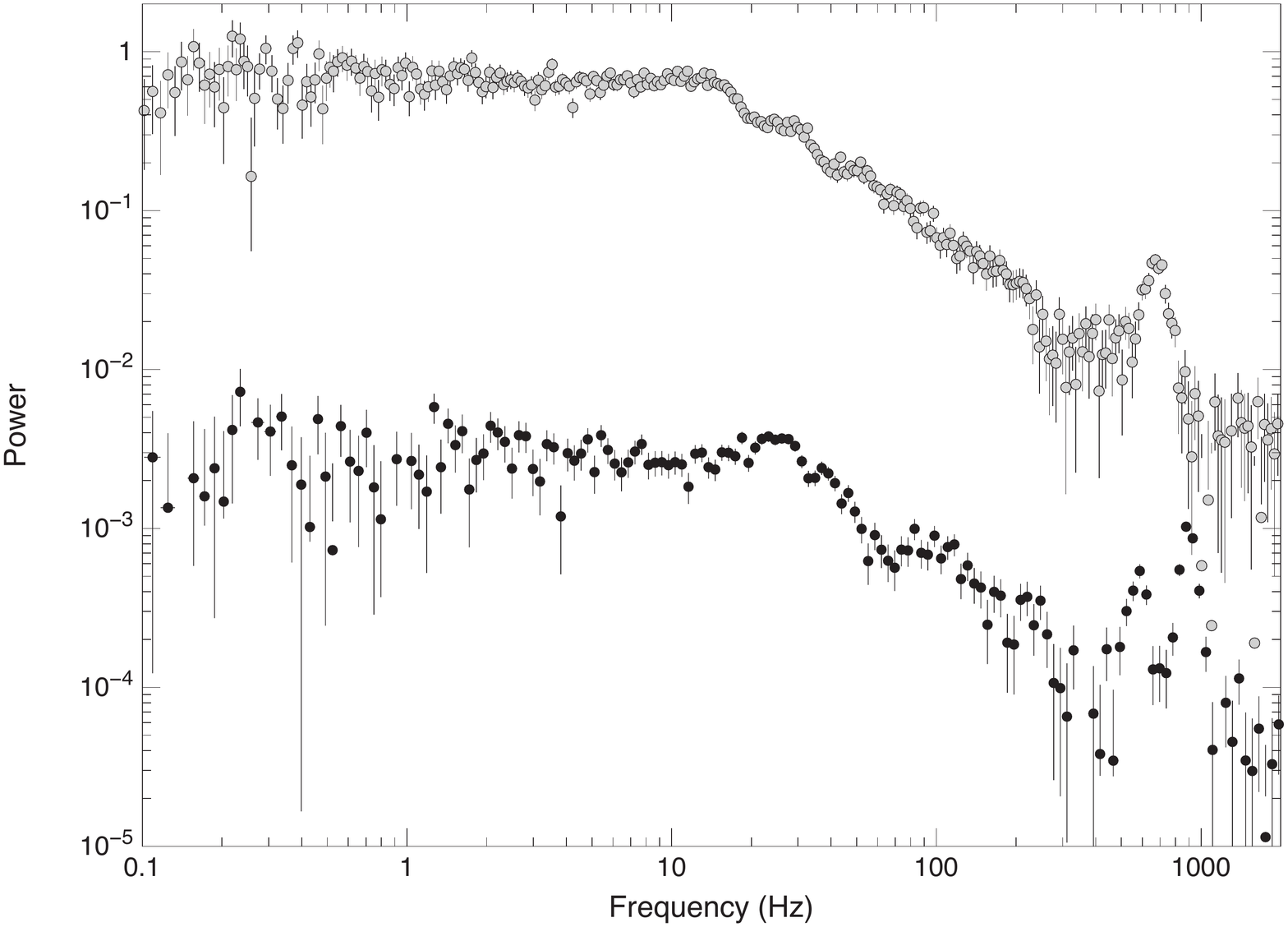}}
\subfloat[]{\label{fig:highlum}
\includegraphics[width=0.34\textwidth, angle=90]{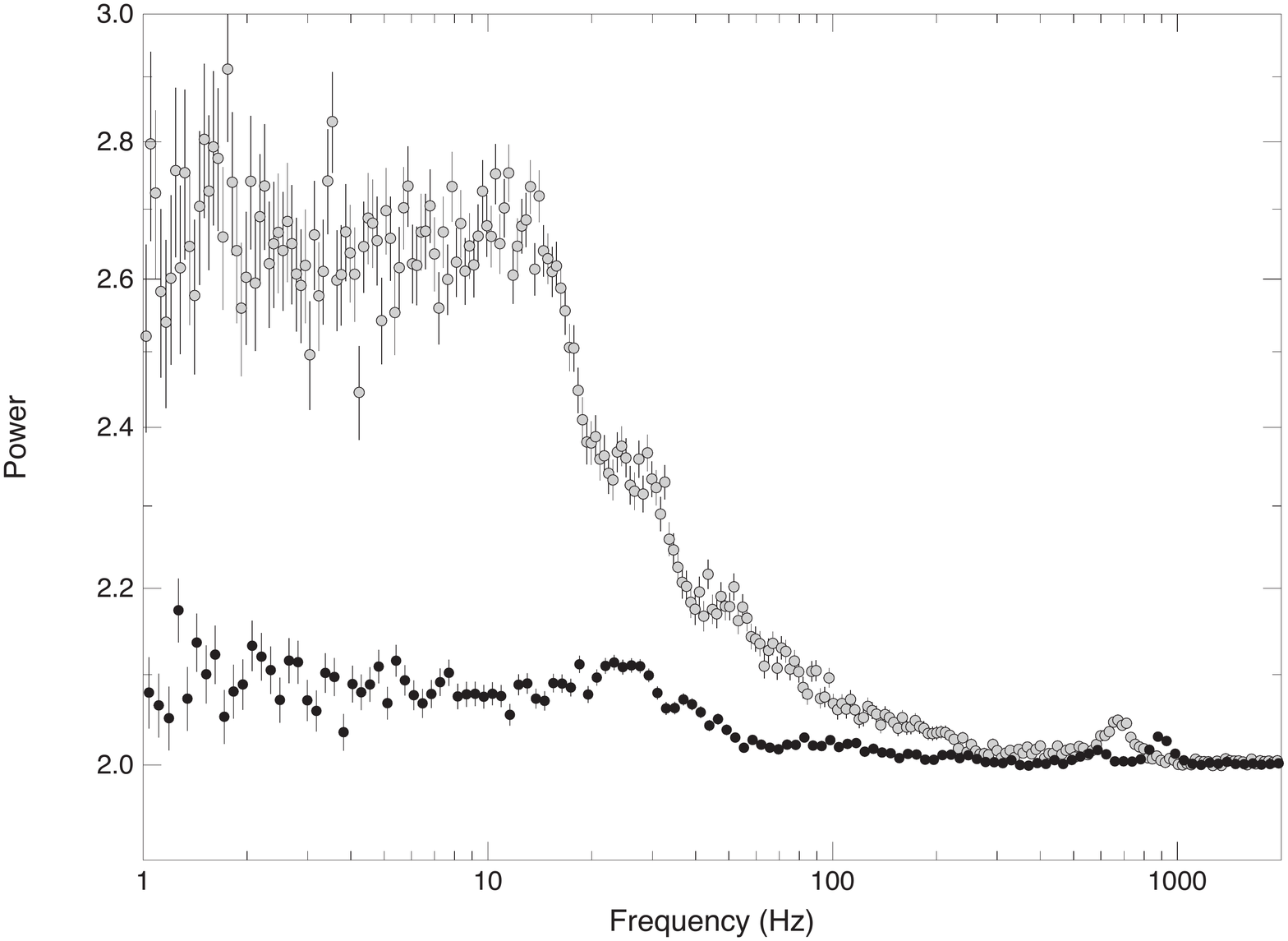}}
\caption{Left: Two power spectra of the low-luminosity source and X-ray burster 1E~1724--3045 in the globular cluster Terzan 2  \citep{Olive-1998, Altamirano-2008b}. Right: Two power spectra of the atoll source 4U~1636--53. The power spectra of 4U~1636--53 were computed on the basis of data published in \citep{Belloni-2007, Sanna-2014}. In both panels the power spectra show the broad-band noise component, extending up to $\sim$$20$$-$$30$ Hz in 1E1724--3045 and up to $\sim$$10$$-$$20$ Hz in 4U~1636--53, some low-frequency QPOs at $\sim$$30$ Hz, and one or two simultaneous kHz QPOs above $\sim$$500$$-$$600$ Hz and up to $\sim$$1000$ Hz. The power spectra on the left have the Poisson noise component subtracted and the power is plotted in units of rms$^2$ per Hz. The power spectra on the right still contain the Poisson noise component and the power is given in units such that the Poisson level is 2.}
\label{fig:classes}
\end{figure}

The power spectrum of the three classes of sources mentioned above show several differences. For example, the amplitude of the variability components in Z sources is generally (but not for all variability components) lower than in atoll and low-luminosity sources. In Figure~\ref{fig:classes} we show examples of the PDS of a low-luminosity (Fig.~\ref{fig:lowlum}; the NS LMXB 1E~1724--3045 in the globular cluster Terzan 2) and an atoll source (Fig.~\ref{fig:highlum}; the NS LMXB 4U~1636$-$53). As it is apparent in that Figure, all the PDS show a broad-band noise component extending up to $\sim$$10$$-$$20$ Hz; above that frequency the power drops as the frequency increases, except for a few relatively narrow features peaking at some specific frequencies. The narrow peaks appearing above $\sim$$400$ Hz are the kHz QPOs. Notice also that the scales in the $y$ axis of the two Figures are different, and that the PDS on the Figure on the right approaches a power level of 2 at high frequencies, whereas the one the left drops to 0 (the $y$ axis on the left panel is in a log scale). The difference is that the powers in Figure~\ref{fig:lowlum} are in units of fractional rms$^2$ per Hz (see above), whereas in Figure~\ref{fig:highlum} the power has not been converted to rms units. Furthermore, in Figure~\ref{fig:lowlum} the contribution of the constant level due to the Poisson nature of the counting process was subtracted from the PDS, leaving only the signal from the source.

The first two sources to show kHz QPOs were the Z source Sco~X-1 \citep{vdk-1996} and the atoll source 4U~1728--34 \citep{Strohmayer-1996}. Both sources displayed (sometimes) two QPOs appearing simultaneously in the PDS at frequencies between $\sim$$700$ Hz and $\sim$$1100$ Hz. The two QPOs were then labeled ``lower'' and ``upper'' kHz QPO according to their frequencies, such that $\nu_{\rm upper} > \nu_{\rm lower}$. If observed frequently enough, most sources with kHz QPOs show two simultaneous QPOs in the PDS, but some (few cases) have so far only showed one. We will come to that below.

\begin{figure}
\centering
\subfloat[]{\label{fig:paral}
\includegraphics[trim=21cm 5cm 0 -1cm, scale=.24, angle=270]{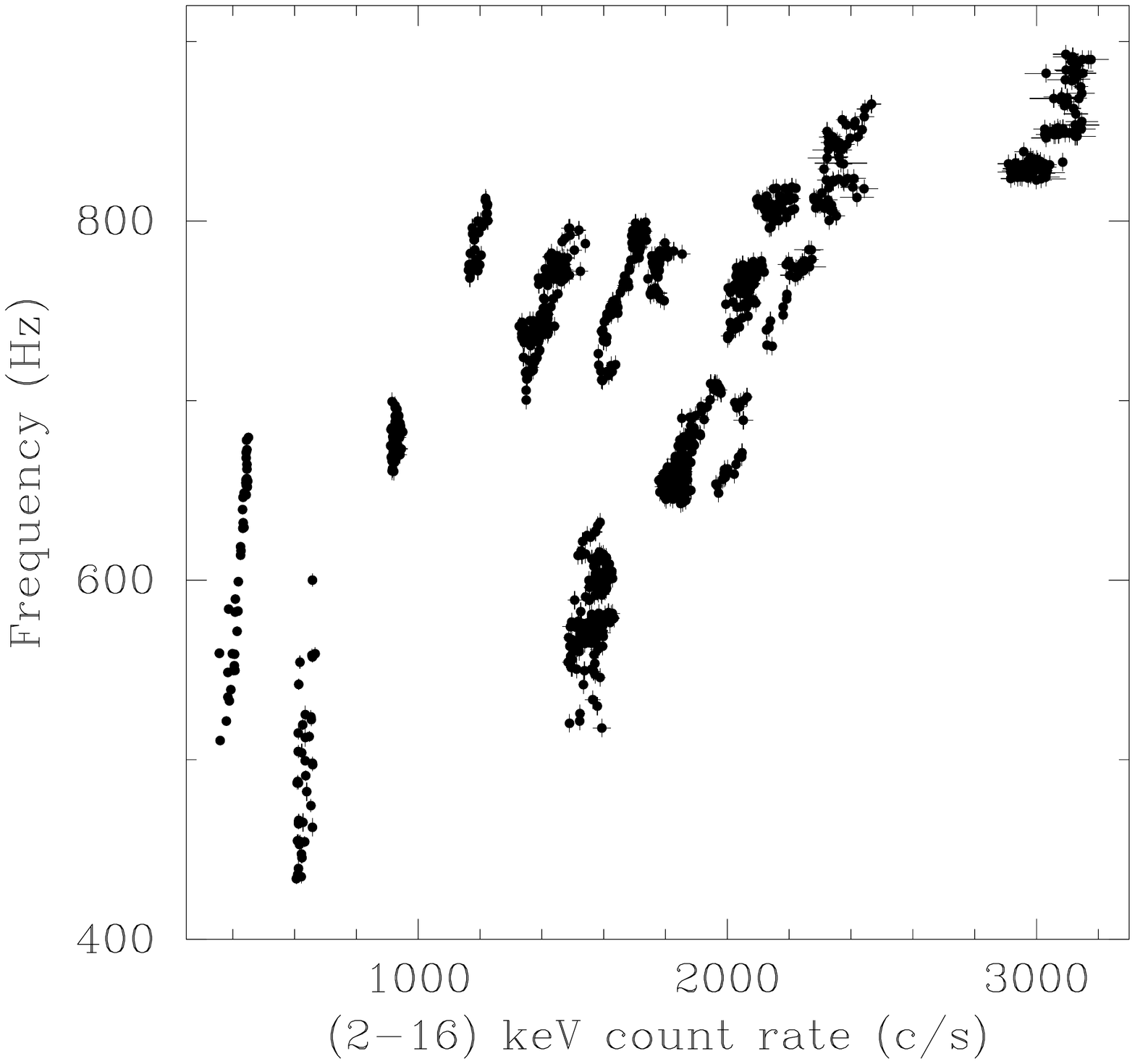}}
\subfloat[]{\label{fig:CCD_atoll}
\includegraphics[trim=5cm -1.3cm 0 1cm, scale=.305, angle=0]{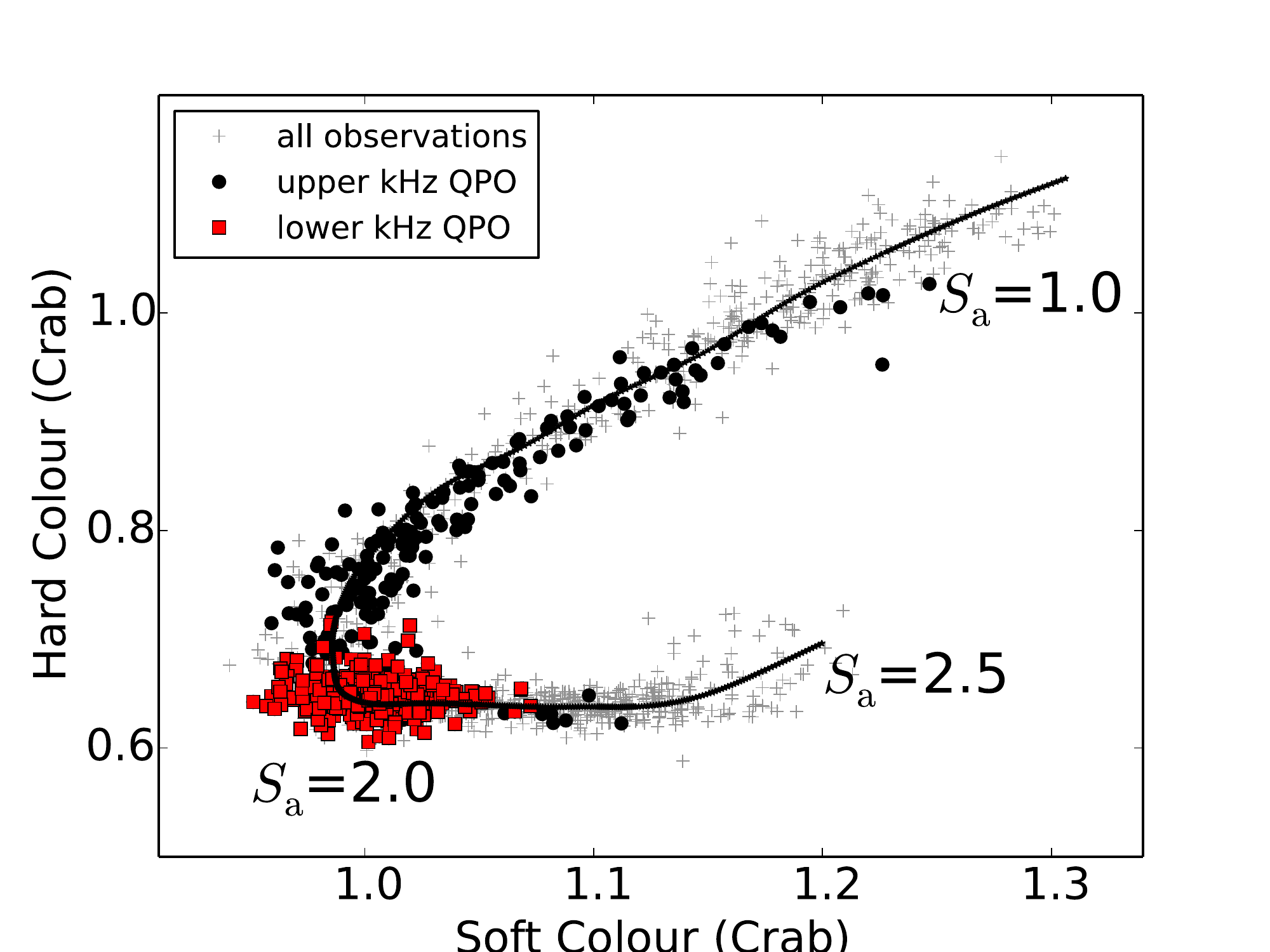}}
\caption{Left panel: Frequency of the lower kHz QPO in 4U~1608--52 vs. the count rate of the source in the $2-16$ keV band. Each point corresponds to a measurement over intervals of $\sim64-128$ s. The frequency and the source count rate are significantly correlated over intervals of a few thousand seconds, producing each of the tracks in the plot, but different observations, made a few days or weeks apart, produce different tracks. This plot shows the so-called parallel tracks \citep[originally published as Figure 2 in][]{Mendez-1999}. Right panel: the distribution of the observations with kHz QPOs on the colour-colour diagram of 4U~1636--53 \citep{Zhang-2017}. Black points correspond to observations in which only the upper kHz QPO was detected, while red points correspond to observations that showed only the lower, or both the lower and the upper, kHz QPO. The grey points indicate observations where no kHz QPO were detected. The parameter $S_a$ gives the length along the solid line, which parametrises the position of the source in this diagram. High values of $S_a$ correspond to high values of inferred $\dot M$. }
\label{fig:parallel}
\end{figure}

Given that the frequency of the upper kHz QPO is consistent with the Keplerian orbital frequency of a test particle at $\sim$$10$$-$$20$ km around a $\sim$1.5-$\Msun$ neutron star (see \S\ref{sec:basic}), the kHz QPOs were immediately associated to motion of matter at the inner parts of the accretion disc. Because of this, and because the bolometric luminosity, and hence the observed flux and intensity, of the source is expected to be proportional to mass accretion rate, $\dot M$, while the inner radius of the accretion disc, $R_{in}$, is expected to decrease as $\dot M$ increases (and vice versa), the expectation was that the QPO frequency would increase with X-ray intensity. While this was the case over short time intervals (a day or less), the long term relation was more complex, with the QPO frequency tracing several, more or less parallel, tracks in a plot of QPO frequency vs. X-ray intensity (Fig.~\ref{fig:paral}). This kind of plots were then, indeed, called parallel tracks.

When enough observations of a single source are collected, a pattern of the detection of the kHz QPOs emerges. In an atoll source, the lower kHz QPO appears in a relatively narrow part of the colour-colour diagram, at an intermediate state, the transitional part of this diagram, between the low-luminosity hard state (called the island state) and the high-luminosity soft state (called the banana; we will try not to use these names here to avoid too much jargon, and we will call these low or hard and high or soft states). The X-ray colours of the source do not change much in the observations in which the lower kHz QPO is present, but the frequency of the QPO appears to correlate with the position of the source in this diagram, with the frequency increasing as the inferred mass accretion rate increases. 

\begin{figure}
\centering
\subfloat[]{\label{fig:HC-nu-Sanna}
\includegraphics[trim= 0cm 3.cm 0 0cm, clip, width=0.38\textwidth,  angle=0]{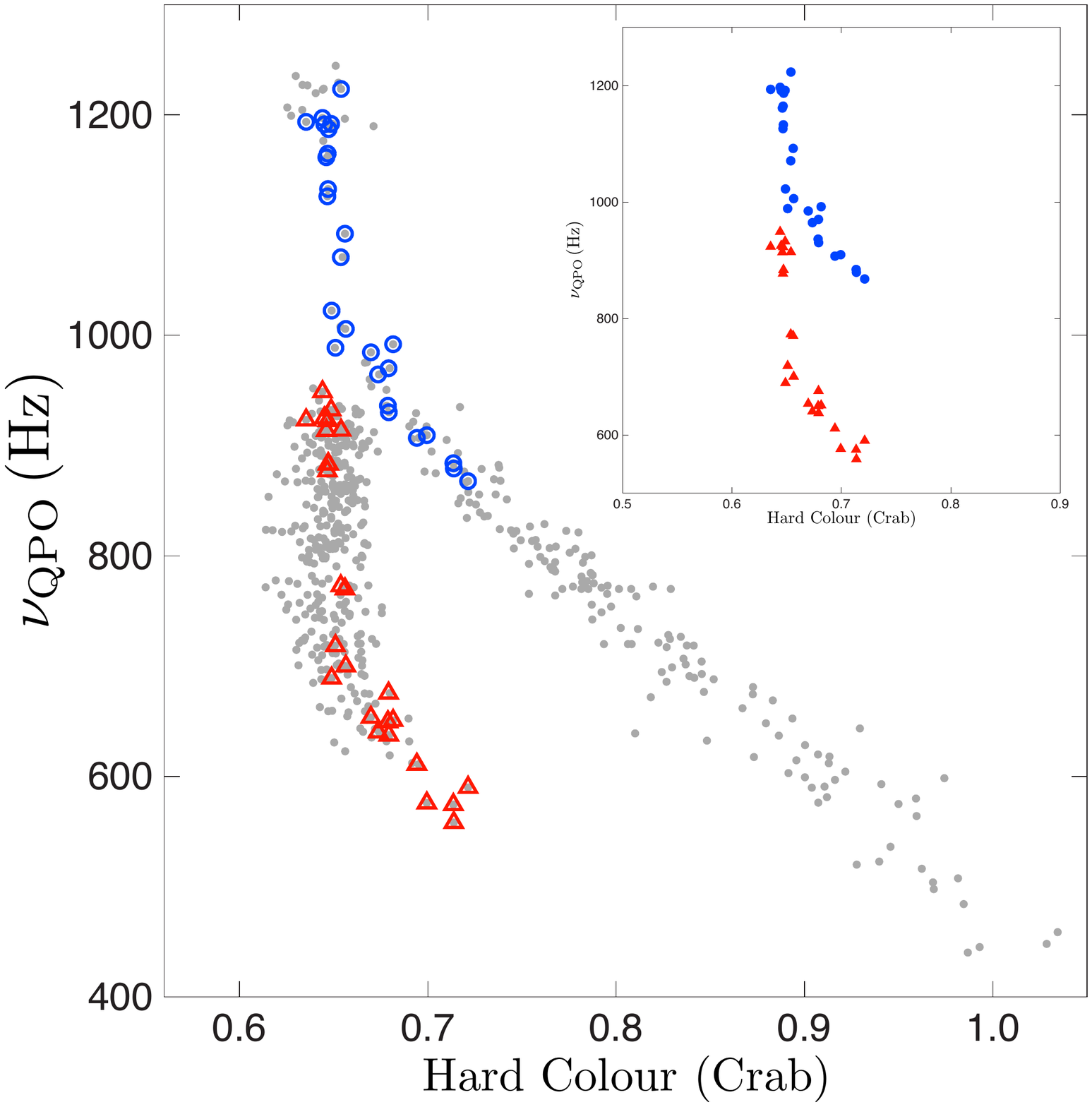}}
\subfloat[]{\label{fig:nu-vs-Sa-Zhang}
\includegraphics[trim= 0.1cm 7cm 0 6cm, clip, width=0.5795\textwidth,  angle=0]{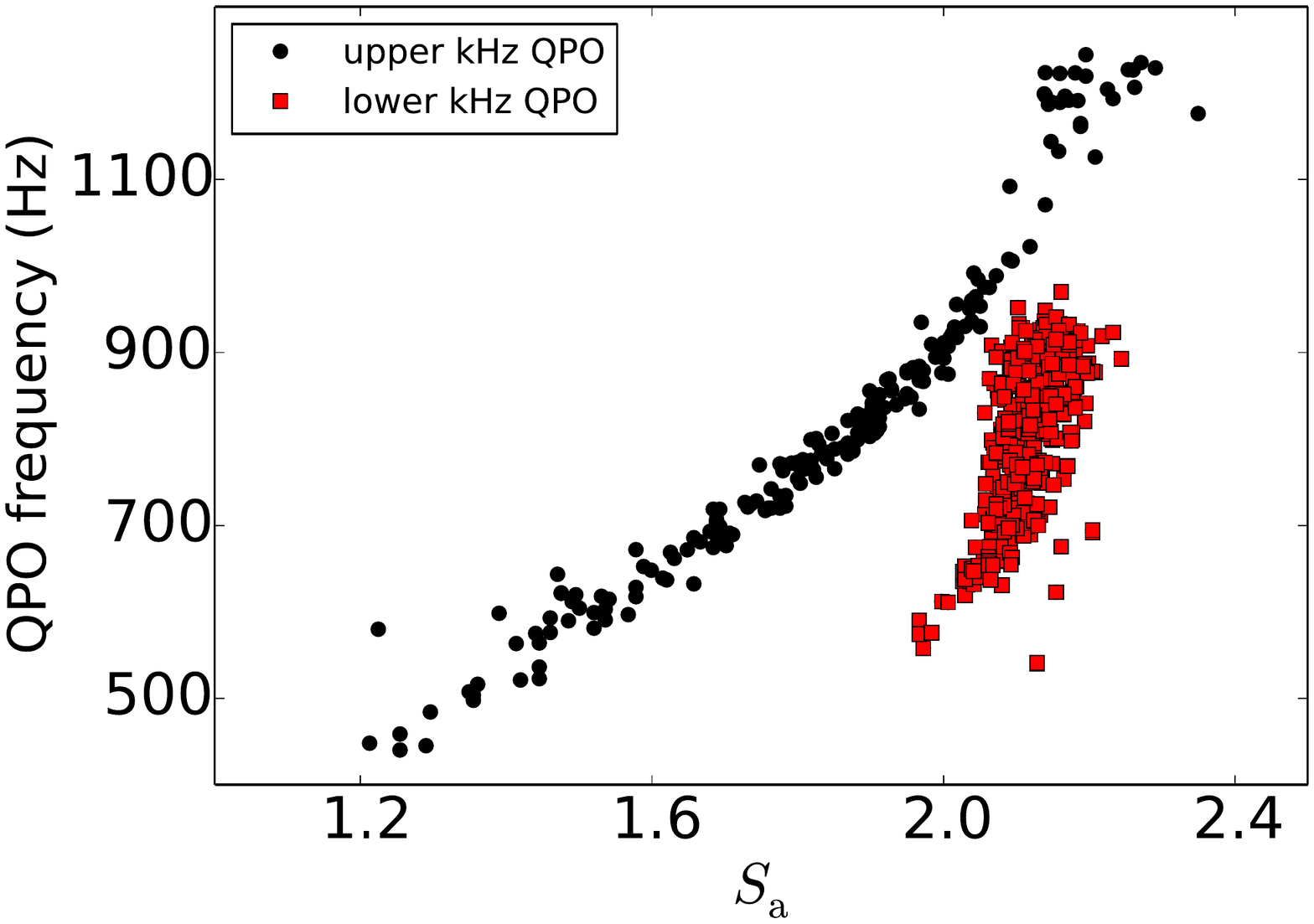}}
\caption{Left: Frequency of the kHz QPOs in 4U~1636--53 vs. the source hard colour (see Fig.~\ref{fig:CCD_atoll}). Grey symbols correspond to observations in which only one of the kHz QPOs was observed; red and blue symbols correspond to QPOs identified, respectively, as the lower and upper kHz QPO. The inset shows only the QPO frequencies of the observations in which two simultaneous kHz QPOs were detected \citep{Sanna-2012}. Right: Frequency of the lower (red symbols) and upper (black symbols) kHz QPOs in 4U~1636--53 vs. $S_a$ \citep{Zhang-2017}. The quantity $S_a$ measures the position of the source along the track in the colour-colour diagram in Figure~\ref{fig:CCD_atoll}.}
\label{fig:nu-vs-HC-Sa-Zhang}
\end{figure}

Figure~\ref{fig:CCD_atoll} shows the colour-colour diagram of 4U~1636--53. The red and black points indicate, respectively, the observations in which the lower and the upper kHz QPOs were detected. The solid line parameterises the position of the source in this diagram, through the variable $S_a$. High values of $S_a$ correspond to high values of inferred $\dot M$. The frequency of the lower kHz QPO increases as $S_a$ increases. The upper kHz QPO, on the contrary, covers a broader range in the colour-colour diagram, with the frequency of the QPO increasing as the source moves from the low-luminosity hard state, via the transitional intermediate state to the high-luminosity soft state (increasing value of $S_a$). This is the sense in which mass accretion rate is inferred to increase in these sources. For completeness, the grey points mark observations in which no kHz QPO was detected. The observations with no kHz QPOs are at the extremes of the C-shaped figure traced by the source in the colour-colour diagram; at the top right the source is in the low-luminosity hard state, where the inferred $\dot M$ is the lowest, while at the bottom right it is in the high-luminosity soft state, where the inferred $\dot M$ is the highest. 

Although not apparent from the plot, there are several observations in which both QPOs were detected simultaneously. This can be seen in Figure~\ref{fig:nu-vs-HC-Sa-Zhang}; the left panel of that Figure shows the frequency of both kHz QPOs as a function of the hard colour, while the right panel shows the frequency of both kHz QPOs as a function of $S_a$. The two kHz QPOs are clearly separated in these two plots, and the parallel tracks of Figure~\ref{fig:paral} collapse into a single track (one for each kHz QPO) when the frequencies of the QPOs are plotted against the hard colour or $S_a$. By the way, since the position of the source in the colour-colour diagram, and therefore hard colour and the value of $S_a$, is driven by changes of the source spectrum, it should be no surprise that the relation of the QPO frequency with the parameters of the models used to fit the energy spectrum also consists of a single track. We will discuss this in \S\ref{sec:spectral-frequency}.

\begin{figure}
\centering
\subfloat[]{\label{fig:HID_Z}
\includegraphics[trim=7cm 4.5cm 10cm 3.8cm, clip, width=0.4655\textwidth, angle=0]{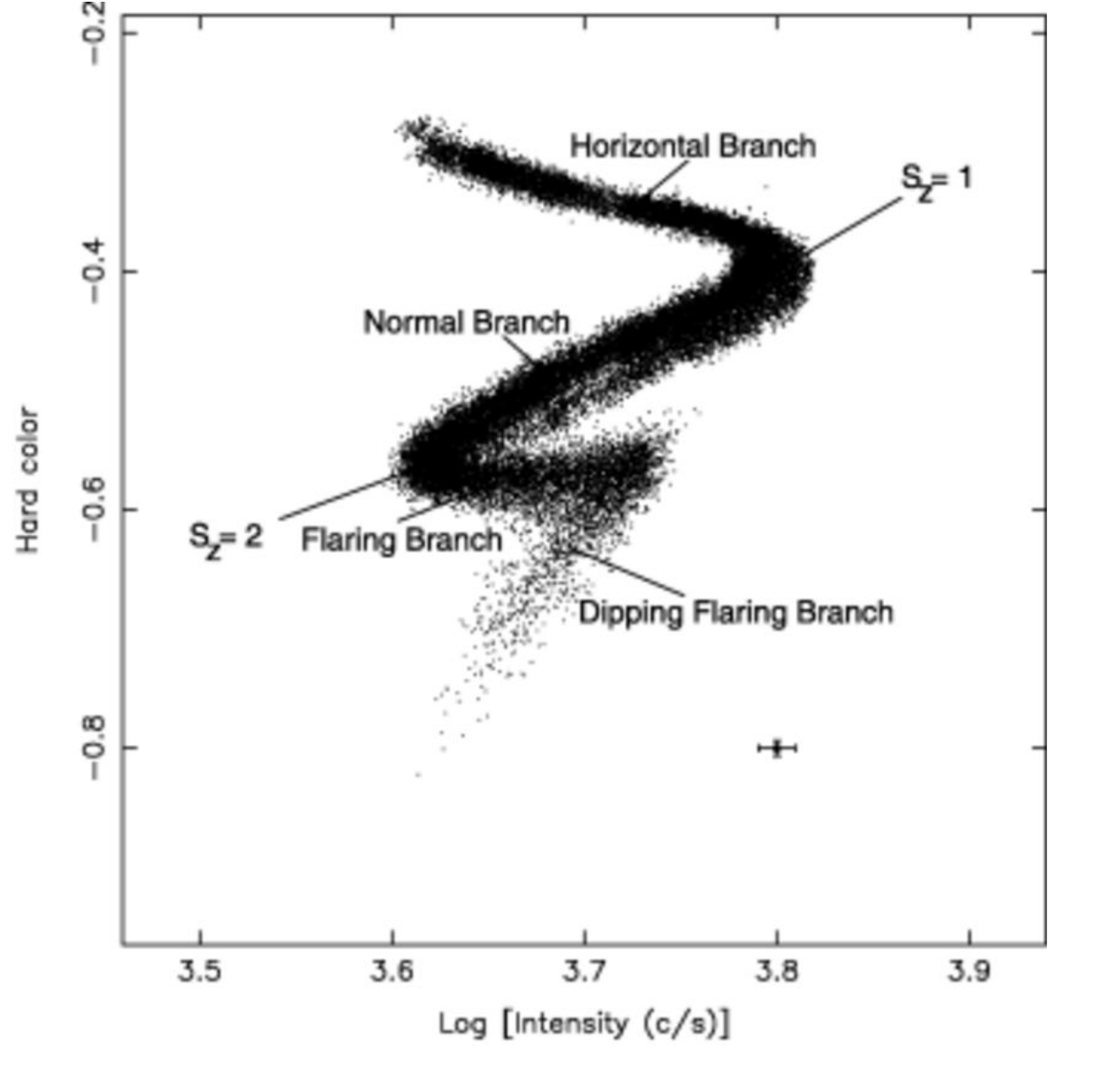}}
\subfloat[]{\label{fig:QPO_Z}
\includegraphics[trim=6.5cm 16.7cm 6cm 4cm, clip, width=0.494\textwidth, angle=0]{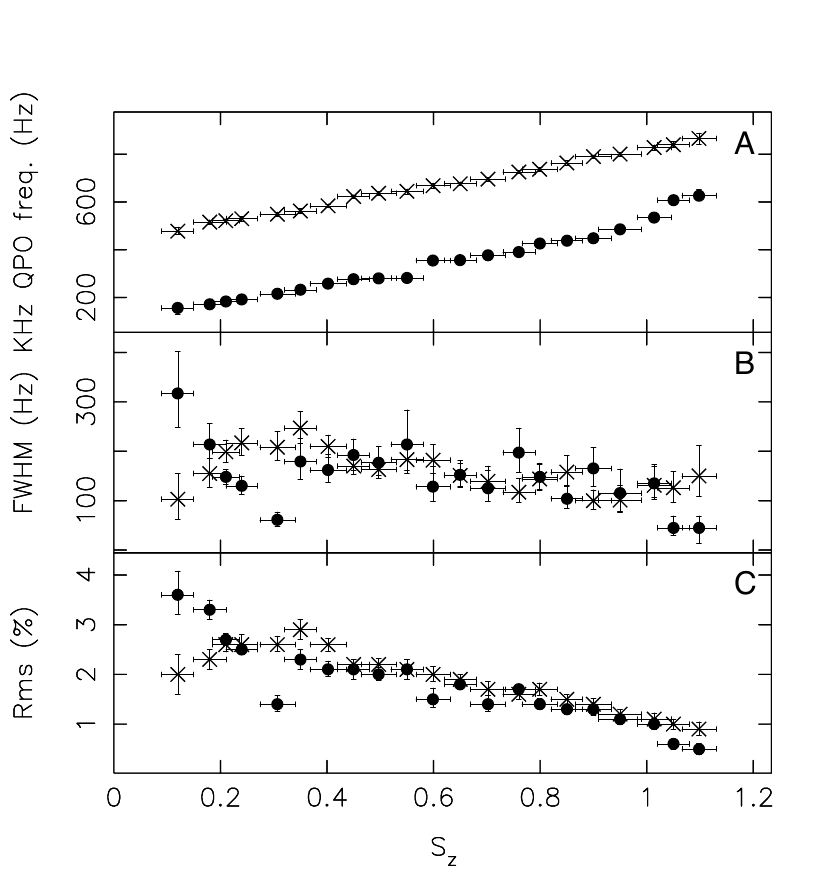}}
\caption{Left panel: Hardness intensity diagram of the Z source GX~5--1 with the different branches of the Z shape indicated \citep[originally published as Figure 1 in][]{Jonker-2002}. The position of the source along the Z-shaped track is parameterised through the quantity $S_{Z}$ which is set to $S_Z$$=$$1$ at the vertex between the Horizontal and the Normal branches, and to $S_Z$$=$$2$ at the vertex between the Normal and the Flaring branches. Right panel: Frequency (top), FWHM (middle) and rms amplitude (bottom) of the kHz QPOs vs. $S_Z$ for GX~5--1 \citep[originally published as Figure 9 in][]{Jonker-2002}. From the values of $S_Z$ in this plot it is apparent that the QPOs are detected in the Horizontal and the initial part of the Normal branches.}
\end{figure}

In Z sources, the kHz QPOs appear mostly in the so-called horizontal and normal branches in the colour-colour diagram (roughly speaking these are, respectively, the top horizontal and diagonal parts of the letter Z that the source traces as it moves in the hardness-intensity diagram), and the QPOs disappear when the source is in the flaring branch (the bottom horizontal part of the letter Z in the hardness-intensity diagram). In Figure~\ref{fig:HID_Z} we show the hardness-intensity diagram of the Z source GX~5--1 with the branches indicated. As for the atoll sources, the parameter $S_Z$ measures the position along the track traced by the source in this diagram, with inferred $\dot M$ increasing when $S_Z$ increases. The frequency of the QPOs increases as the source moves from left to right and then from the top right to the bottom left along the Z shape in the colour-colour diagram, which is the same direction in which, according to work that preceded the discovery of kHz QPOs, $\dot M$ increases in these sources. In Figure~\ref{fig:QPO_Z} we show the frequency, FWHM and fractional rms amplitude of both kHz QPOs in GX~5--1 as a function of $S_Z$. The fact that QPO frequency increases with inferred mass accretion rate (increasing $S_Z$) made the identification of the upper kHz QPO with the Keplerian frequency at the inner disc radius plausible.

\begin{figure}
\centering
\subfloat[]{\label{fig:rms-nu-Ribeiro}
\includegraphics[trim=1cm 1.8cm 2cm 2cm, clip, width=0.48\textwidth, angle=0]{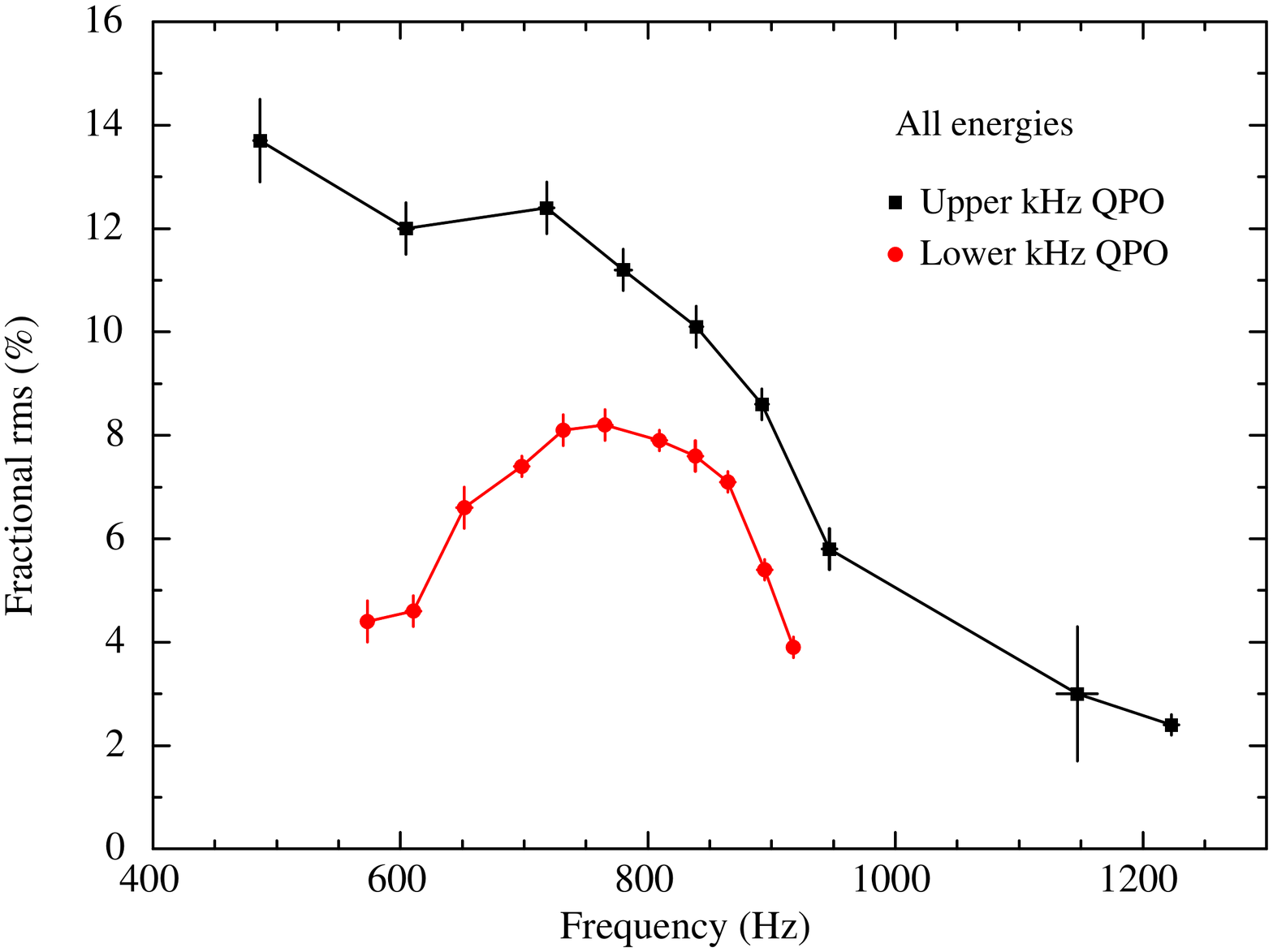}}
\subfloat[]{\label{fig:Q-nu-Barret}
\includegraphics[trim=0 4.5cm 0 10cm, clip, width=0.47\textwidth, angle=0]{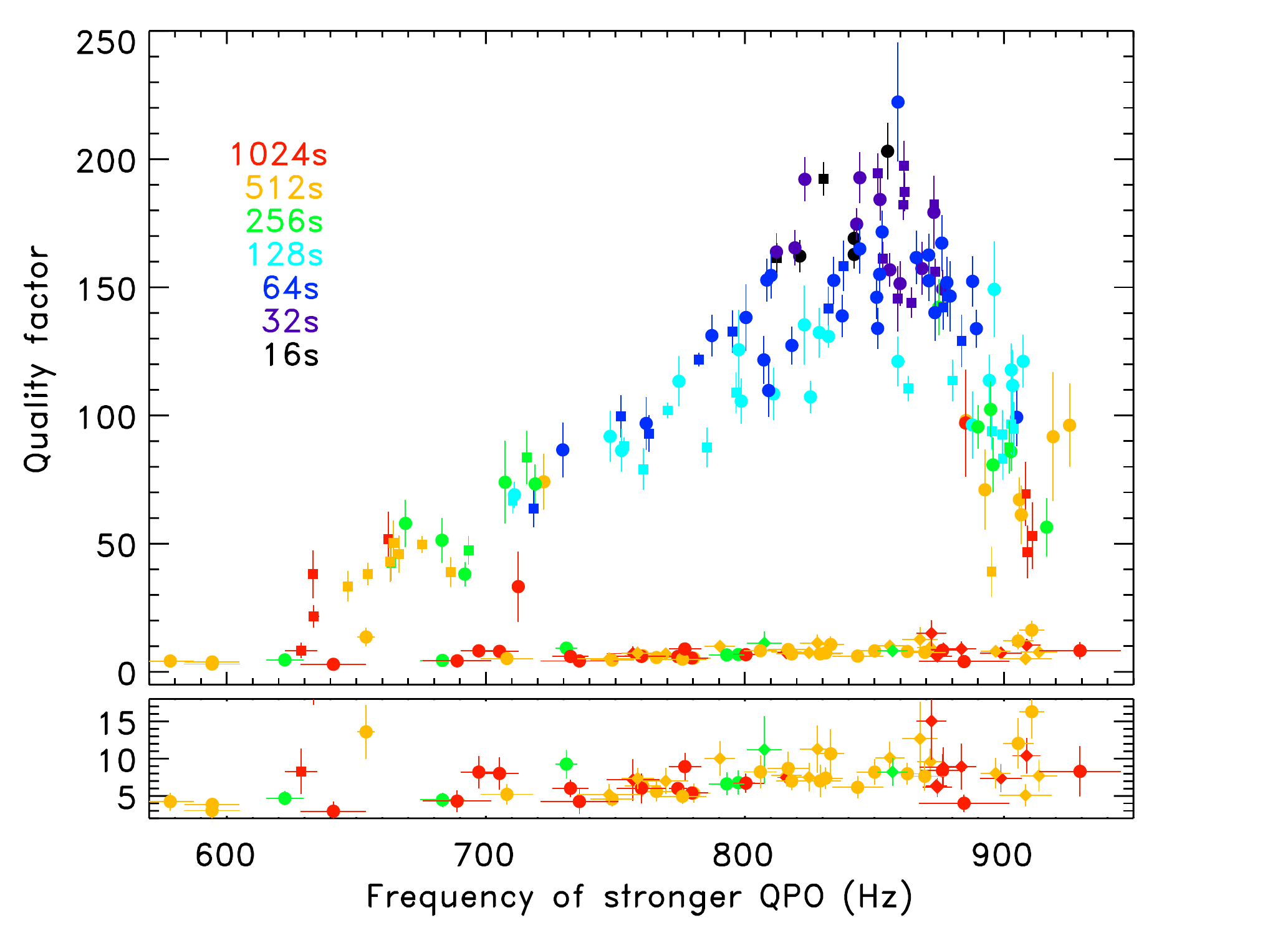}}
\caption{Left: Fractional rms amplitude of the kHz QPOs in 4U~1636--53 vs. the frequency of that same QPO \citep[adapted from][]{Ribeiro-2019}. Right: Quality factor of the kHz QPOs vs. the frequency of the strongest kHz QPO in the power spectrum of 4U~1636--52 \citep{Barret-2005}.}
\end{figure}

The frequency range that makes a QPO a kHz QPO is roughly $400-1200$ Hz. The fact that kHz QPOs are not detected outside this frequency range can be understood from the dependence of the other properties of the kHz QPO upon QPO frequency. The rms amplitude and the $Q$ factor of both kHz QPOs depend upon QPO frequency in a systematic way. The rms amplitude and the $Q$ factor of the lower kHz QPO are maximum when the frequency of the QPO is around $700-800$ Hz, and both the rms and $Q$ decrease when the QPO frequency either increases or decreases. The typical range of fractional rms amplitudes of the lower kHz QPO is $3-15$\% considering photons in the full band covered by RXTE/PCA, nominally from 2 to 60 keV. At the same time, the $Q$ factor of the lower kHz QPO ranges from $\sim$$10$ to $\sim$$50$, and can be as high as $\sim 200-250$ in some sources. For the upper kHz QPO, when the QPO frequency is low the rms amplitude is maximum and remains roughly constant and then drops more or less continuously as the QPO frequency increases whereas, at the same time, the $Q$ factor remains constant or increases slightly. The rms amplitude of the upper kHz QPO is $2-20$\% in the $2-60$-keV band, while $Q$ is usually around 10 or less. In Figures~\ref{fig:rms-nu-Ribeiro} and \ref{fig:Q-nu-Barret} we show, respectively, the rms amplitude and the $Q$ factor, of the lower and  upper kHz QPOs in 4U~1636--53.

The drop of both the rms amplitude and the $Q$ factor limits the detectability of the lower kHz QPO at low and high QPO frequencies below $\sim$$400$ Hz and above $\sim$$950$ Hz. Similarly, the drop of the rms amplitude of the upper kHz QPO limits its detectability at frequencies above $\sim 1200$ Hz, whereas at low QPO frequencies the detectability of the upper kHz QPO is limited by the relatively low $Q$ value and the fact that, when the frequency of the upper kHz QPO goes down to $\sim 400$ Hz the broad-band noise extends up to comparable frequencies such that the upper kHz QPO starts to appear on top of the broad-band noise, and hence it is difficult to detect. All in all, there is a range of frequencies at which the QPOs are the narrowest and the strongest, and hence the most significantly (and hence most often) detected. In the sources in which a single kHz QPO was detected, either the source was not observed for long enough to sample the range of states in which the QPOs are detected, or the source was relatively weak such that sensitivity to detecting kHz QPOs was not sufficient. The fact that the kHz QPOs most often appear in pairs is then a characteristic that needs to be explained.

The frequency of the two QPOs change when other source properties, e.g. the source intensity or colours, change; but an interesting fact is that, as the frequency of the QPOs changes, the difference of the centroid frequency of the two QPO peaks remains more or less constant. When burst oscillations and two simultaneous kHz QPOs were detected in 4U~1728--34, with the frequency separation between the two QPOs consistent with being equal to the frequency of the burst oscillations, a beat-frequency mechanism \citep{Miller-1998} was proposed to explain the double kHz QPOs. In the original model, the upper kHz QPO was identified with the Keplerian frequency at the inner edge of the disc, which is truncated at the sonic radius, the radius at which the radial component of the velocity of the material falling onto the neutron star goes from subsonic to supersonic. The lower kHz QPO was then interpreted as a beat between the oscillation at the Keplerian frequency and the neutron-star spin. Under those conditions, the frequency separation between the two QPOs, which is equal to the neutron-star spin, should remain constant as the frequencies of the QPO move. We will return to this below.

The initial observations of the kHz QPOs in Sco~X-1 had already shown that the frequency separation between the two QPO peaks was not always the same, but decreased systematically, and significantly, by a few percent as the frequencies of the two simultaneous kHz QPOs increased. The beat-frequency model could still explain this behaviour if the material at the inner radius of the disc, where the beating took place, suffered from radiation drag and the beating took place as that material spiralled in towards the neutron star. Being a very luminous source, this effect could be strong in the case of Sco~X-1. But soon after several other less luminous sources, starting with the atoll source 4U~1608--52, showed the same effect. 

The situation got more complicated for this model when burst oscillations and two simultaneous kHz QPOs were detected in 4U~1636--53, with the frequency of the burst oscillations being twice the difference in frequency between the kHz QPOs. The original beat-frequency model could not explain this. The model had to be made more complex, by adding a possible excitation of vertical modes in the accretion disc at a radial distance where the difference between the Keplerian frequency at the inner edge of the disc (that causes the QPO at $\nu_{\rm upp}$) and the neutron-star spin frequency is equal to the vertical epicyclic frequency in the disc. Depending on whether the material in the disc is smooth or clumped, the excited frequency, which produces the lower kHz QPO, would be at $\nu_{\rm upp} - \nu_{\rm spin}$ or $\nu_{\rm upp} - \nu_{\rm spin}/2$.

\begin{figure}
\centering
\includegraphics[trim= -1cm 1cm 0 0, clip, width=0.7\textwidth,  angle=-90]{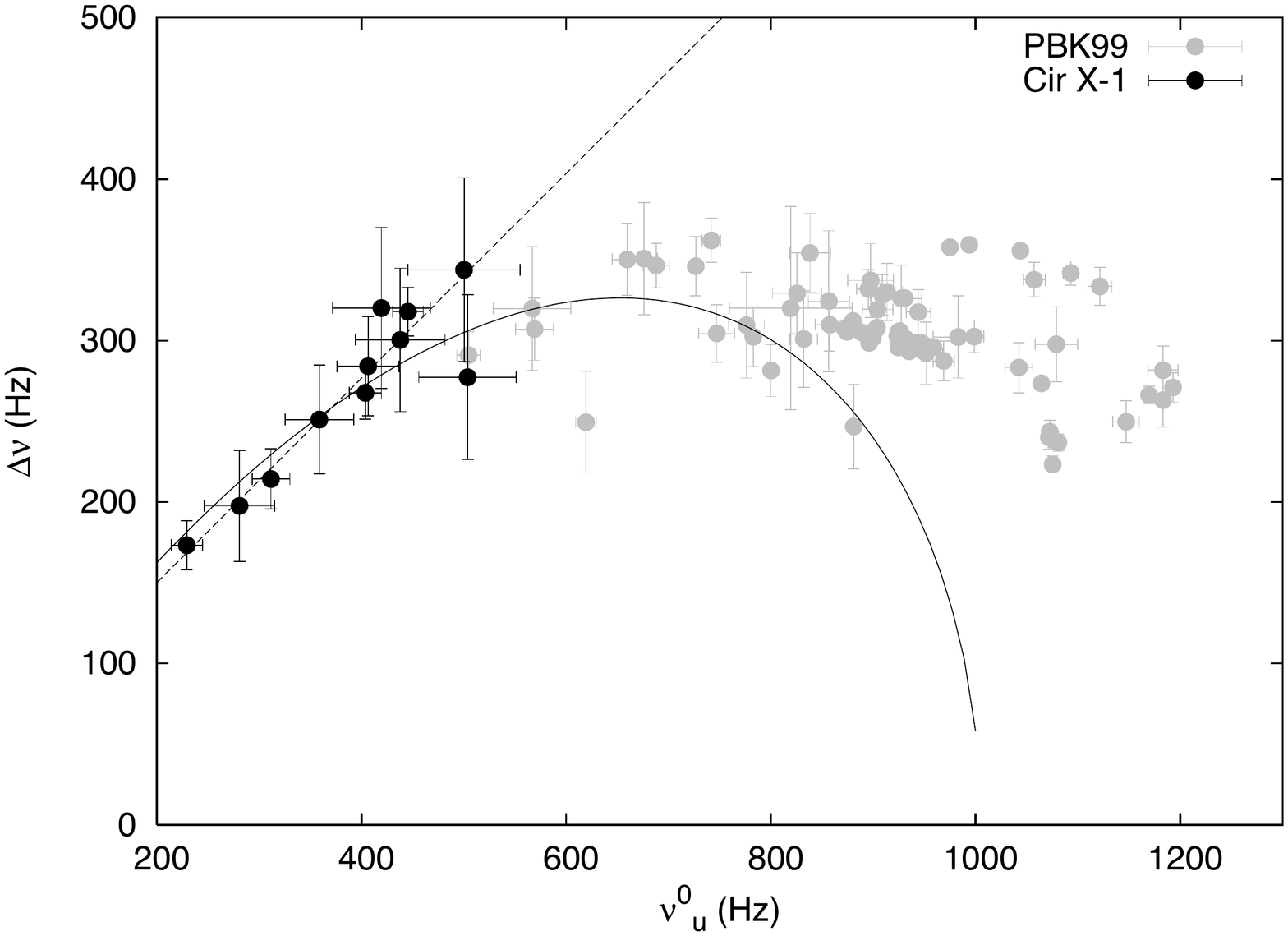}
\caption{Difference between the centroid frequencies of the kHz QPOs, $\Delta\nu$$=$$\nu_{\rm upp}$$-$$\nu_{\rm low}$, as a function of the frequency of the upper kHz QPO in Cir X-1 \citep[originally published as Figure 11 in][]{Boutloukos-2006}.}
\label{fig:CirX-1-Boutloukos}
\end{figure}

These complications for the beat-frequency model triggered other proposals to explain the QPO frequencies. One of them, that could explain the phenomenology rather naturally, was the idea of periastron precession of the innermost parts of the accretion disc. Under this hypothesis, the upper kHz QPO was still identified as the epicyclic azimuthal (Keplerian) frequency of a test particle at the inner edge of the disc, under the influence of the general relativistic (GR) potential of the neutron star. In this model, however, the lower kHz QPO would be the difference between this azimuthal and the epicyclic radial frequency, the so-called periastron precession frequency, at the same spot in the disc. The difference between the azimuthal and the periastron precession frequency in the model changes generally in the same way as in the observations \citep{Stella-1999}, although the calculations do not fit the exact trend of the observations. 

One strong prediction of this model was that the difference between the frequency of the two QPOs should not only decrease at high, but also at low QPO frequencies, something that was later on observed in the system Cir X-1 (Fig.~\ref{fig:CirX-1-Boutloukos}). To be fair to history, the relativistic-precession model, as this model was called, came about as an extension of the Lense-Thirring model \citep{Stella-1998} that was proposed a year earlier to explain the correlation between the frequency of the upper kHz QPO and a low-frequency QPO in neutron-star LMXBs. The Lense-Thirring and the relativistic-precession models became one consistent model for both the low- and the high-frequency variability. Notice, also, that in this model there is no relation between the frequencies of the kHz QPOs and the spin of the neutron star, therefore this model was also applicable to QPOs in black-hole systems. 

\begin{figure}
\centering
\subfloat[]{\label{low-high-PDS}
\includegraphics[trim= -1cm 1cm 0 2cm, clip, width=0.4\textwidth,  angle=90]{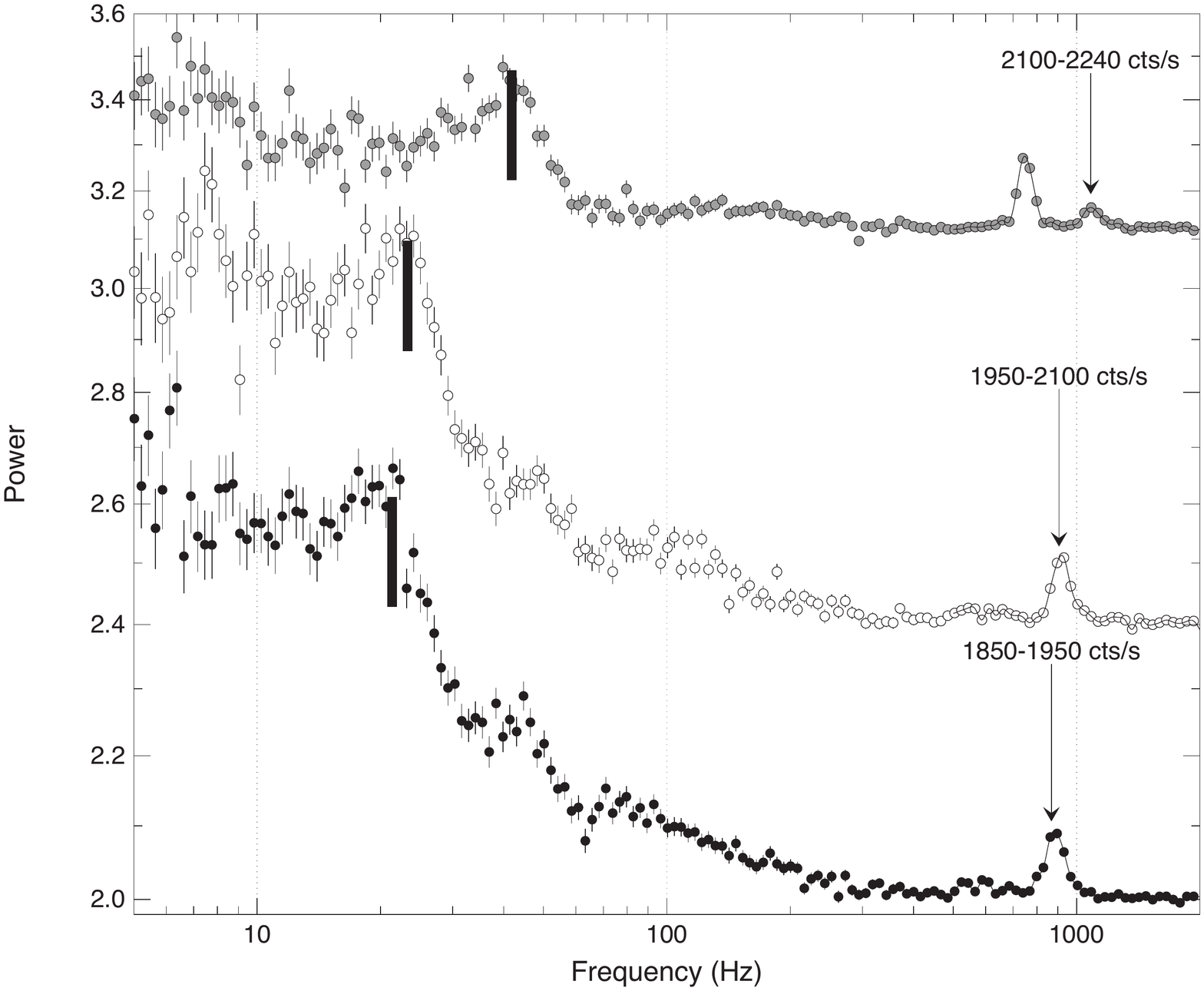}}
\subfloat[]{\label{fig:low-high-freq}
\includegraphics[trim=0 4cm 0 0, clip, width=0.403\textwidth, angle=0]{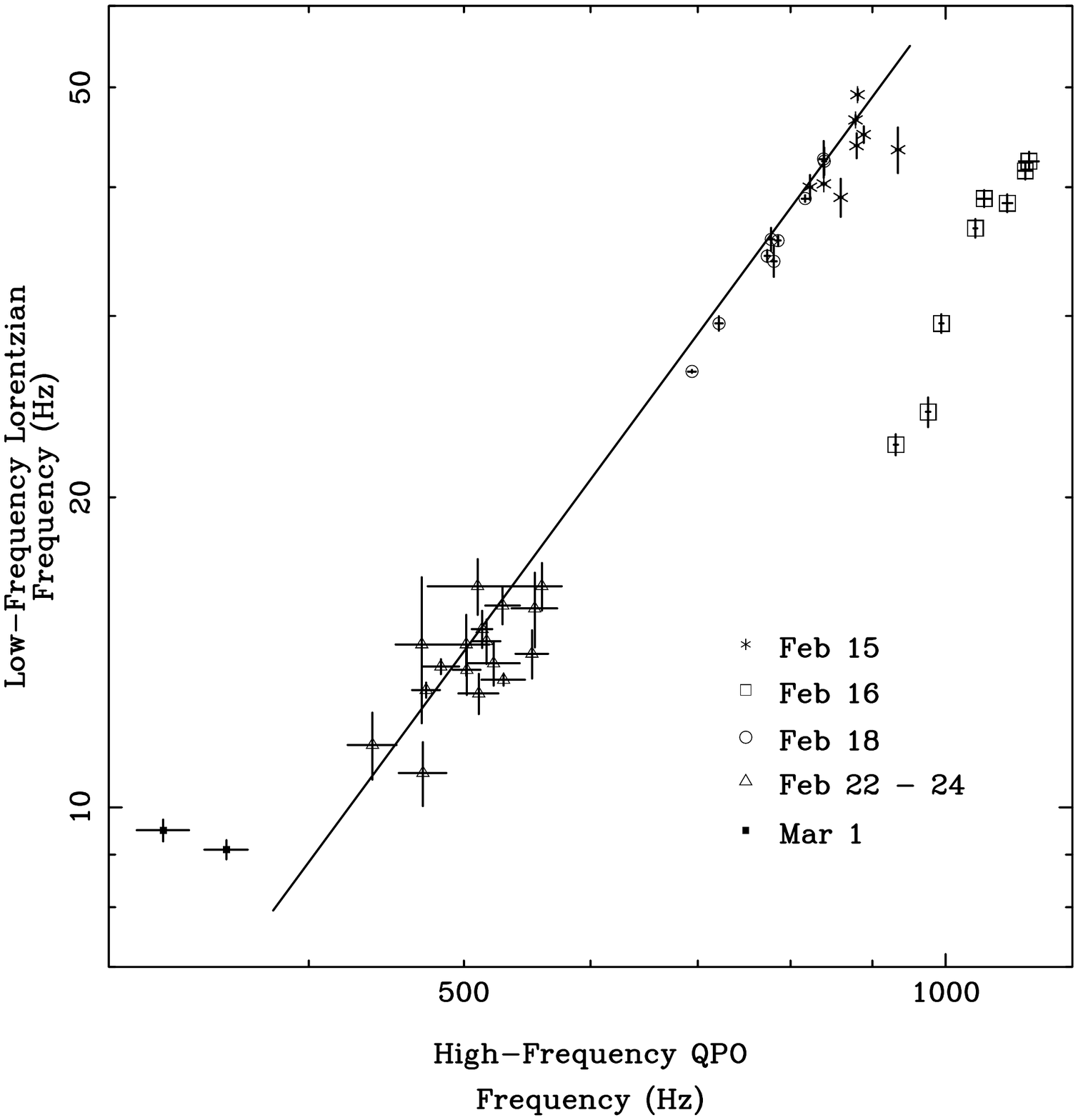}}
\caption{Left: Power spectra of three separate observations of 4U 1728--34 with the low-frequency and upper kHz QPO indicated \citep[the plot is based on the data published in][]{Strohmayer-1996}. Right: Plot of the frequency of the low-frequency QPO vs. that of the upper kHz QPO in 4U 1728--34. The solid line is the best-fitting power-law relation to the data \cite[originally published as Figure 2 in][]{Ford-1998}.
\label{fig:ratedep}}
\end{figure}

If the kHz QPOs and the low-frequency QPOs are all GR frequencies in the disc (but notice that the calculations assume test particles, so no disc hydrodynamics), another prediction of this model is that the frequency of the low-frequency QPO should be proportional to the square of the frequency of the upper kHz QPO. Figure~\ref{low-high-PDS} shows the PDS of three separate observations of 4U~1728--34 in which the low-frequency and upper kHz QPOs are marked with vertical lines. Figure~\ref{fig:low-high-freq}, on the other hand, shows the relation between the frequency of the low-frequency QPO and that of the upper kHz QPO in this same source, with the line corresponding to the best-fitting power to the data with index of $2.11\pm0.11$. We will expand on models in \S\ref{sec:models}

One can extract useful information about the mechanism that causes the QPOs from the $Q$ factor, if one has a model to explain its behaviour with QPO frequency. As shown in Figure~\ref{fig:Q-nu-Barret}, in 4U~1636--53 the $Q$ factor of the lower kHz QPO first increases as the QPO frequency increases, it reaches a maximum at $\nu_{\rm low}$$\sim$$800-850$ Hz, and drops rather abruptly as the frequency of the QPO continues to increase. This same behaviour was observed in all sources for which enough data were available. The rapid drop at high frequencies was interpreted as the inner radius of the accretion disc reaching closer and closer to the ISCO (see \S\ref{sec:basic}), where the faster and faster radial drift of the material in the disc towards the neutron star causes a drift of the QPO frequency over the lifetime of the process that produces that QPO, hence broadening the observed QPO peak, and reducing $Q$. We will discuss this effect, and other alternatives, in \S\ref{sec:vs-frequency}.

\begin{figure}
\centering
\includegraphics[width=0.6\textwidth, trim=2cm 2cm 4cm 3cm, clip, angle=0]{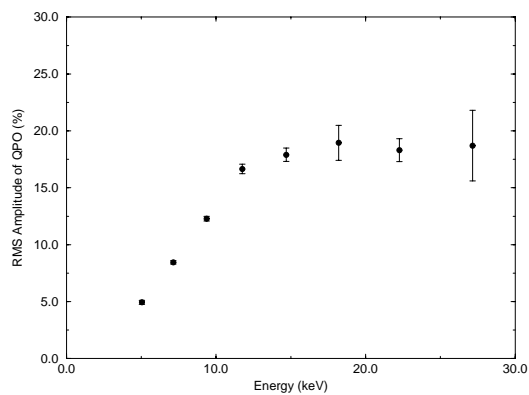}
\caption{Fractional rms amplitude as a function of energy for the lower kHz QPO in 4U~1608--52 \citep[originally published as Figure 4 in][]{Berger-1996}.}
\label{fig:rms-E-Berger}
\end{figure}

The other parameter of the Lorentzian function in eq.~\ref{eq:lor} is the rms amplitude, equal to $\sqrt{N}$ in that equation. Already the initial observations showed that the spectrum of the variability is hard. In other words, the fractional rms amplitude of both kHz QPO increases with energy. For instance, in 4U~1608--52 and 4U~1636--53, the rms amplitude of the lower kHz QPO at $\sim 25$ keV is $\sim 20$\%, while the rms of the upper kHz QPO in these two sources increases a bit less steeply with energy, reaching $\sim 12$\% at $\sim 20$ keV. In Figure~\ref{fig:rms-E-Berger} we show the rms spectrum of the lower kHz QPO in 4U~1608--52.

The soft thermal component, which is the combined emission from the neutron-star surface and the accretion disc, in the time-averaged X-ray energy spectrum of these sources peaks at $\sim$$3$$-$$6$ keV and drops very rapidly as the energy increases. Therefore, the contribution of the disc and the neutron-star surface to the total emission at energies higher than $\sim$$10$$-$$15$ keV is always negligible and, even if the disc or the neutron-star surface were oscillating with an rms amplitude of 100\%, their contribution to the observed fractional rms amplitude at and above those energies would be totally negligible. This shows that, while the {\em dynamical} process that determines the frequency of the QPOs could take place in the disc, like in the models described above, the {\em radiative} process that modulates the source emission at the QPO frequency cannot come from either the neutron star or the disc. At those energies, the dominant spectral component is the corona, in which highly energetic electrons transfer energy to the soft photons emitted form the neutron star and the disc via inverse Compton scattering, redistributing those photons into a power-law shaped component in the energy spectrum. We will discuss this further in \S\ref{sec:rms-amplitude}.

Finally, a property of the kHz QPOs (and any other variable signal) that is not represented in eq.~\ref{eq:lor} is the energy-dependent phase lag (or, equivalently, time lag) of the signal. To understand the phase lag one needs to go back to the Fourier analysis of a signal. The power spectrum that we described at the beginning of this section, is the modulus square of the complex Fourier transform of the light curve of the source as a function of frequency or, equivalently, the product of the Fourier transform of the signal by its complex conjugate. If, instead, one multiplies the Fourier transform of a signal by the complex conjugate of another signal, both functions of frequency, the result is the cross-spectrum. In the same way that the power spectrum measures the variance of the signal per unit frequency (through Parseval's theorem), the modulus and the argument of the cross-spectrum measure, respectively, the covariance per unit frequency and the phase difference, also called phase lag, $\Delta \phi$, between the two signals as a function of Fourier frequency. In the same way that the power spectrum gives the degree of correlation of a light curve with itself, the autocorrelation of the light curve, the cross-spectrum gives the degree of correlation of one light curve with the other, the cross-correlation between the two light curves.  
If the signals are uncorrelated, at each Fourier frequency the covariance and the phase lag will be on average 0. (Notice that uncorrelated signals give a 0 phase lag, but a 0 phase lag does not imply that the signals are uncorrelated.) At any given frequency, $\nu$, the phase lag can be converted into a time lag, $\displaystyle \Delta t(\nu) = \frac{\Delta\phi(\nu)}{2\pi\nu}$. The phase lags are defined\footnote{Phase lags equal to $\Delta \phi \pm 2n\pi$, with $n$ any integer number, cannot be distinguished from a phase lag $\Delta \phi$.} from $-\pi$ to $\pi$, while the time lags run between $-1/(2\nu)$ and $1/(2\nu)$. 
Since both quantities are related, depending on the context, we will either use the term phase or time lags to refer to the delay between the two light curves in the Fourier space.

If the two light curves used to compute the cross-spectrum come from two different energy bands, the time lag at each Fourier frequency represents the time delay between the light curves in those two energy bands at each Fourier frequency. For a QPO (and any other somewhat broad component) with a centroid frequency $\nu_0$ and a FWHM $\Delta$, we call the phase (or time) lag of the QPO to the average of the phase (or time) lags over a frequency range around the centroid frequency of the QPO, e.g. from $\nu_0-\Delta$ to $\nu_0+\Delta$. 
It is customary to take the light curve at the lowest energy band as the {\em reference band} and to measure the phase lag, with respect to reference band, of the light curve in the bands, called {\em subject bands}, at energies above the energy of the reference band. Under this convention, a positive phase/time lag, also called hard lag, indicates that the hard light curve lags (follows after) the soft one, whereas a negative phase/time lag, when the soft light curve leads (comes before) the hard light curve, is called soft lag. Alternatively, one can use the full band as the reference band, and narrow bands within the full band to measure the lags, provided that one corrects for the correlation introduced by the part of the signal that is both in the subject and the reference bands. In the end one obtains the energy dependent phase lags, of the subject bands with respect to the reference band, over the frequency range in which the QPOs dominate the variability of the source. 

Because the lower kHz QPO is usually narrower and, therefore usually more significantly detected, than the upper, the first measurements of lags where obtained for the lower kHz QPO. The magnitude of the time lags of the lower kHz QPO in 4U~1608--52 \citep{Vaughan-1997b, Vaughan-1998} and 4U~1636--53 \citep{Kaaret-1999} was $\Delta$$t$$\sim$$20$$-$$25$$\mu$s, constraining the size of the region where the lags are produced to $c\Delta$$t \simless 10$ km. A remarkable fact of those detections was that the lags of the lower kHz QPO in these two sources were soft, contrary to the expectation if the lags were produced by inverse Compton scattering in the corona, since in that case the low-energy photons escape from the system before the photons that are up-scattered in the corona to high energies.

Fifteen years passed before new measurements of the lags of the kHz QPOs were published. In that period the number of sources with kHz QPOs, and the number of detections of QPOs in individual sources, covering a broad range of frequencies, allowed for more detailed studies of the lags as a function of energy and QPO frequency \citep{deAvellar-2013, Barret-2013}. At the same time, this also allowed to measure, for the first time, the lags of the upper kHz QPO \citep{deAvellar-2013, Barret-2013, Peille-2015, deAvellar-2016, Troyer-2018}.

\begin{figure}
\centering
\subfloat[]{\label{fig:lags-E-avellar}
\includegraphics[width=0.49\textwidth, trim=3cm 2cm 5cm 2.5cm, clip, angle=0]{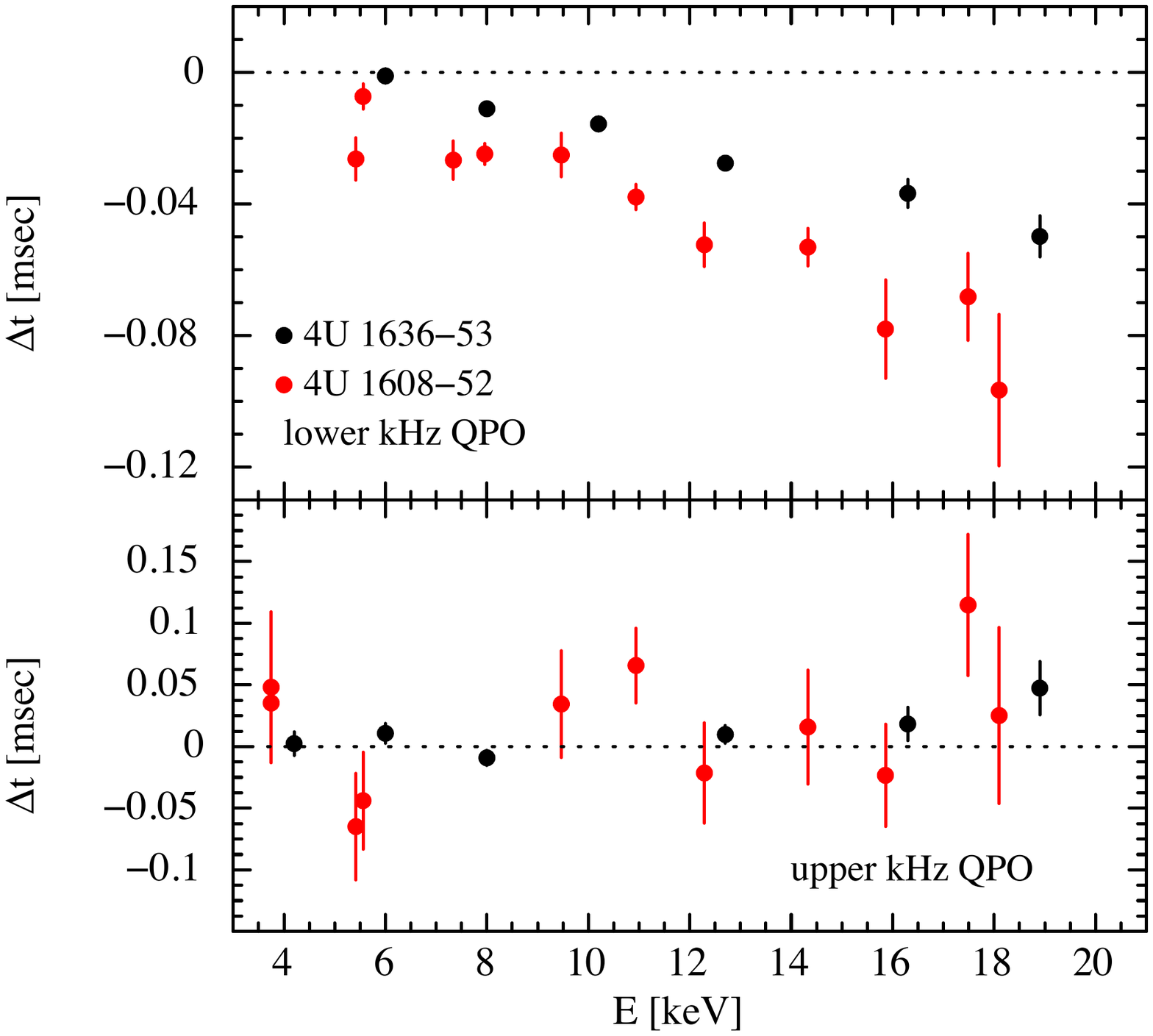}}
\subfloat[]{\label{fig:lags-nu-avellar}
\includegraphics[width=0.49\textwidth, trim=3cm 2cm 5cm 2.5cm, clip, angle=0]{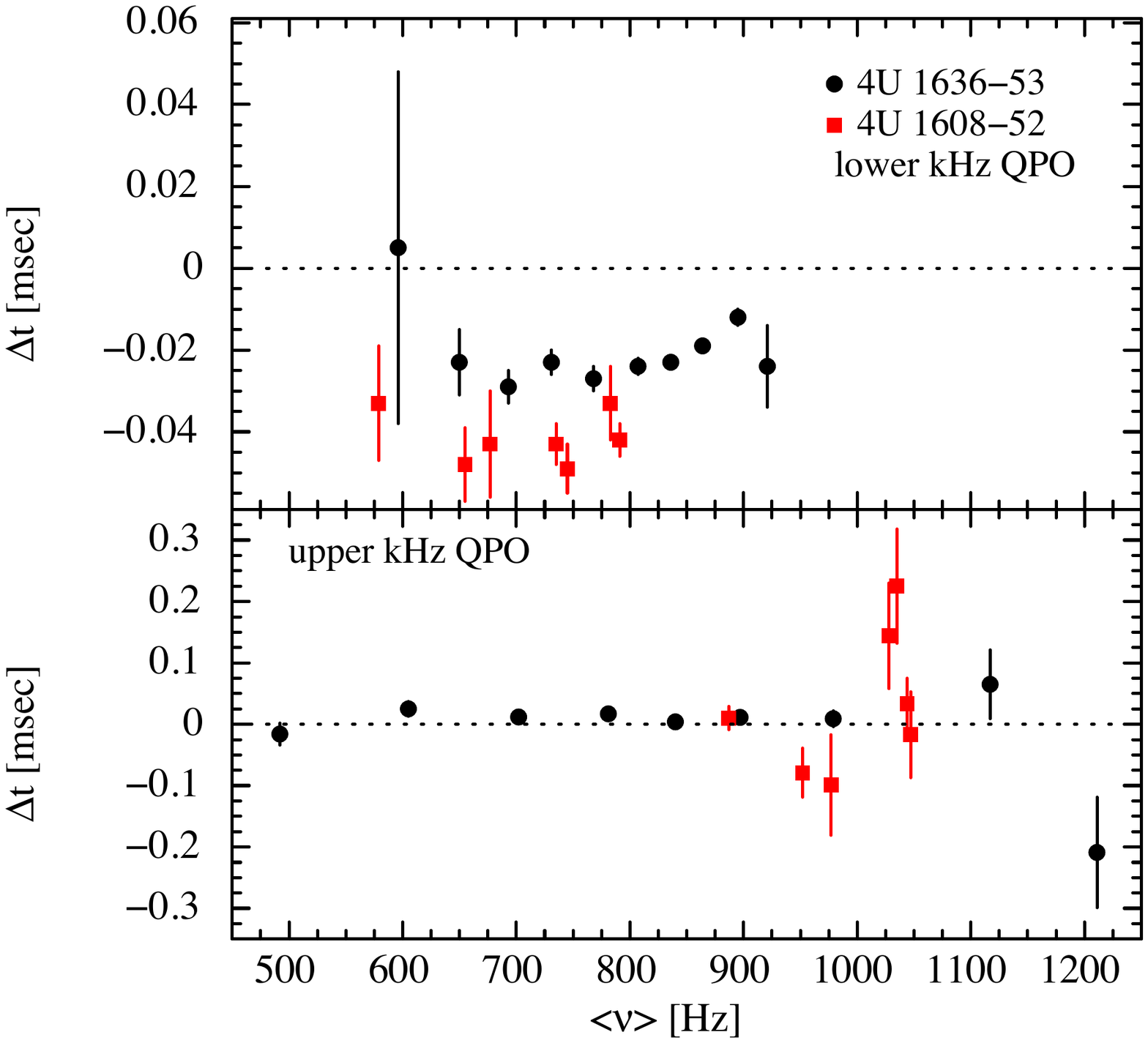}}
\caption{Time lags as a function of energy (left) and frequency (right) for the lower (top panels) and the upper (bottom pane4ls) kHz QPOs in 4U~1636--53 and 4U~1608--52 \citep[adapted from][]{deAvellar-2013}}.
\label{fig:lags-avellar}
\end{figure}

Figure~\ref{fig:lags-E-avellar} shows the lags of the lower and the upper kHz QPO in 4U~1608--52 and 4U~1636--53 as a function of energy. The lags of the lower kHz QPO in both sources are soft and become softer as the energy increases, whereas the lags of the upper kHz QPO are either consistent with zero or increase slightly with energy. 
Figure~\ref{fig:lags-nu-avellar} shows the lags measured between two broad energy bands for the lower and the upper kHz QPO in the same two sources as a function of frequency. The magnitude of the lags of the lower kHz QPO in 4U~1636--53 first increases and then decreases as the frequency of the QPO increases, whereas the lags of the upper kHz QPO remain more or less constant at zero. The lags of the lower and upper kHz QPOs in 4U~1608--52 have larger error bars, but their dependence upon QPO frequency is consistent with that of 4U~1636--53.
These two plots show that two different radiative mechanisms operate to produce the lags (and, as we saw, the rms amplitude) of the lower and the upper kHz QPO, and this, in turn, provides valuable information for models that try and explain these phenomena. We will come back to this in \S\ref{sec:lags}.


\section{Linking observed frequencies with theoretical expectations} 
\label{sec:models}

The initial reports of the detections of the first kHz QPOs in Sco~X-1 and 4U~1728--34 already put forward the suggestion that the observed frequency could be the Keplerian frequency at the inner edge of the disc. The IAU Circulars with those reports stated: ``The high-QPO frequency, and its increase with mass-transfer rate, suggest that we may be seeing the keplerian frequency at the inner edge of the disk near the magnetospheric boundary, or its beat frequency with a slower (about 100 Hz) pulsar.'' \citep{vanderKlis-1996},  and ``Explanations in terms of either keplerian frequencies or a beat-frequency model cannot yet be ruled out, although no evidence has yet been seen for a coherent pulsar frequency in the same data." \citep{Strohmayer-1996a}.

About two months later the first detection of coherent pulsations in a neutron star during an X-ray burst, the so-called burst oscillations (see the Chapter by Patruno \& Watts in this book for more details on this phenomenon), was announced. The IAU Circular with that report said: ``Of the seven bursts that we have observed from 4U~1728--34 during a recent campaign with RXTE, five show oscillations with a frequency of 363 Hz.'', and continued: ``We have also found in the same data set two simultaneously-present kHz quasiperiodic oscillations (QPOs), one of which has been reported earlier (IAUC 6320). The centroid frequencies of the two QPOs change with intensity and time, but their difference appears to be always near 363 Hz. These observations are consistent with a neutron star spin period of 2.75 ms.'' \citep{Strohmayer-1996b}.

These results set up the stage for the idea \citep{Miller-1998} of a beat-frequency model of the kHz QPOs. (Although published in 1998, the idea was first presented in detail by the same authors in a preprint in 1996, arXiv:astro-ph/9609157.) In this model, called the sonic-point beat-frequency model or, for short, the sonic-point model, the upper kHz QPO is a beaming oscillation produced by a hot spot on the neutron-star surface. This spot is the footprint of a stream of matter falling from material at the sonic radius (see \S\ref{sec:qpo101}) onto the neutron star. When this stream hits the star it heats a small area producing a footprint that rotates around the surface of the star at the same frequency as that of the material in the disc, at the sonic radius, where the stream originates. When the neutron star rotates, radiation from the pole(s) illuminates periodically the part in the disc where the stream starts and, because of radiation drag, increases momentarily the rate of mass that is injected into the stream and falls onto the neutron star. When this extra amount of material hits the neutron-star surface at the footprint of the stream, the temperature of the spot increases. The emission from the footprint is therefore modulated at a frequency that is equal to the Keplerian frequency at the sonic radius minus the neutron-star spin frequency. In this model, the luminosity modulation of the hot spot produces the lower kHz QPO. \citep[Please read][for the full explanation of the model]{Miller-1998}.

One obvious conclusion of this scenario is that the frequency difference between the kHz QPOs, which is equal to the neutron-star spin frequency\footnote{As explained in \S\ref{sec:qpo101}, in a modified version of the sonic-point model the frequency difference between the kHz QPOs can also be equal to half the neutron-star spin frequency \citep{Lamb-2003}.}, has to remain constant when the QPO frequencies move (\S\ref{sec:qpo101}). This was the case for most of the sources in which kHz QPO had been detected, except for Sco~X-1 \citep{vdk-1997}, in which the frequency difference decreased systematically as the QPO frequencies increased together. While this result posed a problem to the sonic-point model, the situation could be explained if the clumps in the disc, where the stream originates, spiralled in due to the strong radiation drag in this bright source \citep{Lamb-2001}. In this case, the frequency difference could be less than the neutron-star spin frequency and decrease as the QPO frequencies increased. The situation became even more difficult for the sonic-point model when this effect was observed in more sources, all much weaker than Sco~X-1 \citep{Mendez-1998, Mendez-1998b, Mendez-1999b} and, especially, when the difference of the QPO frequencies in some observations of 4U~1636--53 \citep{Jonker-2002b} turned out to be larger than half the neutron-star spin frequency in this source, something that could not be explained in the sonic model and its subsequent extensions.

Almost at the same time, a different model that could explain the dependence of $\nu_{\rm upp} - \nu_{\rm low}$ vs. the frequency of the QPO was proposed. As in the sonic-point model, this relativistic-precession model \citep{Stella-1998, Stella-1999} considered that the frequency of the upper kHz QPO is the Keplerian frequency at the inner edge of the disc; but differently from the previous model, in this case the lower kHz QPO would be the periastron precession frequency, equal to the difference between the Keplerian and epicyclic radial frequencies at the inner edge of the disc. In this model the frequency difference between the kHz QPOs is independent of the neutron-star spin and, as can be readily seen from the identification of the lower kHz QPO, should be equal to the epicyclic radial frequency at the inner radius of the disc. This epicyclic frequenciy is 0 at the ISCO, first increases as the radial distance in the disc increases, and then decreases again as the radial distance continues increasing. This implies that the frequency difference between the kHz QPOs should decrease both at high and low QPO frequencies, corresponding to small and large radial distances in the disc. As indicated, this explained the observed decrease of  $\nu_{\rm upp} - \nu_{\rm low}$ with QPO frequencies as the QPO frequencies increase in Sco~X-1 \citep{vdk-1997, Mendez-2000} and other sources \citep{Mendez-1998, Mendez-1998b, Mendez-1999b, Jonker-2002b}, but also predicted a trend at low kHz QPO frequencies for which there were no data at the time. A few years later, the neutron-star LMXB Cir X--1 \citep{Boutloukos-2006} showed exactly that (Fig.~\ref{fig:CirX-1-Boutloukos} in \S\ref{sec:qpo101}) and, since other predictions of the model for low-frequency variability had already been validated (Fig.~\ref{fig:low-high-freq} in \S\ref{sec:qpo101}), all this lent support to this model. Notice, however, that the model relies on frequencies of test particles around the neutron star, and therefore does not consider the hydrodynamical effects in the disc that may affect those frequencies. We will come to this again in \S\ref{sec:other}.

A third class of models considers wave patterns in the disc as the cause of the kHz QPOs. These models also rely upon the three basic GR epicyclic frequencies discussed in the previous models (and sometimes also upon the neutron-star spin), but in this case those are not the frequencies of test particles orbiting the neutron star, but characteristic frequencies in a hydrodynamical flow that determine how pressure and gravity waves travel in the disc and, sometimes, lead to other frequencies that are resonances of the basic ones \citep[some examples of those ideas can be found in][but the list is much longer]{Kato-1980, Kato-1990, Nowak-1997, Wagoner-1999, Kluzniak-2004, Kluzniak-2005, Lai-2009}. Among these models, one that received some attention \citep{Abramowicz-2001, Abramowicz-2003} argued that a resonance in the disc appears when the ratio of two of the epicyclic frequencies discussed above is the ratio of two small integer numbers, e.g. $2$$:$$3$. Such a preferred frequency ratio was reported for the kHz QPOs in Sco~X-1 \citep{Abramowicz-2003}, and the model gained popularity because, in two cases in which two simultaneous high-frequency QPOs were observed in black-hole systems, those QPOs appear at frequencies that are in a $2$$:$$3$ ratio \citep[e.g., at $300$ Hz and $450$ Hz in the black-hole LMXB GRO~J1655--44; see][]{Remillard-1999, Strohmayer-2001}. In essence, this resonance model is equivalent to the example of a double pendulum discussed in books of Mechanics \citep[e.g.][]{Landau-1976} with, in this case, a mechanism that couples two oscillating phenomena in the disc. The report of a $2$$:$$3$ frequency ratio of the kHz QPOs in Sco~X-1 has been subsequently disputed \citep{Miller-2004, Belloni-2005, Belloni-2007b}, but the model is still considered for high-frequency QPOs in black-hole LMXBs. 

The models described in this section, and most of the models of the kHz QPOs that appeared in the last 20 years, aim at explaining the frequencies of the oscillations and, in that sense, are dynamical models of the QPO phenomenon. Very few models have attempted to give an explanation of the other, radiative, properties of the QPOs. We will discuss those radiative properties of the kHz QPOs in \S\ref{sec:beyond}, and we will also mention some of the latest attempts to try and explain those properties.


\section{QPO frequency correlations}
\label{sec:freq-freq}

\begin{figure}
\centering
\includegraphics[width=0.6\textwidth, clip]{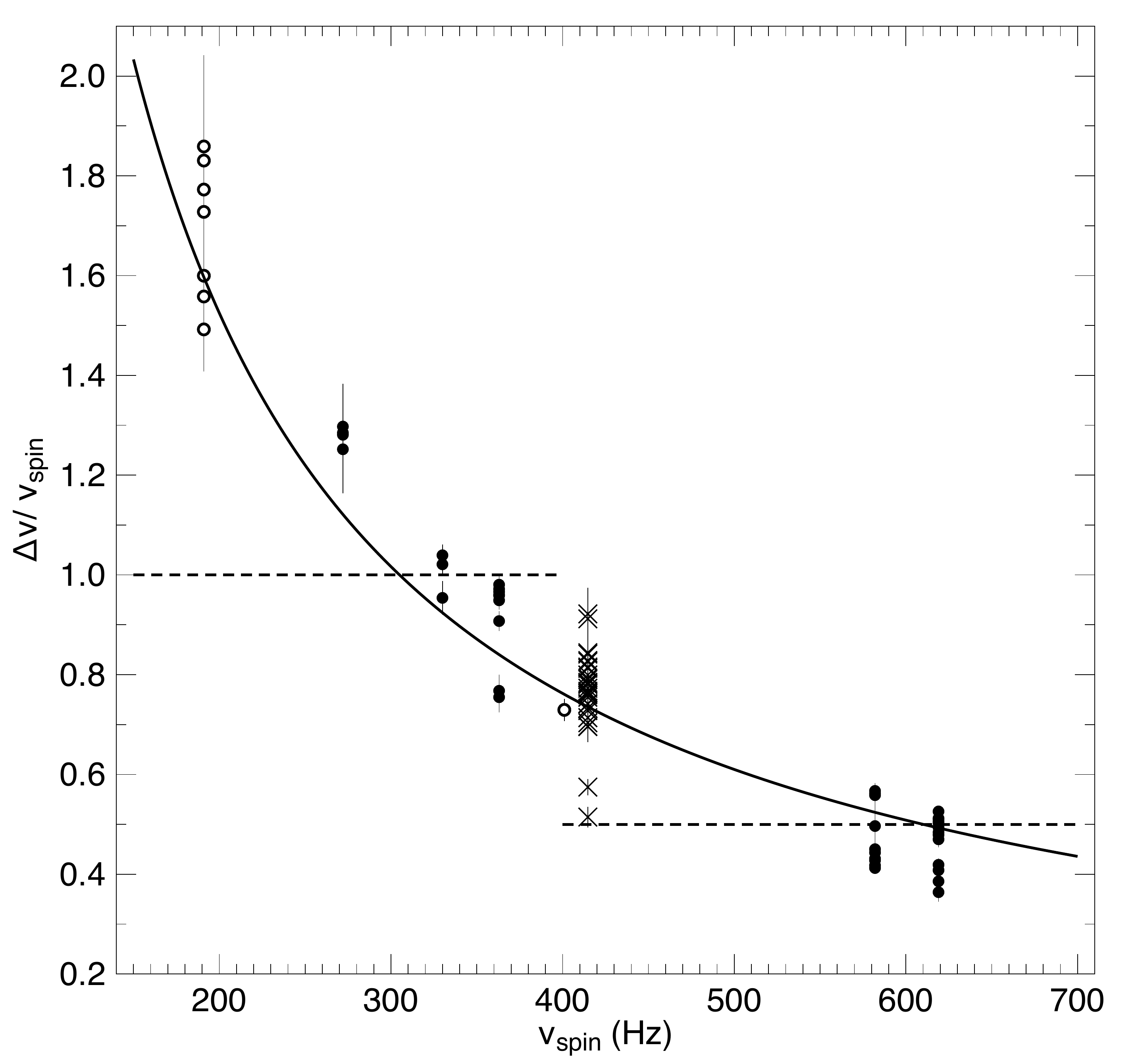}
\caption{$\Delta\nu / \nu_{\rm spin}$ vs.$\nu_{\rm spin}$ for all sources for which both kHz QPOs and pulse periods are available \citep[adapted from][]{Mendez-2007}. The $\Delta\nu$ values for accreting millisecond pulsars (empty circles) have been multiplied by 1.5 (see text). The X symbols correspond to 4U 0614+09. The solid line corresponds to a constant $\Delta\nu$$=$$305$ Hz, the dashed line is 1 until 400 Hz, then 0.5.
}
\label{fig:deltanunu}
\end{figure}

Although the beat frequency model is unable to explain all observations, it is reasonable to think that the kHz QPOs could in some way be connected to the rotation of the neutron star. When the first source for which both burst oscillations and kHz QPOs were discovered, 4U 1728-34, it was realised that $\Delta\nu$ was around the same value as the burst oscillation frequency \citep{Strohmayer-1996b}. However, the next source with a double detection was 4U 1636-63, where $\Delta\nu$$\sim$$270$ Hz and the burst oscillation frequency was 581 Hz, close to twice that value. After then, every time a new source showed kHz QPOs and had an estimate of the spin period either through burst oscillations or through a direct detection in the case of accreting millisecond pulsars, it turned out that the latter were close to $\Delta\nu$ or half of it. More specifically, if the $\nu_{spin}$ was slower than $\sim$$400$ Hz, $\Delta\nu$$\sim$$\nu_{\rm spin}$, if it was faster $\Delta\nu$$\sim$$\nu_{\rm spin}/2$. The symbol $\sim$ here is to be intended as ``close to'', since $\Delta\nu$ is not constant for any particular source, but varies over a range. However, it was later realised that the data are also compatible with $\Delta\nu$ being essentially constant around 305 Hz \citep{Mendez-2007}, especially after multiplying the kHz QPO frequencies of accreting millisecond pulsars by 1.5, as suggested by an offset in the correlation with the low-frequency QPO frequencies \citep{vanStraaten-2005, Linares-2005}. The situation can be seen in Fig. \ref{fig:deltanunu}. Notice that the spin period of 4U 0614+09 was discovered after the original version of this plot was published and its $\Delta\nu$ values fall on the constant-$\Delta\nu$ track rather than the $\Delta\nu$$\sim$$\nu_{\rm spin}/2$ one.

\begin{figure}
\centering
\includegraphics[width=0.8\textwidth, clip]{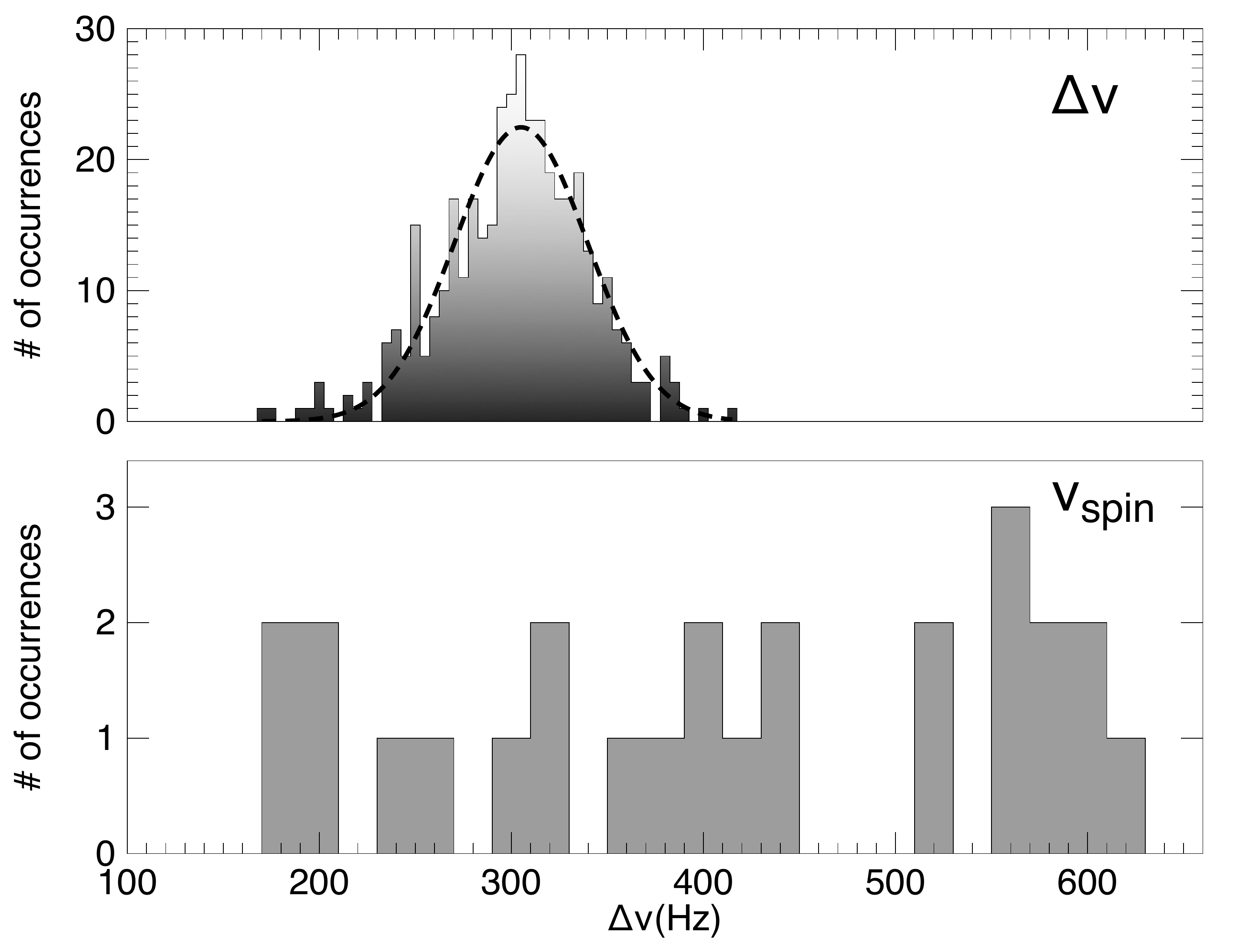}
\caption{Top: distribution of all available $\Delta\nu$ tabulated in the literature. The dashed line is a Gaussian fit, which yields a centroid of 305 Hz. 
Bottom: distribution of pulse periods for neutron-star LMXBs, both from accreting millisecond pulsars and burst oscillations. An earlier version of this figure can be found in \citep{Mendez-2007}.
}
\label{fig:deltanu_comparison}
\end{figure}

It is interesting to compare the distribution of all $\Delta\nu$ values available in the literature and the distribution of detected (or derived from burst oscillations) spin periods (see chapter by Patruno \& Watts, this book) as shown in Fig. \ref{fig:deltanu_comparison}. The distribution of $\Delta\nu$ values, coming from a large number of sources, peaks around 300 Hz and is well approximated by a Gaussian with centroid 305 Hz. The distribution of pulse periods, obviously less populated, is rather flat between 200 Hz ad 600 Hz. From these data, it appears that the kHz QPOs are not related to the spin period of the neutron star, although in a number of sources $\Delta\nu$ does increase towards $\nu_{\rm spin}$ with decreasing $\nu_{\rm upp}$  \citep[e.g.][but see \citep{Jonker-2002b}]{Mendez-1998, Mendez-1998b, Mendez-1999b} . 

\begin{figure}
\centering
\includegraphics[width=0.6\textwidth, clip]{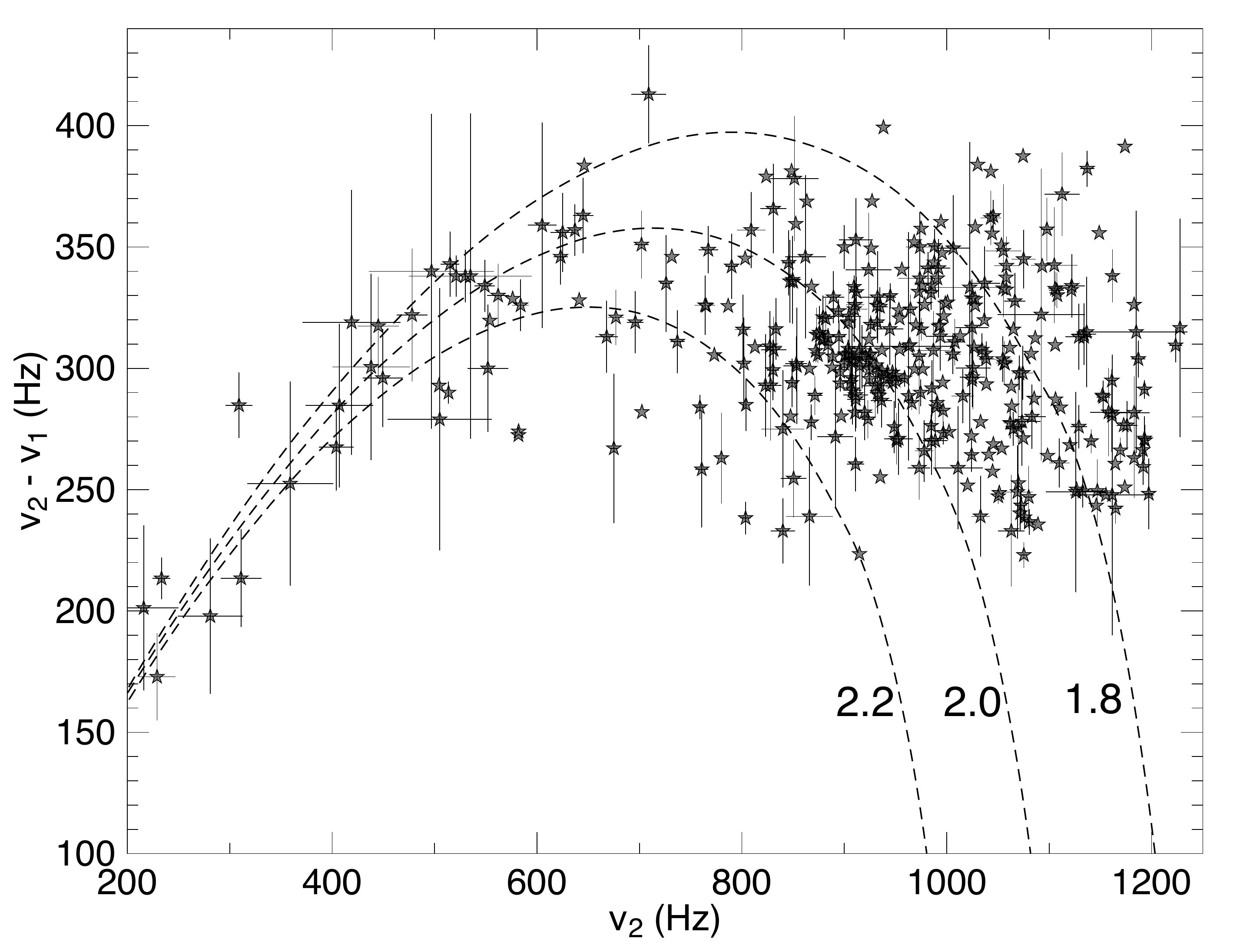}
\caption{$\Delta\nu$ vs. $\nu_{\rm upp}$ for all values published in the literature. The three curves correspond to the prediction of the relativistic-precession model for a mass of 1.8, 2.0 and 2.2 solar masses.
}
\label{fig:new_stella_vietri}
\end{figure}

Going back to the correlations between kHz QPO frequencies and theoretical models, it is interesting to produce an updated version of the plot shown in Figure \ref{fig:CirX-1-Boutloukos} \citep[originally shown for a few sources in][]{Stella-1999}, which gives $\Delta\nu$ vs. $\nu_{\rm upp}$, where all published values from RXTE are shown (the same values used for the top panel of Fig. \ref{fig:deltanu_comparison}). They can be seen in Figure \ref{fig:new_stella_vietri}. Notice that a prediction of the relativistic-precession model is that, for these masses, $\Delta\nu$ should not exceed $\sim$$400$ Hz, which indeed is what is observed. 

\begin{figure}
\centering
\includegraphics[width=0.6\textwidth, clip]{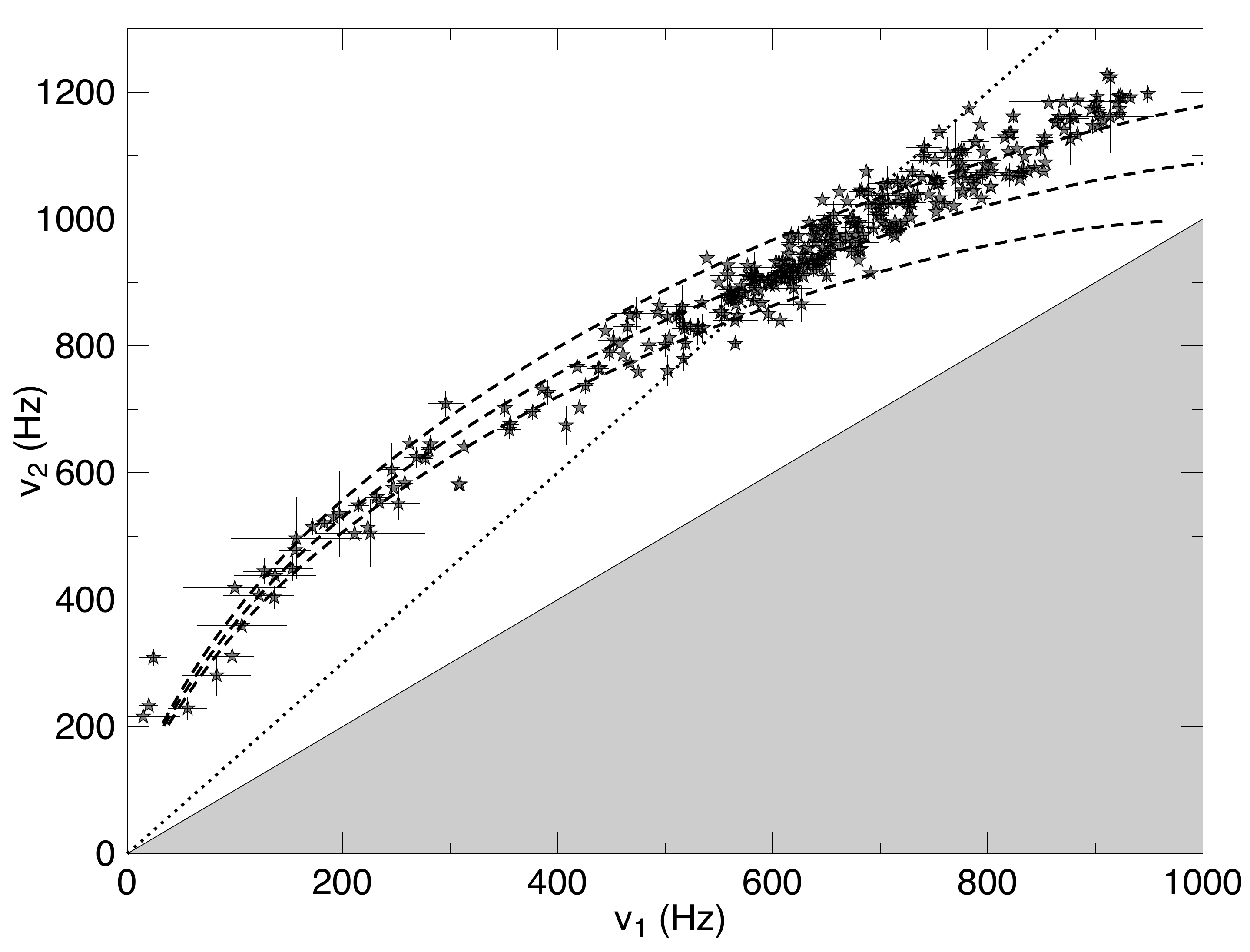}
\caption{The frequency of the upper kHz QPO, $\nu_{\rm upp}$, vs. that of the lower kHz QPO, $\nu_{\rm low}$, for all values published in the literature. The gray area is obviously not allowed. The dashed lines are the predictions of the relativistic-precession model for a neutron-star mass of 1.8, 2.0 and 2.2 solar masses. The dashed line is $\nu_{\rm upp}$$=$$1.5 \nu_{\rm low}$.
}
\label{fig:nucorr}
\end{figure}

However, when dealing with pairs of values, in this case $\nu_{\rm low}$ and $\nu_{\rm upp}$, it is best to plot them one versus the other. This was done in \citep{Mendez-2007}; in Figure~\ref{fig:nucorr} we show a new version of that plot with all published values included (the same values used for Fig.~\ref{fig:new_stella_vietri}). The predictions of the relativistic-precession model for a neutron-star mass of 1.8, 2.0 and 2.2 solar masses (dashed lines) fit rather well the distribution of points at low frequencies, but diverge slightly at high frequencies, as can also be seen from Figure~\ref{fig:new_stella_vietri}. Moreover, a constant 3:2 ratio, shown by the dotted line, fails to represent the data. What is important to note is that, despite the fact that the plot contains points from a number of different sources, the overall correlation is rather good. This suggests that the process that gives rise to the signal at these frequencies is not strongly dependent on other parameters of the sources, like the spin period.

\section{Relation between properties of the kHz QPOs and parameters of the energy spectrum}
\label{sec:spectral-frequency}

If kHz QPOs are produced in the accretion flow close to the neutron star, one would expect that properties of that accretion flow will affect the properties of the QPOs. For instance, the frequencies would be related to the radius of the disc, obtained from spectral fits, if the QPOs reflect the Keplerian frequency at that radius while, if the photons that oscillate at the QPO frequency are up-scattered in the corona, the fractional rms amplitude and the phase lags of the QPO would depend on the optical depth and the electron temperature of the corona. One should keep in mind that, to recover a potential relation between timing and spectral parameters, one needs to study both the energy and the power spectrum of a source over time scales that are comparable to (or preferably shorter than) the time scales over which the properties of the accretion flow change. 

\begin{figure}[t]
\centering
\includegraphics[width=1\textwidth, trim=1cm 2cm 1cm 0cm, clip, angle=0]{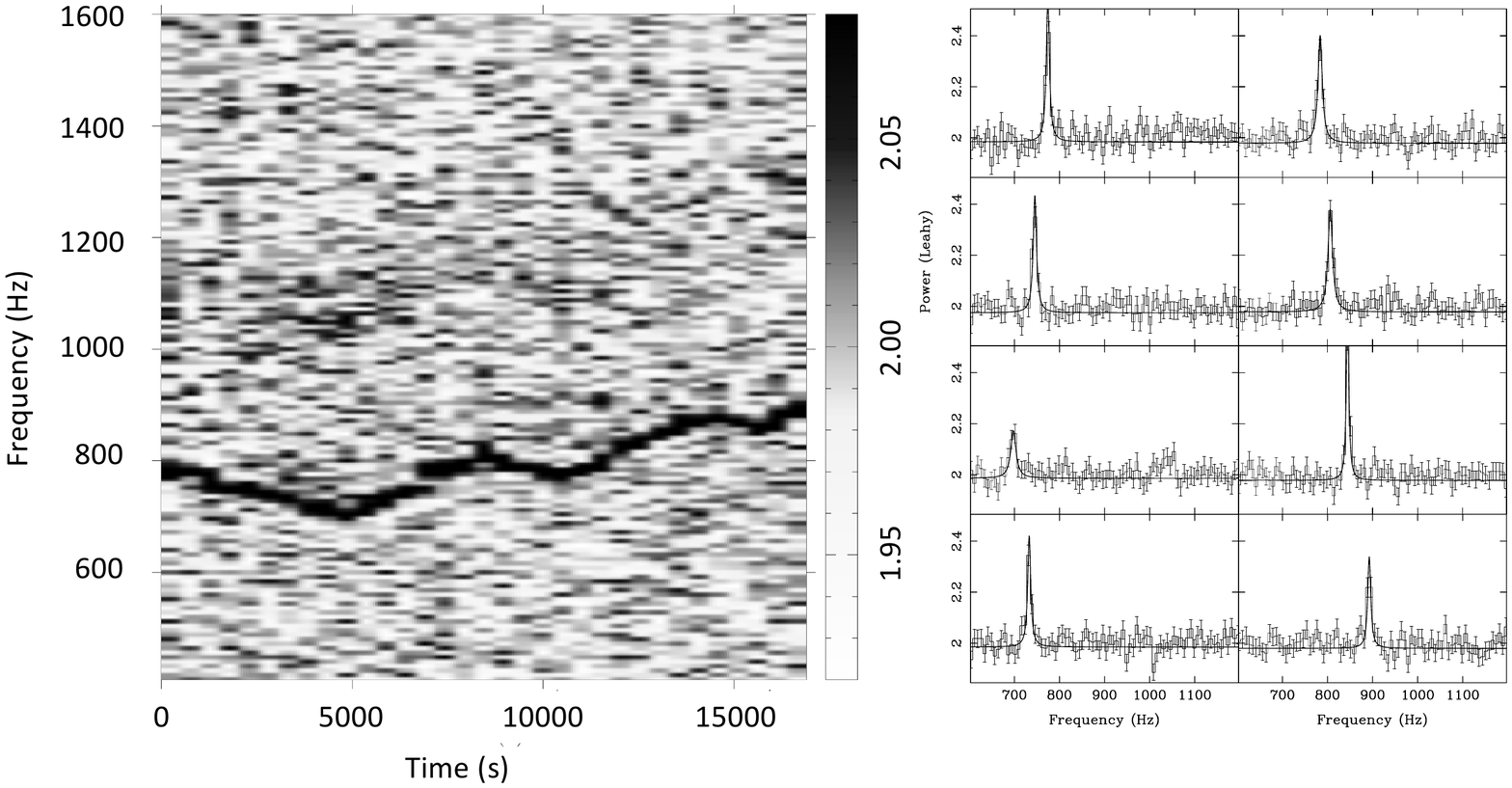}
\caption{Dynamical power spectrum of an observation of 4U~1728--34 (left), and six power spectra from that same observation (right) computed over intervals of $\sim$$2000$ seconds \citep[adapted from][]{Mendez-1999}.}
\label{fig:Dyn-PDS_Mendez}
\end{figure}

Following the discussion in \S\ref{sec:basic}, a perturbation in the accretion disc travels through the disc over the viscous time scale, which in these systems is of the order of hundreds of seconds \citep[e.g.][]{Berger-1996}. This is also the time scale over which the frequency of the kHz QPOs was observed to change in the power spectrum of some of this sources. For instance, the left panel of Figure~\ref{fig:Dyn-PDS_Mendez} shows the dynamical power spectrum of an observation of 4U~1728--34 \citep{Mendez-1999}. In a dynamical power spectrum one plots time in the $x$ axis (in this case $t$$=0$ corresponds to the start of the observation), Fourier frequency in the $y$ axis (the plot shows only the frequencies above $\sim$$400$ Hz to focus on the kHz QPOs), and the power density in the $z$ coordinate (plotted with colours). The dark track in the dynamical power spectrum is the lower kHz QPO in this source. As it is apparent in the plot, the frequency of the QPO changes by $\sim$$100$ Hz over time scales of a few thousand seconds. The right panel of Figure~\ref{fig:Dyn-PDS_Mendez} shows power spectra of six contiguous time intervals within that same observation, with the changes of the QPO frequency, going from $\sim$$700$ Hz to $\sim$$900$ Hz over the period of the observation, visible in the individual power spectra. If one wants to compare, for instance, the frequency of the QPO with the inner radius of the accretion disc, one needs to match the length of the observations used for the comparison with intervals over which the QPO frequency is more or less constant.

\begin{figure}
\centering
\includegraphics[width=0.55\textwidth, trim=0cm 2.5cm 0.5cm 2cm, clip, angle=90]{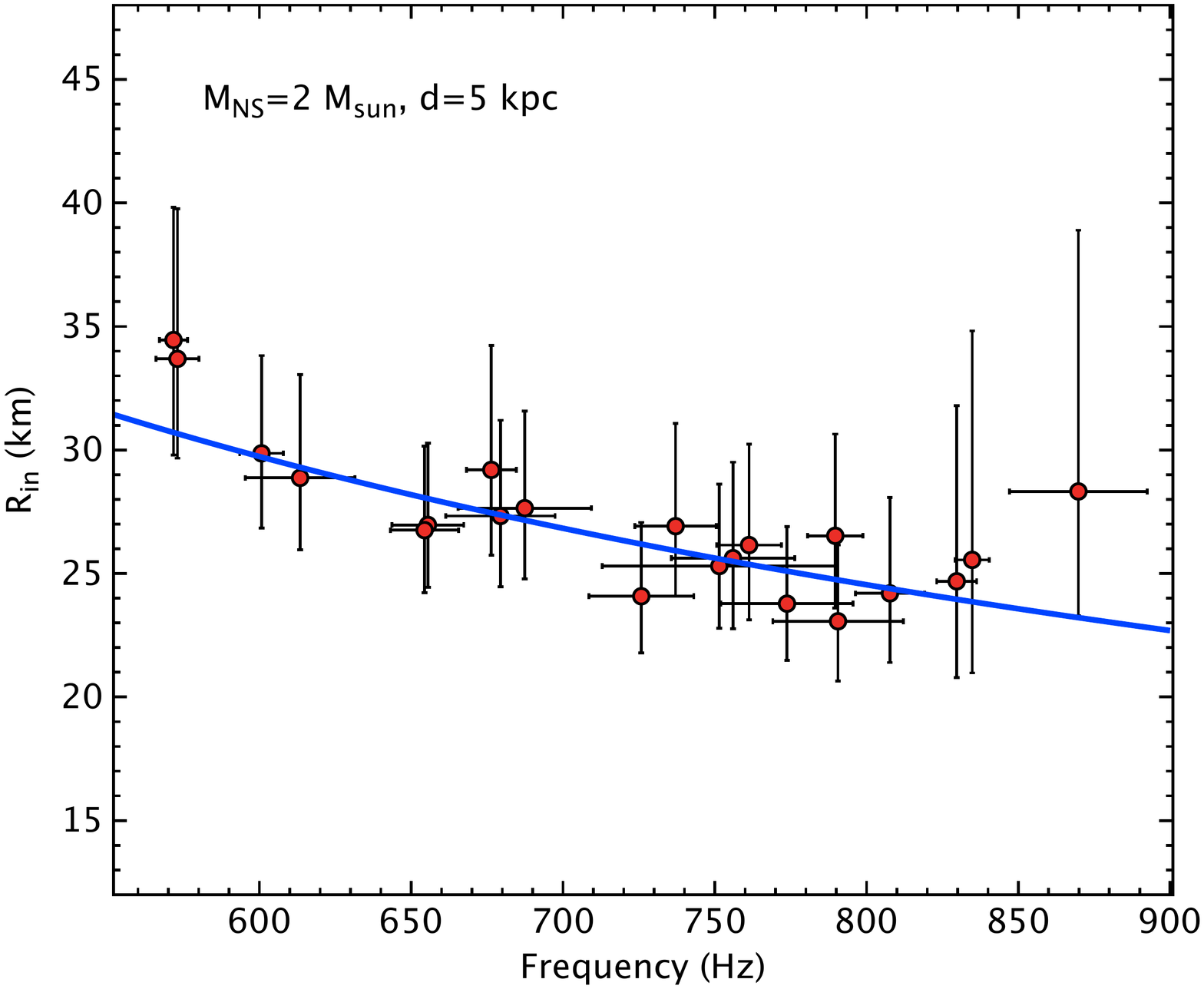}
\caption{Inner radius of the accretion disc, from fits to the energy spectra, as a function of the frequency of the lower kHz QPO, from fits to the power spectra, in 4U~1608--52 \citep[originally published as Figure 5 in][]{Barret-2013}.}
\label{fig:R_in-nu_Barret}
\end{figure}

Figure~\ref{fig:R_in-nu_Barret} shows the inner radius of the accretion disc as a function of the frequency of the lower kHz QPO in 4U~1608--52 \citep{Barret-2013}. The solid line in the plot is the radius as a function of the Keplerian orbital frequency around a $2$-$\Msun$ neutron star. As expected, if the QPO frequency is equal to the orbital frequency at that radius, the radius decreases as the QPO frequency increases. Notice, however, that the match of the orbital frequency as a function of the radius with the QPO frequency in that Figure would imply that, contrary to what most models propose (see \S\ref{sec:qpo101}), the lower, not the upper, kHz QPO would reflect the Keplerian frequency at the inner disc radius.

\begin{figure}
\centering
\includegraphics[width=0.9\textwidth, trim=0cm 0cm 0 0cm, clip, angle=0]{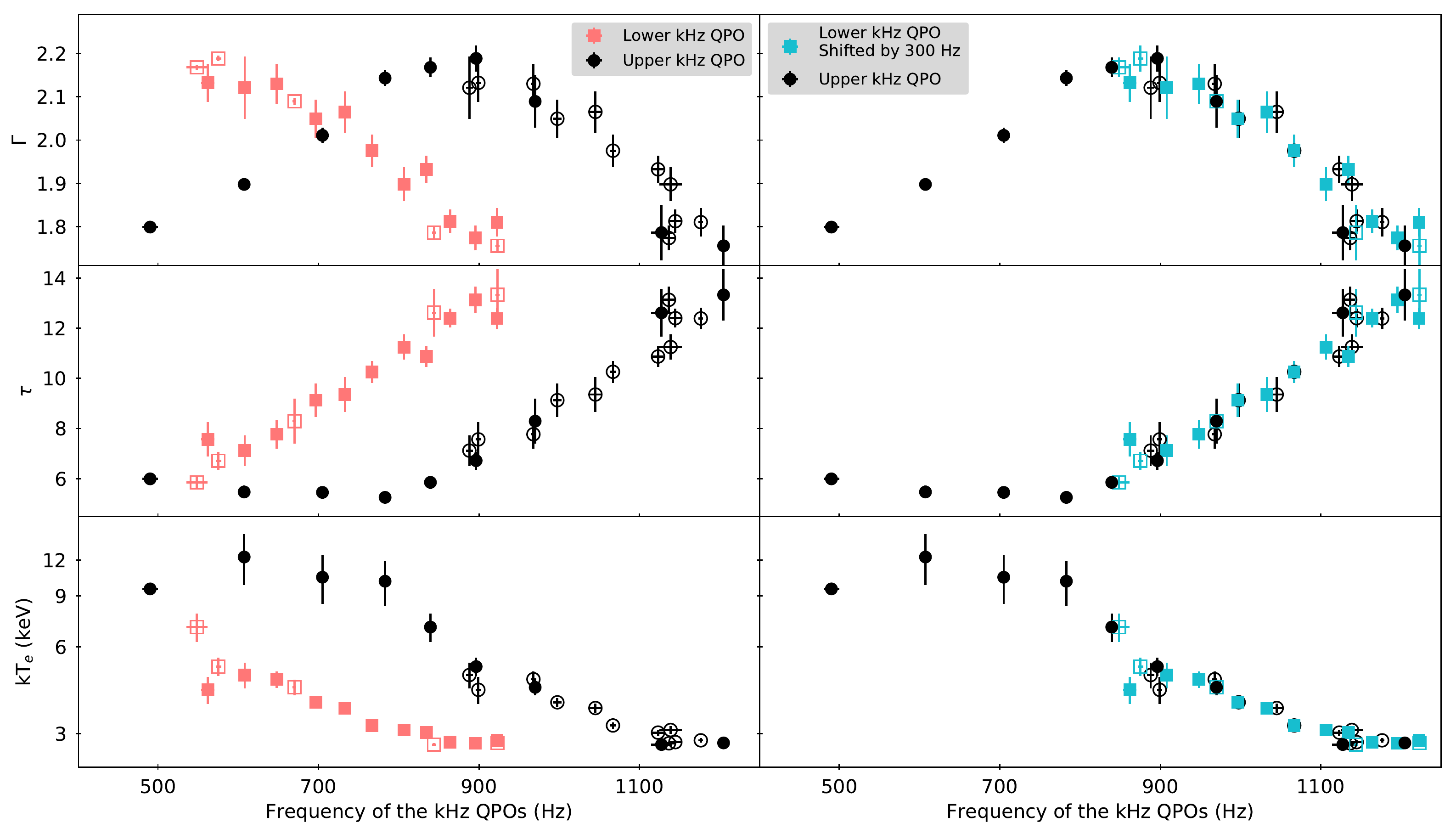}
\caption{Left: Spectral parameters of the lower and the upper kHz QPOs in 4U~1636--53 as a function of the QPO frequency. Right: Same as left panel, but with the frequency of the lower kHz QPO shifted up by 300 Hz \citep{Ribeiro-2017}.}
\label{fig:spectral-pars_QPO-freq-Ribeiro}
\end{figure}

Figure~\ref{fig:spectral-pars_QPO-freq-Ribeiro} shows the spectral parameters of the X-ray corona as a function of the kHz QPO frequencies in 4U~1636--53 \citep[][see also \citep{Kaaret-1998}]{Ribeiro-2019}. The fits to the energy spectra yield $\Gamma$ and $kT_e$, the power-law index and the electron temperature of the Comptonised component, respectively, whereas the optical depth, $\tau$, of the corona is a function of the other two parameters \citep{Sunyaev-1980}. In the left panel the red and black points correspond to, respectively, the lower and the upper kHz QPO. The right panel shows the same parameters but with the frequency of the lower kHz QPO shifted up by 300 Hz. From this Figure it is apparent that there is a smooth relation of the frequency of the QPOs and the parameters of the corona. Given that the corona is driven by the soft photons in the disc, it is no surprise that both the inner disc radius \citep{Barret-2013} and the corona parameters \citep{Kaaret-1998, Ribeiro-2019} change with QPO frequency in a systematic way. The dependence of the rms amplitude of the lower and upper kHz QPOs upon the spectral parameters of the corona \citep[see plots in][]{Ribeiro-2017}, however, do not match in the same way; in other words, one cannot apply a shift to the relation of the rms amplitude of one of the kHz QPOs vs. any of the spectral parameters and make it match the same plot of the other kHz QPO \citep{Ribeiro-2017}. As we discuss below, the same applies to the quality factor and phase lags. The fact that, except for a frequency shift, the relation of the parameters of the corona vs. the QPO frequency is the same for both QPOs, whereas the relation of the rms amplitude is different, indicates that the dynamical mechanism that drives the frequency of both kHz QPOs can be the same, whereas the radiative mechanisms that modulate the QPO signals must be different.

\section{Beyond QPO frequencies}
\label{sec:beyond}
\label{sec:vs-frequency}

\subsection{The fractional rms amplitude of the kHz QPOs}
\label{sec:rms-amplitude}

\begin{figure}
\centering
\subfloat[]{\label{fig:rms-E-EXO}
\includegraphics[width=0.341\textwidth, trim=0 4cm 0 1cm, clip, angle=0]{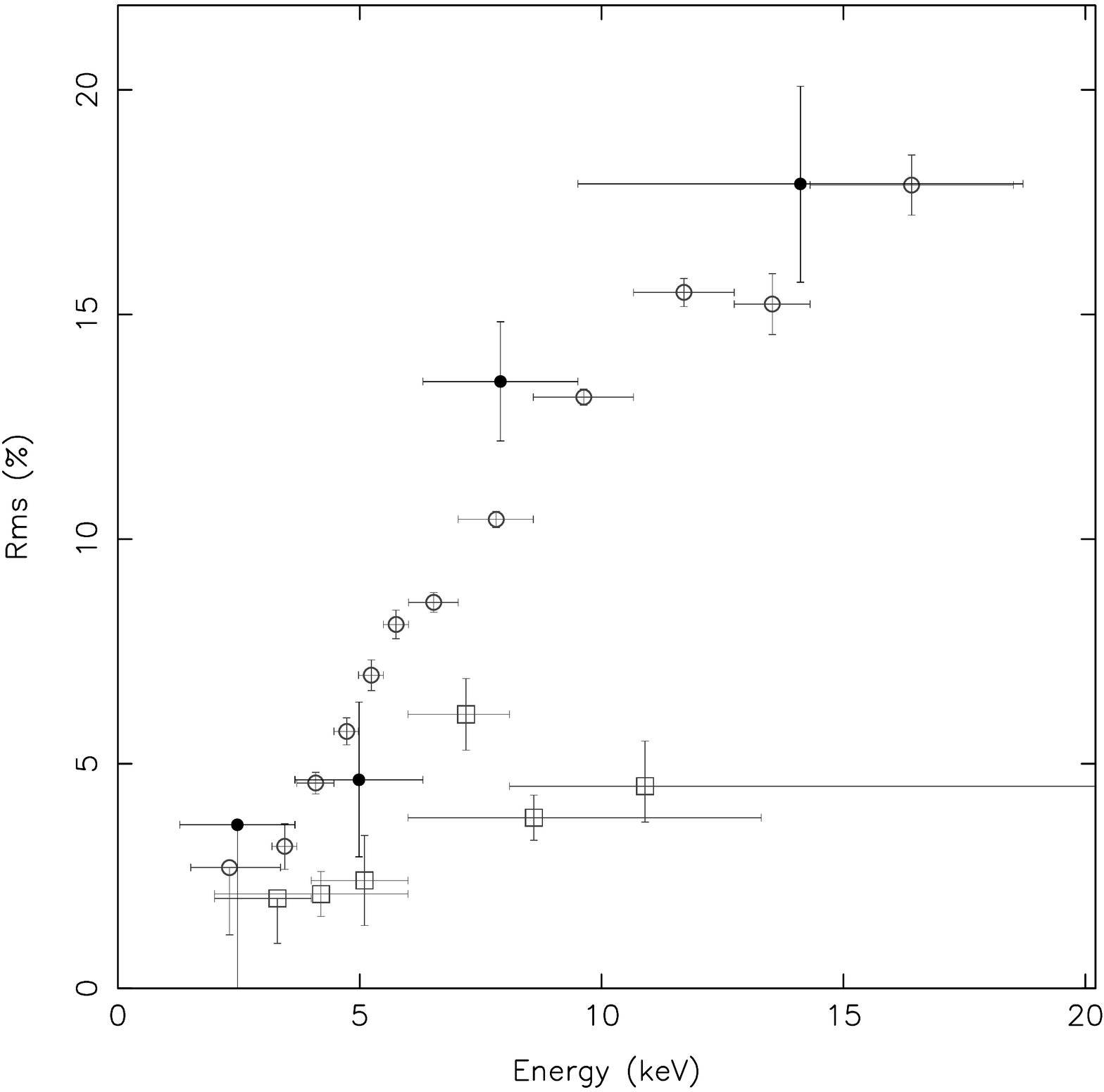}}
\hspace{0.2cm}
\subfloat[]{\label{fig:rms-E-upp-Troyer}
\includegraphics[width=0.605\textwidth, trim=0 15cm 0 0, angle=0]{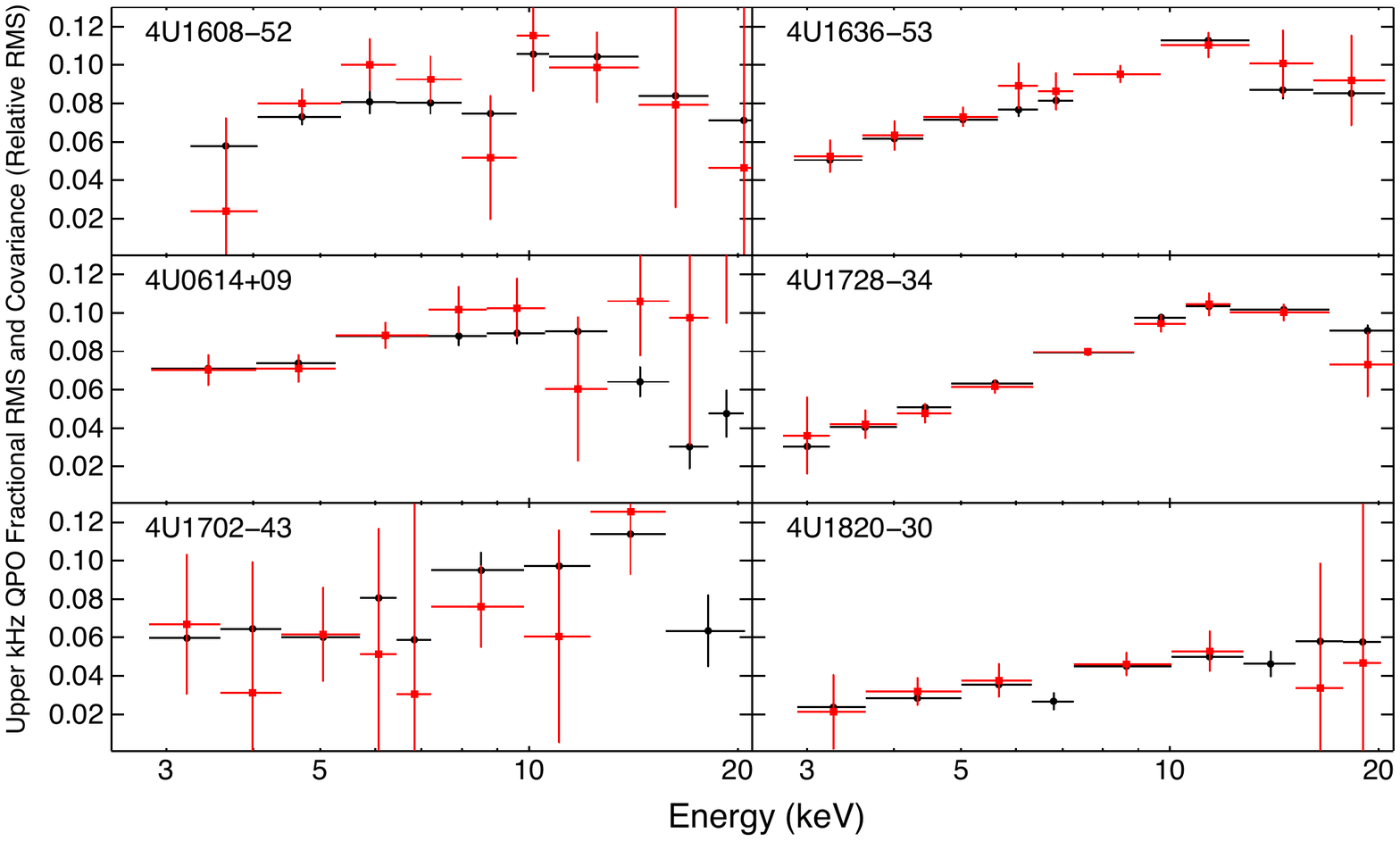}}
\caption{Left: rms spectrum of the single kHz QPO in EXO~0748--676 \citep[originally published as Figure 4 in][]{Homan-2000}, and the lower and upper kHz QPOs in 4U~1608--52 \citep{Berger-1996, Mendez-2001}. Right: rms (red) and covariance (black) spectrum of the upper kHz QPO in six sources \citep[originally published as Figure 5 in][]{Troyer-2018}.}
\label{fig:rms-E}
\end{figure}

For both kHz QPOs, the spectrum of the fractional rms amplitude of the variability is hard. For instance, in 4U~1608--52 the fractional rms amplitude of the lower kHz QPO increases from $\sim 5$\% at $\sim 3$ keV up to $\sim 20$\% at $20-25$ keV \citep{Berger-1996, Gilfanov-2003, Mendez-2001}. A similar trend is seen for the lower kHz QPOs of 4U~1728--34 and Aql~X-1 \citep{Mendez-2001, Mukherjee-2012}, 4U~1636--53 \citep{Ribeiro-2019}, and the only kHz QPO in EXO~0748--676 \citep[see][and Fig.~\ref{fig:rms-E-EXO}]{Homan-2000}. For the upper kHz QPO the trend is similar, although the increase of the fractional rms amplitude with energy is less steep (Figs.~\ref{fig:rms-E-EXO} and \ref{fig:rms-E-upp-Troyer}; but notice that, as we show below, the total rms amplitude and the slope of the rms spectrum of both QPOs depend upon QPO frequency, and therefore one has to consider that to draw conclusions from the comparisons). For instance, for 4U~1608--52 the amplitude of the upper kHz QPO increases from $\sim 2-4$\% at $\sim 3$ keV to $\sim 20$\% at $\sim 20$ keV \citep{Berger-1996, Mendez-2001}. On the other hand, there is no evidence that the frequency or the width of either of the kHz QPOs change with energy \citep{Mukherjee-2012}. 

\begin{figure}[t]
\centering
\includegraphics[width=1\textwidth, trim=0 0 0 0, clip, angle=0]{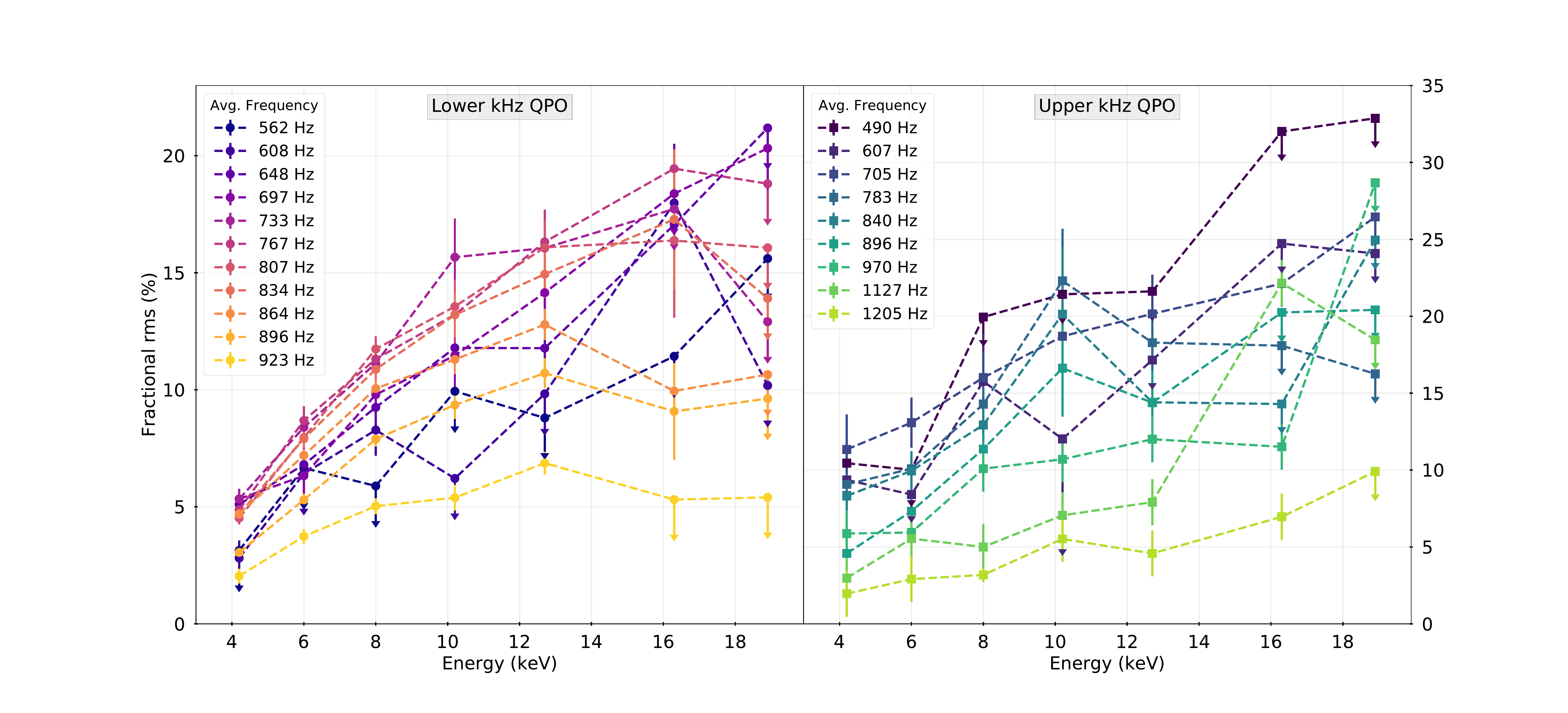}
\caption{Fractional rms amplitude of the lower (left) and upper (right) kHz QPOs as a function of energy for different frequencies of the QPOs. From the plots it is apparent that the slope of the rms-energy relation changes in a systematic way as a function of the frequency of the QPO \citep{Ribeiro-2019}.}
\label{fig:rms-nu-E-Ribeiro}
\end{figure}

\begin{figure}
\centering
\includegraphics[width=0.8\textwidth, trim=0 0 0 0.1cm, clip, angle=0]{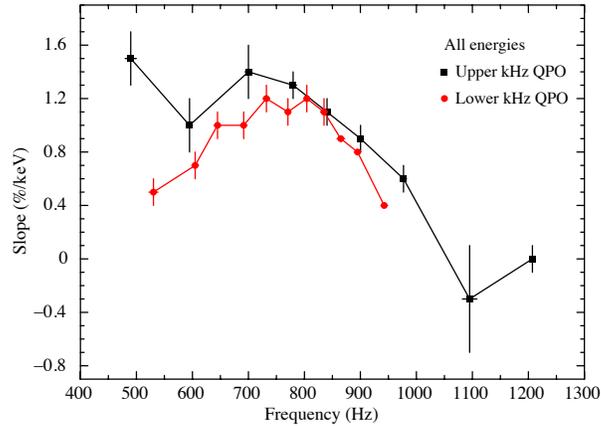}
\caption{Slope of the rms vs. energy relation of the lower (red) and upper (blue) kHz QPO in 4U~1636--53 as a function of the QPO frequency. The slopes were obtained from fits to the data in Figure~\ref{fig:rms-nu-E-Ribeiro}. Compare this Figure with Figure~\ref{fig:rms-nu-Ribeiro} that shows, for the same data, the total rms amplitude vs. QPO frequency \citep[adapted from][]{Ribeiro-2019}.}
\label{fig:rms-slopes-Ribeiro}
\end{figure}

Figure~\ref{fig:rms-nu-E-Ribeiro} shows the rms spectrum of the lower and the upper kHz QPO plotted in different colours for different QPO frequencies. For both kHz QPOs the rms increases with energy (remember Fig.~\ref{fig:rms-E-Berger} showing the rms spectrum of the lower kHz QPO in 4U~1608--52) with, at least for the lower kHz QPO, the rate of increase being faster at low than at high energies. It is also apparent from this Figure that, for the lower kHz QPO, as the QPO frequency increases the slope of the rms spectrum first increases, and then decreases. For the upper kHz QPO the slope of the rms spectrum decreases more or less steadily as the frequency of the QPO increases. 

This is seen more clearly in Figure~\ref{fig:rms-slopes-Ribeiro}, which shows the slope of the rising part of the rms spectrum of both QPOs as a function of the frequency of each QPO. The dependence of the slope of the rms spectrum with QPO frequency is almost exactly the same as that of the total fractional rms (for all energies combined) as a function fo QPO frequency shown in Figure~\ref{fig:rms-nu-Ribeiro}. From this comparison one can conclude that the change of the rms amplitude of the QPOs with frequency is driven by a non-monochromatic change of the rms spectrum, rather than by an energy-independent shift of the rms amplitude of the QPOs at all energies \citep[see][for more details]{Ribeiro-2019}.

\begin{figure}
\centering
\includegraphics[width=0.6\textwidth, trim=0 7cm 0 3.5cm, clip, angle=0]{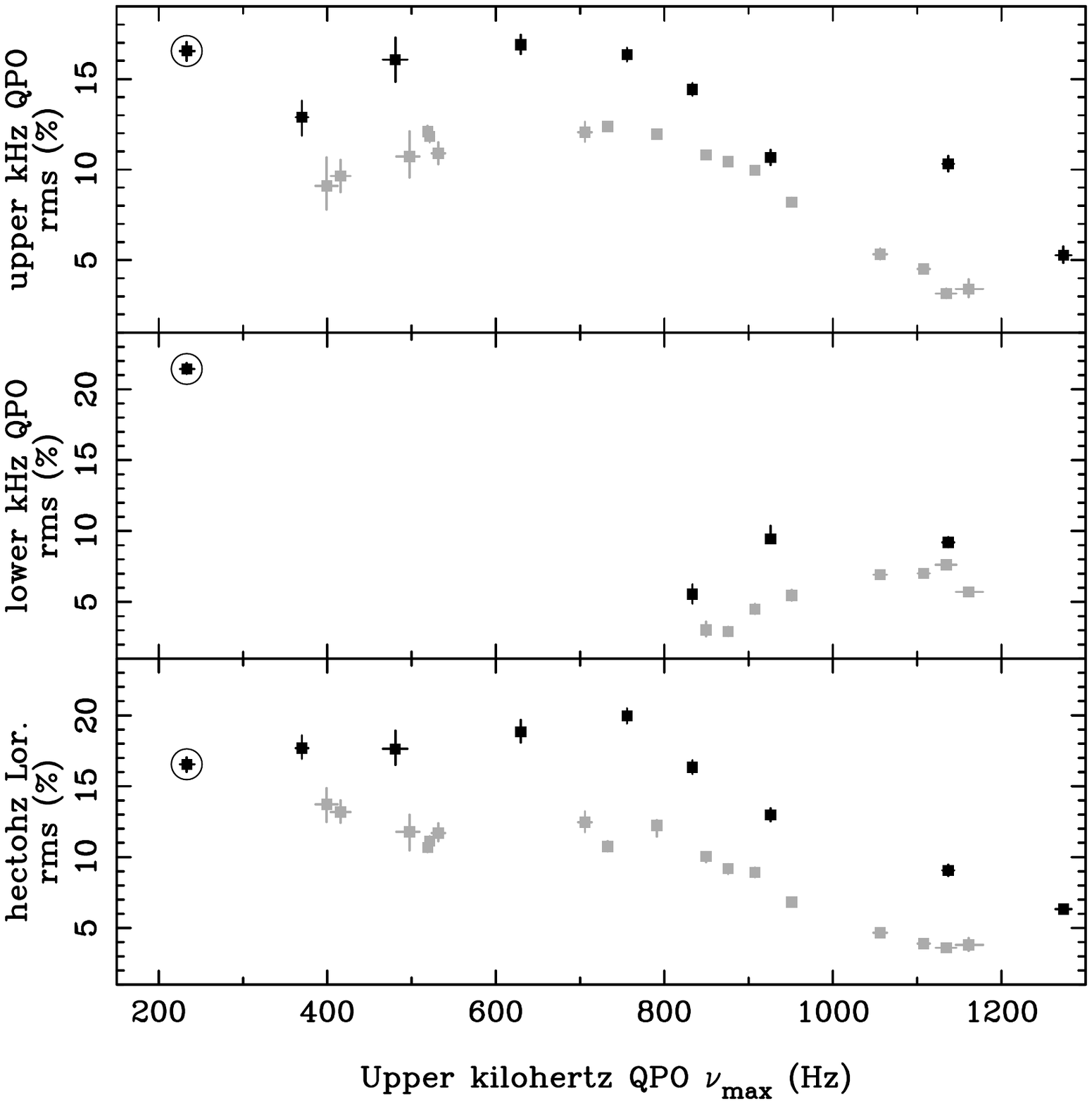}
\caption{Fractional rms amplitude of the lower (upper panel) and lower (middle panel) kHz QPOs as a function of the frequency of the upper kHz QPO in 4U~0614+09 (black points) and 4U~1728--34 (grey points). The bottom panel shows the rms amplitude of the hectohertz (hHz) QPO as a function of the frequency of the upper kHz QPO in these tow sources \citep[originally published as Figure 4 in][]{vanStraaten-2002}.}
\label{fig:rms_vs_freq_vanStraaten}
\end{figure}

On the other hand, the fractional rms amplitude of both kHz QPO changes in a systematic way with the frequency of the QPO. Figure~\ref{fig:rms_vs_freq_vanStraaten} shows the behaviour of the total (for all energies combined) rms amplitude of both kHz QPOs as a function of the frequency of the upper kHz QPO in the atoll sources 4U~0614+09 and 4U~1728--34 \citep{vanStraaten-2002}. Figure~\ref{fig:rms-nu-Ribeiro} showed the same for another atoll source, 4U~1636--53. (Notice that in that Figure the rms amplitude of each kHz QPO is plotted as a function of the frequency of the corresponding QPO itself, whereas in Fig.~\ref{fig:rms_vs_freq_vanStraaten} the rms amplitude of both QPOs is plotted as a function of the frequency of the upper kHz QPO).

As shown in Figure~\ref{fig:QPO_Z}, the case of Z sources is similar, albeit in those cases the custom is to plot all the QPO parameters as a function of $S_Z$, the variable that measures the position of the source along the branches in a colour-colour or hardness-intensity diagram. Bus since the QPO frequencies are correlated with $S_Z$ (see upper panel of Figure~\ref{fig:QPO_Z}), it follows that the rms amplitude of the QPOs in Z sources follows a similar trend as in atoll sources.

\begin{figure}[ht]
\centering
\includegraphics[width=0.7\textwidth, trim=0 2cm 0 1cm, clip, angle=0]{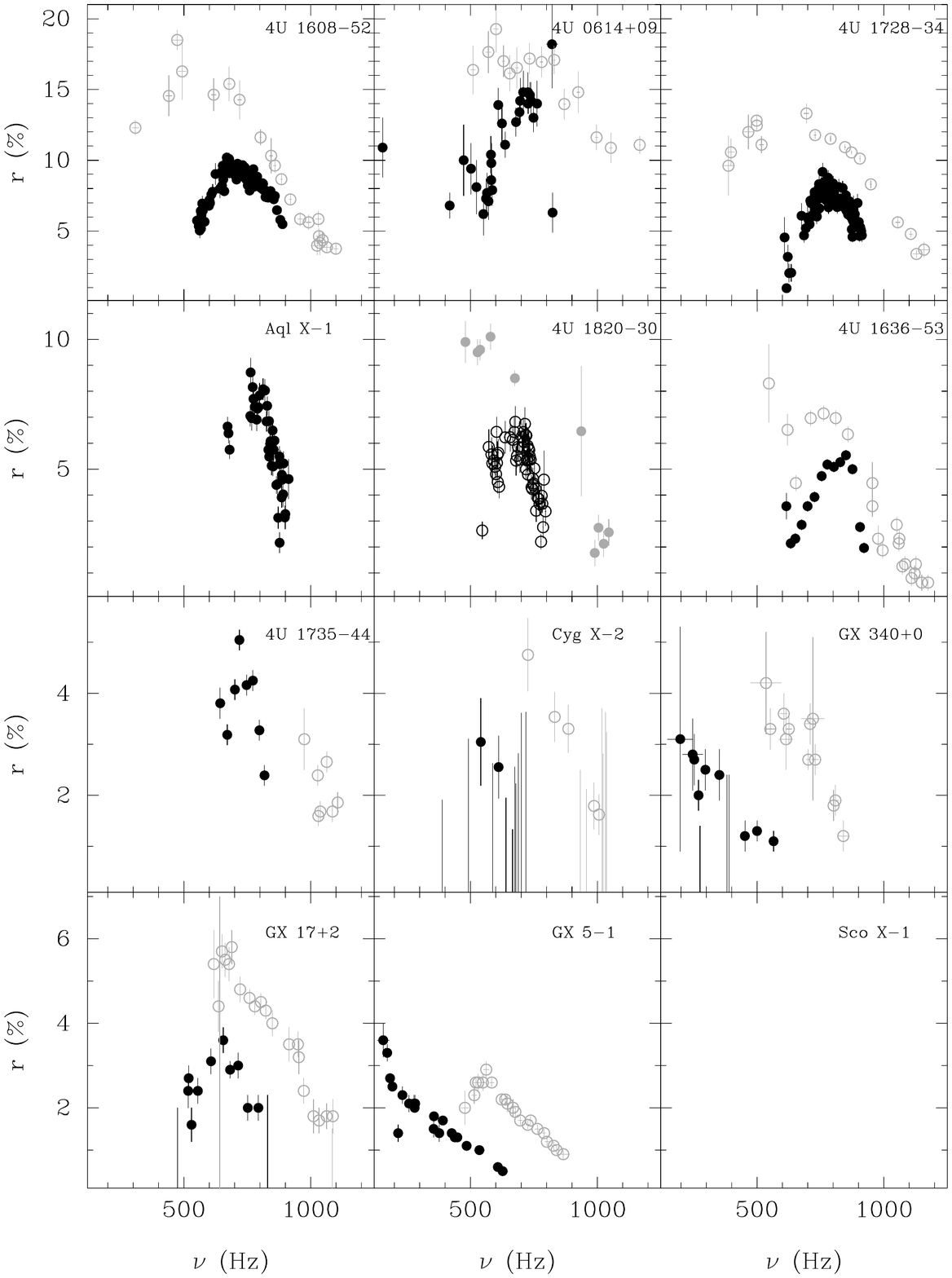}
\caption{Fractional rms amplitude of the lower (filled symbols) and the upper (open symbols) kHz QPOs as a function of the QPO frequency for  seven atoll sources, 4U~1608--52, 4U~0614+09, 4U~1728--34, Aql~X-1, 4U~1820--30, 4U~1636--53, and 4U~1735--44, and four Z sources, Cyg~X-2, GX~340+0, GX~17+2 and GX~5--1 \citep{Mendez-2006}. Because Sco~X-1 is so bright, the detector's dead time made it impossible to measure the fractional rms amplitude of the kHz QPOs in this source. Notice that the scale in the $y$ axis is different for the atoll and the Z sources, because the maximum fractional rms amplitude of the QPOs is generally larger in the former than in the latter.}
\label{fig:rms_all_sources-Mendez}
\end{figure}

At first, at low QPO frequencies, the rms amplitude of the upper QPO increases slightly or stays more or less constant as the QPO frequency increases, and then decreases more or less steadily as the frequency increases further. For the lower kHz QPO the trend is similar, but the rising part when the frequency of the QPO increases, at low QPO frequencies, is steeper than that of the upper kHz QPO. 

In fact, a similar behaviour has been observed in all sources for which enough measurements of the QPOs are available \citep{vdk-1997, Wijnands-1997a, Wijnands-1997b, Ford-1997, Wijnands-1998, Ford-1998, vanStraaten-2000, Jonker-2000, Mendez-2001, DiSalvo-2001, Homan-2002, Jonker-2002, vanStraaten-2002, DiSalvo-2003, vanStraaten-2003, vanStraaten-2005, Altamirano-2005, Barret-2005, Barret-2006, Mendez-2006, Altamirano-2008, Boutelier-2009, Sanna-2010, Barret-2011, deAvellar-2016, Ribeiro-2017, vanDoesboergh-2017, Ribeiro-2019, vanDoesburgh-2019}. Figure~\ref{fig:rms_all_sources-Mendez} shows the fractional rms amplitude of the lower and upper kHz QPO as a function of the QPO frequency for seven atoll and four Z sources \citep{Mendez-2006}. While the trend is the same, the data are noisier in the case of the Z sources, because the QPOs in those cases are weaker (have lower fractional rms amplitude; notice the scale of the $y$ axis in the different panels) and generally broader (see \S~\ref{sec:Q-factor}) than in the atoll sources. Since the spectrum of the Z sources is in general softer than that of the atoll sources \citep[e.g.][]{Christian-1997}, the difference between the rms amplitude of the kHz QPO in the Z and atoll sources suggests that the same mechanism that modulates the oscillations at the QPO frequency sets the shape of the emitted spectrum.   

\begin{figure}
\centering
\subfloat[]{\label{fig:low-rms_E_nu-Ribeiro}
\includegraphics[width=0.49\textwidth, trim=3cm 0cm 0cm 4cm, clip, angle=0]{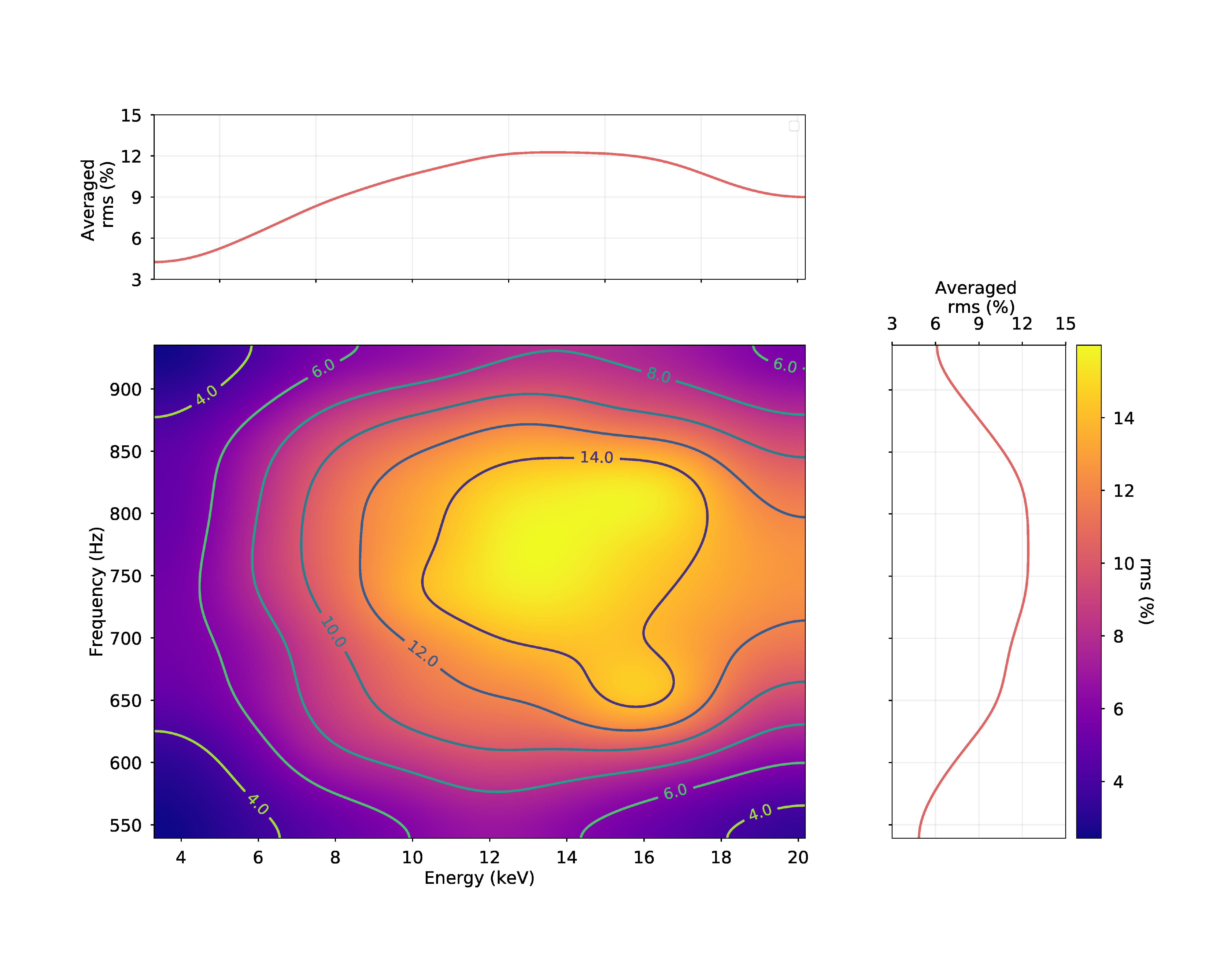}}
\subfloat[]{\label{fig:upp-rms_E_nu_Ribeiro}
\includegraphics[width=0.49\textwidth, trim=3cm 0cm 0cm 4cm, clip, angle=0]{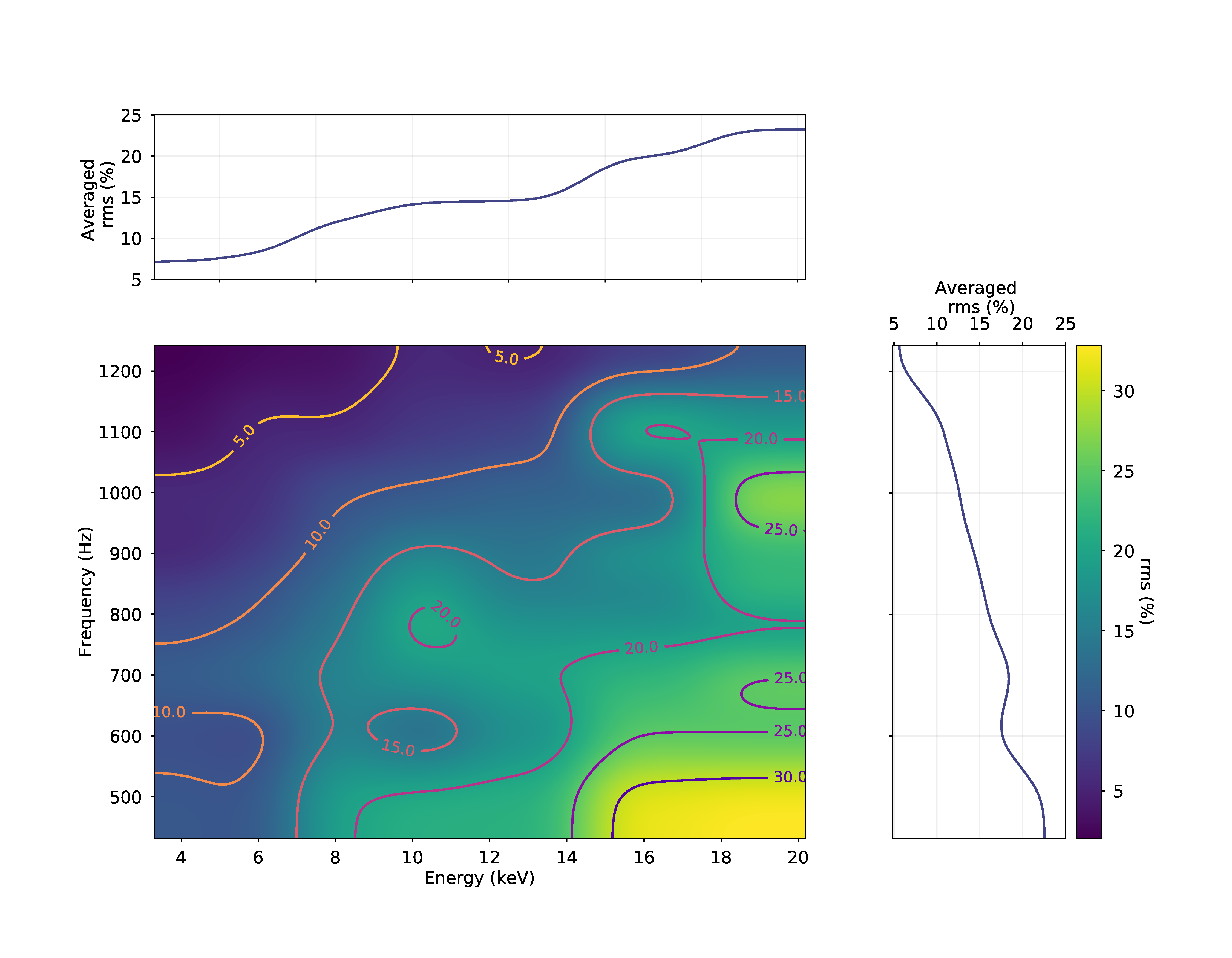}}
\caption{Fractional rms amplitude of the lower (left) and upper (right) kHz QPO in 4U~1636--53 as a function of both energy ($x$ axis) and QPO frequency ($y$ axis). The colours ($z$ coordinate) represent the rms amplitude of the QPO. The top and right panels next to each Figure show the rms amplitude of the QPOs marginalised over, respectively, QPO frequency and energy \citep{Ribeiro-2019}.} 
\label{fig:rms_E_nu-Ribeiro}
\end{figure}

The conclusion from the results presented above is that the fractional rms amplitude of the kHz QPOs depends both on energy and QPO frequency. So far we have shown either the rms amplitude vs. QPO frequency, marginalised over energy, the rms amplitude vs. energy, marginalised over QPO frequency, or the rms vs. energy for a given frequency (the conditional plots). In Figure~\ref{fig:rms_E_nu-Ribeiro} we show the rms amplitude of the lower and upper kHz QPO plotted vs. both energy and QPO frequency \citep[the joint plots;][]{Ribeiro-2019}. 

The rms amplitude of the lower kHz QPO in 4U~1636--53 is maximum at $\nu_{\rm low}$$\approx$$800$ Hz and $E$$\approx$$13$ keV, while the rms amplitude of the upper kHz QPO increases both as the QPO and the energy increase.

From spectral modelling of accreting neutron-star systems, the temperature of the accreting gas at the inner edge of the accretion disc is typically $\sim 0.3 - 2$ keV, depending on the state of the source, while the temperature of the neutron-star itself is $\sim$$1-2$ keV  \citep{Gierlinski-2002, Sanna-2013, Lyu-2014}. This implies that the emission from the disc and the neutron star components peaks at $\simless$$1-6$ keV, and drops quickly at energies higher than that, such that at energies above $\sim$$10-15$ keV the spectrum of accreting neutron stars is dominated by a power-law like component which is usually ascribed to inverse Compton scattering in a corona (with unspecified geometry) of highly-energetic electrons \citep{Sunyaev-1980, White-1988}. The total contribution of the disc or the neutron-star surface at $\sim$$20-25$ keV, where the amplitude of the QPOs is $\sim$$10 - 25$\%, is between $10^{-3}$ and $10^{-6}$ of the total flux of the source at those energies \citep[e.g.][]{Barret-1994, DiSalvo-2001b}. Therefore, even if the kHz QPOs may represent variations of a dynamical property of the accretion disc, e.g., one of the epicyclic frequencies in the relativistic precession model \citep{Stella-1998, Stella-1999}, a beat between the Keplerian frequency at the inner edge of the accretion disc and the neutron-star spin \citep{Miller-1998, Lamb-2001}, or a perturbation wave in the disc \citep{Abramowicz-2003, Lee-2004}, the mechanism that modulates the QPO signal cannot be at the disc itself, but must be connected to the corona. We will return to this below.

\subsection{The width of the kHz QPOs}
\label{sec:Q-factor}

\begin{figure}
\centering
\includegraphics[width=0.7\textwidth, trim=0cm 4cm 2cm 8cm, clip, angle=0]{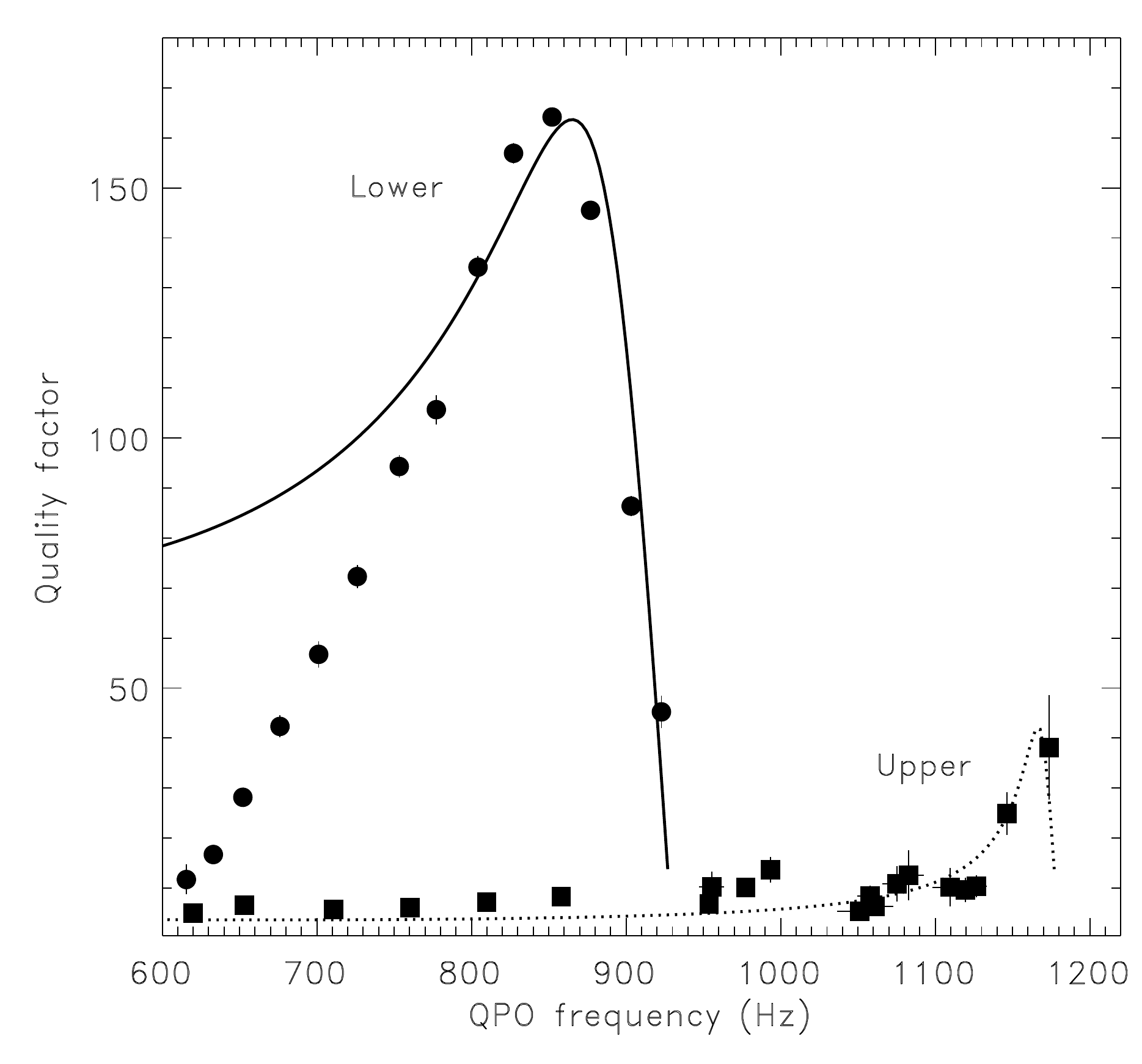}
\caption{Quality factor of the lower (filled circles) and upper (filled squares) kHz QPOs in 4U~1636--53. The solid line shows the trend expected from a simplified calculation of the effect of broadening of the profile of the QPO by: (i) radial drift of the inner parts of the disc as the gas falls onto the neutron star, (ii) the range of radii in the disc at which the QPO is produced, and (iii) the lifetime of the blobs that produce the oscillations \citep{Barret-2006}.}
\label{fig:Q-model-Barret}
\end{figure}

The quality factor, or equivalently the FWHM, of the kHz QPOs depends upon the QPO frequency \citep{vdk-1997, Wijnands-1997a, Wijnands-1997b, Ford-1997, Wijnands-1998, Ford-1998, vanStraaten-2000, Jonker-2000, Mendez-2001, DiSalvo-2001, Homan-2002, Jonker-2002, vanStraaten-2002, DiSalvo-2003, vanStraaten-2003, vanStraaten-2005, Barret-2005, Barret-2005b, Barret-2006, Mendez-2006, Altamirano-2008, Boutelier-2009, Sanna-2010, Barret-2011}. For the lower kHz QPO, the quality factor first increases slowly as the frequency of the QPO increases, and after reaching the maximum value it drops more or less abruptly as the QPO frequency continues increasing. This can be seen in Figure~\ref{fig:Q-model-Barret} for the case of 4U~1636--53 \citep[][see also \citep{DiSalvo-2001}]{Barret-2006}. Notice also in this Figure the difference of the quality factors of the lower and the upper kHz QPOs: The lower kHz QPO is almost always narrower than the upper one, and can be as narrow as $\sim$$5$ Hz. To reduce the errors in the measurements, this plot shows the quality factor averaged over intervals of $\sim$$20$ Hz in frequency for the lower kHz QPO, and combines measurements for many separate observations, taken at different epochs, of the source. When measured over short intervals in single observations, the quality factor in 4U~1636--53 can be as large as $200-220$ (see Fig.~\ref{fig:Q-nu-Barret}) which, at $\sim$$850$ Hz means that the lower kHz QPO can be as narrow as  $\sim$$4$ Hz FWHM \citep[see also][for the source EXO~1745--248]{Mukherjee-2011}. The quality factor of the upper kHz QPO, on the other hand, is much less than that of the lower kHz QPO, and it remains more or less constant or increases slowly as the frequency of the QPO increases. Translated into a width, the FWHM of the upper kHz QPO goes from $\Delta$$\sim$$40$ Hz at $\nu_{\rm upp}$$\sim$$1150$ Hz to $\Delta$$\sim$$120$ Hz at $\nu_{\rm upp}$$\sim$$600$ Hz.

The more or less abrupt drop of the quality factor of the lower kHz QPO has been interpreted \citep{Barret-2005b, Barret-2005, Barret-2006} as an indication that the inner radius of the accretion disc where, according to most models, the QPOs are formed, starts to reach the ISCO (\S\ref{sec:basic}). Close to the ISCO, three effects contribute to the width of the QPO peak: First, because of the rapid increase of the radial component of the velocity of a particle orbiting the neutron star when it approaches the ISCO, the gas will move a certain radial distance $\Delta r_{\rm drift}$ during the lifetime of the oscillation, leading to a change $\displaystyle \Delta\nu_{\rm drift}\approx \frac{3}{2}\frac{\Delta r_{\rm drift}}{r_{\rm orb}} \nu_{\rm orb}$ of the frequency of the oscillations during this time. Second, the region in the disc where the QPO is formed is stretched in the radial direction, leading to a range of frequencies of the oscillations $\displaystyle \Delta\nu_{\rm orb}\approx \frac{3}{2}\frac{\Delta r_{\rm orb}}{r_{\rm orb}} \nu_{\rm orb}$. Finally, the finite lifetime of the oscillations adds to the width of the QPO profile. 

The solid curve in Figure~\ref{fig:Q-model-Barret} shows a simplified calculation of these effects combined, for a set of parameters that roughly reproduce the data, and that yield a neutron-star mass of $\sim$$2\Msun$ or higher. In this model, at frequencies above $\nu_{\rm low}$$\approx$$800$ Hz the main contribution to the width of the QPO is from the drift of the QPO frequency, $\Delta\nu_{\rm drift}$. If this interpretation is correct, the drop of the quality factor of the lower kHz QPO in this and other sources would provide a constrain to the mass of the neutron star, and a direct evidence of the existence of the ISCO around the neutron star in these systems. 

\begin{figure}
\centering
\includegraphics[width=0.6\textwidth, trim=0cm 1cm 0.5cm 1cm, clip, angle=90]{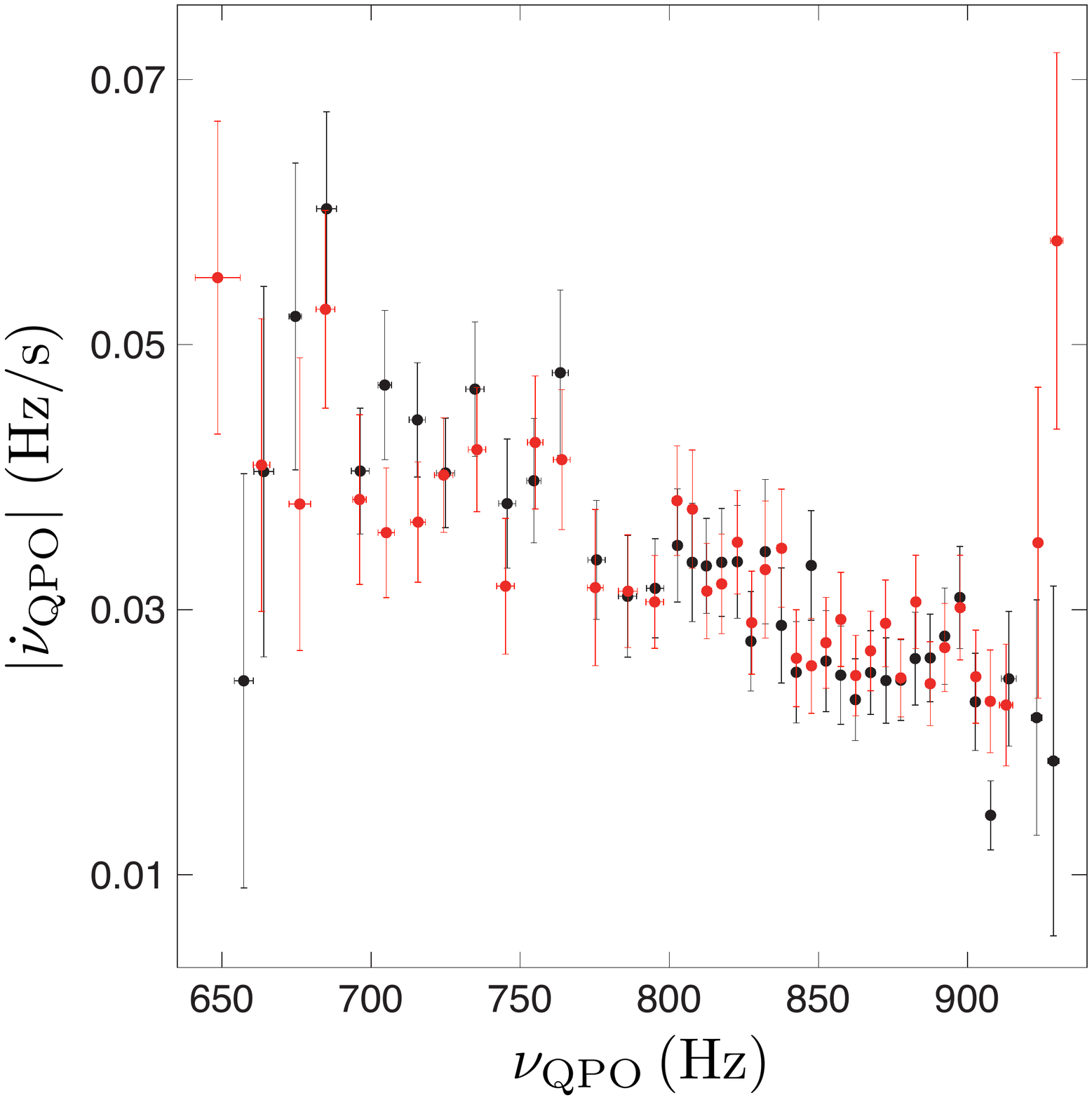}
\caption{Absolute values of the positive (red points) and negative frequency derivative (black points) for the lower kHz QPO in 4U~1636--53 as a function of the QPO frequency measured on time-scales of 64 s or shorter \citep{Sanna-2012}.}
\label{fig:nu_dot-nu-Sanna}
\end{figure}

Measurements of the rate at which the QPO frequency changes as a function of the QPO frequency, however, appear to question this interpretation. As shown in Figure~\ref{fig:nu_dot-nu-Sanna} \citep{Sanna-2012}, the rate of change of the QPO frequency is the largest at low QPO frequencies, decreases as the QPO frequency increases, and is the smallest at the highest QPO frequency. On the contrary, under the interpretation that the drop of $Q$ is due to the ISCO, at high QPO frequencies, when the radius of the disc is closest to the ISCO, the effect of $\Delta\nu_{\rm drift}$ and the rate of change of the frequency of the lower kHz QPO should be the largest.

\begin{figure}
\centering
\includegraphics[width=0.7\textwidth, trim=0 2cm 0 1cm, clip, angle=0]{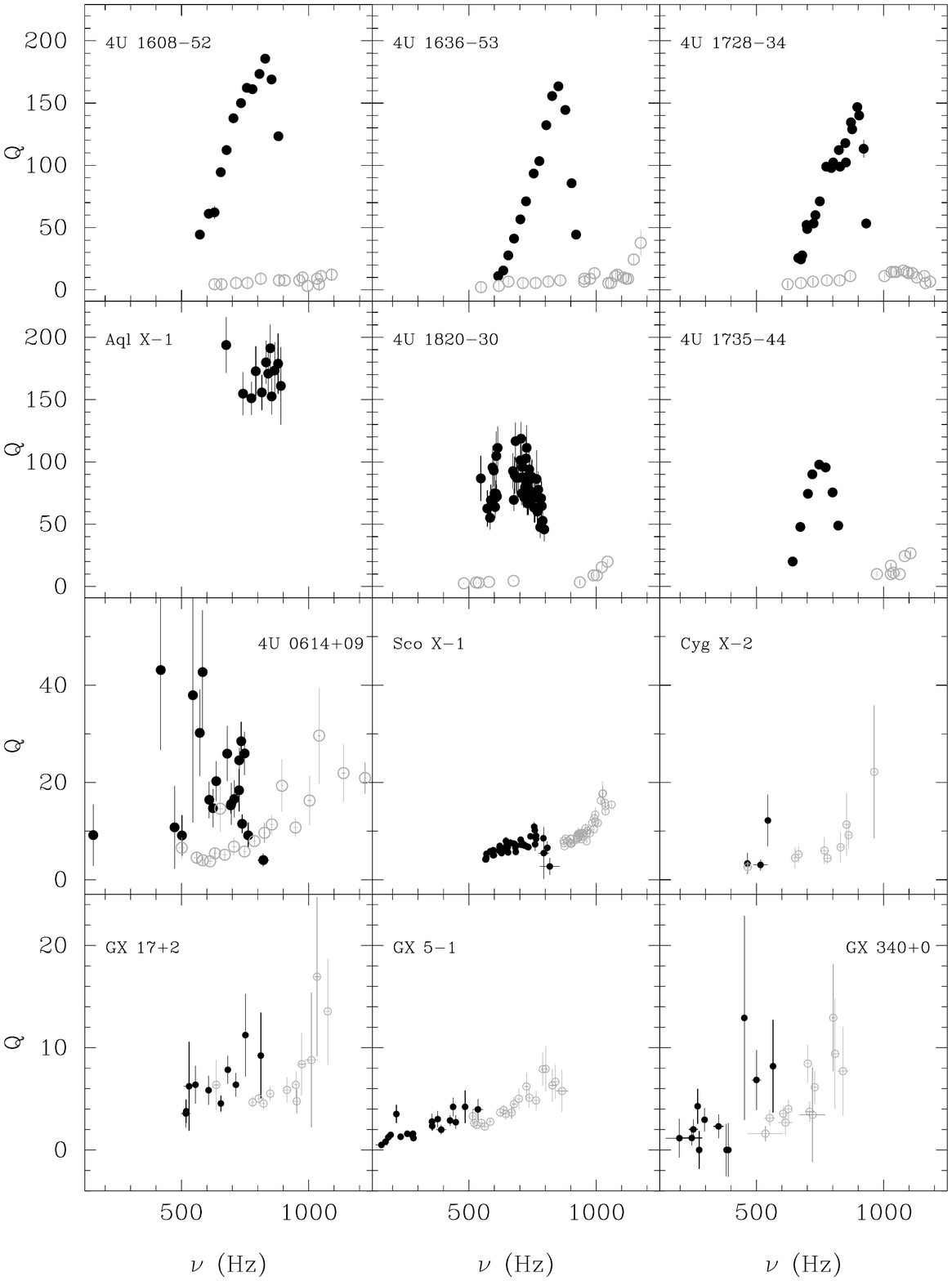}
\caption{Quality factor of the lower (filled symbols) and the upper (open symbols) kHz QPOs as a function of the QPO frequency for seven atoll sources, 4U~1608--52, 4U~1636--53, 4U~1728--34, Aql~X-1, 4U~1820--30, 4U~1735--44, and 4U~0614+09, and five Z sources, Sco~X-1, Cyg~X-2, GX~17+2, GX~5--1, and GX~340+0 \citep{Mendez-2006}. Notice that, as with the fractional rms (Fig.~\ref{fig:rms_all_sources-Mendez}), the scale of the $y$ axis is different for atoll and Z sources, reflecting the fact that the maximum value of the quality factor is larger in the former than in the latter.}
\label{fig:Q_all_sources-Mendez}
\end{figure}

In Figure~\ref{fig:rms_all_sources-Mendez} we saw that the rms amplitude of the kHz QPOs is lower in Z than in atoll sources. In fact, also the quality factor of the kHz QPO, and in particular that of the lower kHz QPO, is in general lower in Z
\citep{vdk-1997, Wijnands-1998, Zhang-1998, Jonker-2000, Homan-2002, O'Neill-2002} than in atoll sources \citep{Wijnands-1997a, Wijnands-1997b, Ford-1997, Ford-1998, vanStraaten-2000, Mendez-2001, DiSalvo-2001, vanStraaten-2002, DiSalvo-2003, vanStraaten-2003, vanStraaten-2005, Altamirano-2005, Barret-2005, Barret-2006, Mendez-2006, Altamirano-2008, Boutelier-2009, Sanna-2010, Barret-2011, deAvellar-2016, Ribeiro-2017, vanDoesboergh-2017, Ribeiro-2019, vanDoesburgh-2019}. This is shown in Figure~\ref{fig:Q_all_sources-Mendez} in which the quality factor, both of the lower and the upper kHz QPO, in seven atoll and 5 Z sources is plotted as a function of the frequency of the QPO \citep{Mendez-2006}. From this Figure it is apparent that the lower kHz QPO is narrower in the atoll than in the Z sources and that, within the atoll sources, the minimum width that the QPO can attain is different for different sources. The situation is less clear for the upper kHz QPO because the measurements, especially of the Z sources, have larger errors. The Z sources are more luminous and have a softer spectrum than the atoll sources, and this differences are likely due to the difference of the total mass accretion rate in these two classes of sources (see above in this section). It is, therefore, possible that the rms amplitude and quality factor of the kHz QPOs depend upon mass accretion rate, reflecting properties of the accretion flow that produces the X-ray spectrum in these sources.

\begin{figure}
\centering
\subfloat[]{\label{fig:low-Q-rms-Sanna}
\includegraphics[width=0.4\textwidth, trim=0 4cm 1cm 1cm, clip, angle=-90]{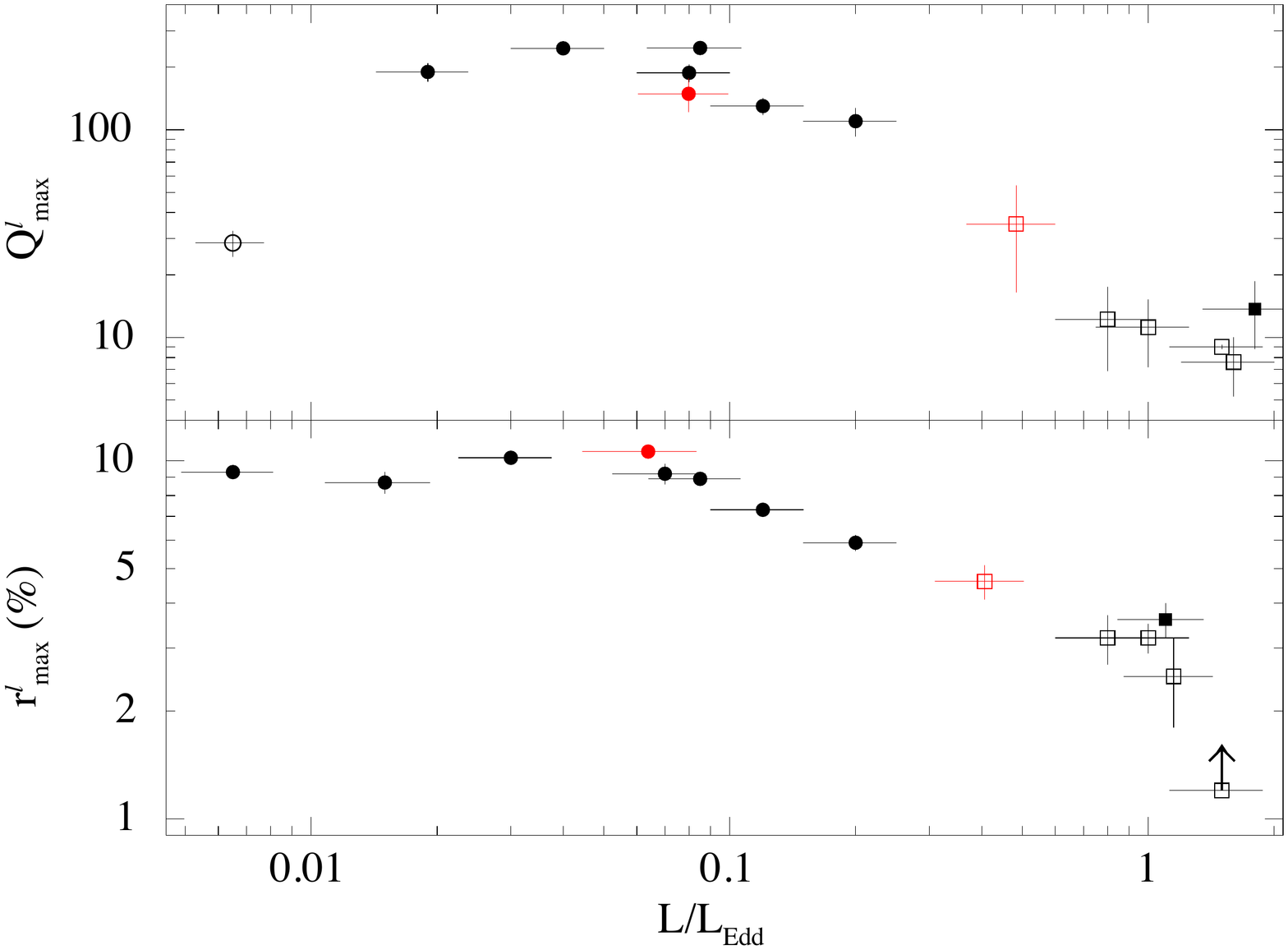}}
\subfloat[]{\label{fig:high-Q-rms-Mendez}
\includegraphics[width=0.4\textwidth, trim=0 4cm 1cm 1cm, clip, angle=-90]{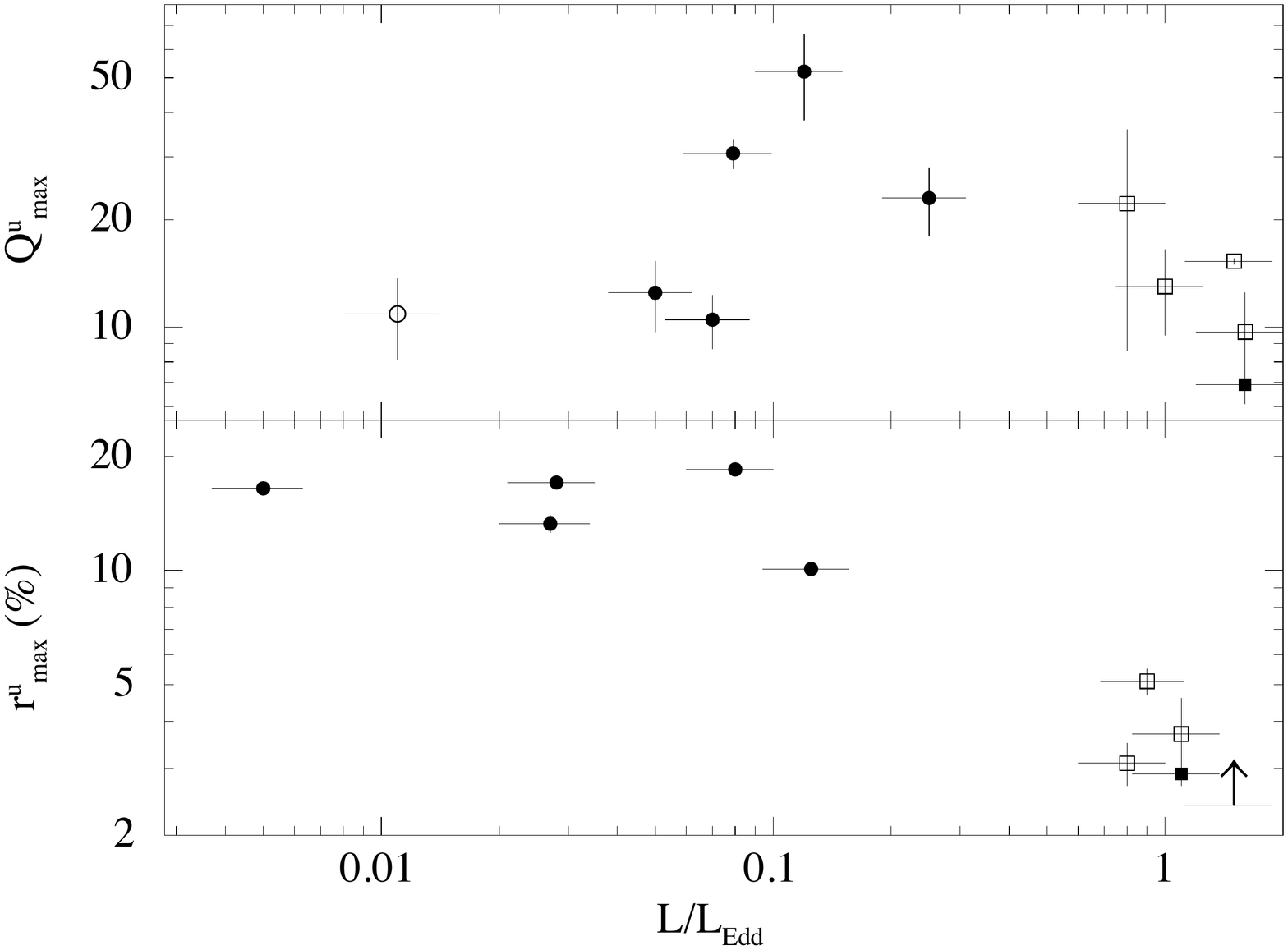}}
\caption{Maximum quality factor (top panels), obtained from Figure~\ref{fig:Q_all_sources-Mendez}, and maximum fractional rms amplitude (lower panels), obtained from Figure~\ref{fig:rms_all_sources-Mendez}, of the lower (left) and upper (right) kHz QPOs of seven atoll (circles) and five Z sources (squares) as a function of the source luminosity \citep{Mendez-2006}. The red points in the left panel correspond to the measurements of the quality factor and rms amplitude of the lower kHz QPO in the transient LMXB XTE~J1701--462. The red circles and red squares show the maximum quality factor (upper panel) and maximum rms amplitude (lower panel) of the lower kHz QPO in XTE~J1701--462 when the source was, respectively, in the atoll and the Z phases of the outburst \citep{Sanna-2010}.}
\label{fig:low-Q-rms}
\end{figure}

Figures~\ref{fig:low-Q-rms-Sanna} and \ref{fig:high-Q-rms-Mendez} summarise these points. The black symbols in those Figures show the maximum quality factor and maximum fractional rms amplitude of, respectively, the lower and the upper kHz QPO as a function of the source luminosity (in Eddington units), for the seven atoll and five Z sources in Figures~\ref{fig:rms_all_sources-Mendez} and 
\ref{fig:Q_all_sources-Mendez}. (We will discuss the red points below.) The relation between the maximum quality factor of the lower kHz QPO and the luminosity of the source in this Figure resembles the relation between the quality factor and the frequency of the lower kHz QPO in 4U~1636--53 \citep{Barret-2006} and other sources (see Figure~\ref{fig:Q-model-Barret}). The same is true for the dependence of the maximum rms amplitude of the lower kHz QPO with luminosity in the set of sources seen in this Figure, and the relation between rms amplitude and QPO frequency for the lower kHz QPO in individual sources (e.g., Figs.~\ref{fig:rms-nu-Ribeiro} and \ref{fig:rms_vs_freq_vanStraaten}). Since in individual sources the QPO frequency generally increases with luminosity (see Fig.~\ref{fig:paral}; despite being partially affected by the parallel-track phenomenon, this is generally the case), this suggests that the same mechanism is responsible for the drop of quality factor and rms amplitude of the lower kHz QPO with QPO frequency in 4U~1636--53 and other sources, as well as for the drop of the maximum QPO quality factor and maximum QPO rms amplitude with luminosity in the set of sources. 
This would imply that the drop of the quality factor in 4U~1636--53 is not driven by the inner edge of the disc approaching the ISCO, but to changes in the properties of the accretion flow, e.g. optical depth and temperature of the boundary layer \citep{Gilfanov-2003} or the corona \citep{Mendez-2006}, where the signal of the QPO is likely modulated (see below). 

The fact that the maximum rms amplitude and quality factor of the lower kHz QPO in the set of sources are lower in the Z than in the atoll sources offers the possibility to test these ideas. If the scenario in which the drop of the quality factor and rms amplitude of the lower kHz QPO in 4U~1636--53 and other sources is driven only by the inner edge of the disc approaching the ISCO was correct, one would expect that, if a source ever switched from atoll to Z, or vice versa, and continued showing kHz QPOs both in the atoll and Z phases, at the same QPO frequency, hence the same inner disc radius, the quality factor and rms amplitude of the lower kHz QPO would be the same. The reason for this is that the radius of the ISCO depends only on the mass, spin and equation of state of the neutron star, which do not change when the source switches from one class to the other. On the other hand, if the quality factor and rms amplitude of the lower kHz QPO were driven (at least in part) by the properties of the accretion flow, since the properties of the accretion flow are different in Z and atoll sources, the average quality factor and rms amplitude of the lower kHz QPO between the Z and atoll phases would change. 

In 2006, when this was proposed \citep{Mendez-2006}, such a source did not exist. But in April of 2007, the transient source XTE~J1701--462 \citep{Remillard-2006}, which started its outburst as a Z source, underwent a transition and switched into an atoll source \citep{Homan-2007, Lin-2009, Homan-2010}. This source also showed kHz QPOs both in the Z \citep{Homan-2006} and atoll phases \citep{Homan-2007b}. As proposed in the second scenario, at the same QPO frequency, the quality factor and rms amplitude of the lower kHz QPO in the Z phase of XTE~J1701--462 were, respectively, $\sim$$7-8$ and $\sim$$3$ times larger in the atoll than in the Z phase \citep{Sanna-2010}. Not only that, but the quality factor and rms amplitude of the lower kHz QPO in the Z phase of XTE~J1701--462 were also similar to those of the other Z sources, and in the atoll phase of XTE~J1701--462 they were similar to those of the other atoll sources. 
The red points in Figure~\ref{fig:Q_all_sources-Mendez} show the quality factor and rms amplitude of the lower kHz QPO in XTE~J1701--462 in the Z (open square) and atoll (filled circle) phase, perfectly in line with the other sources in that plot. This shows that in XTE~J1701--462, and other sources, the quality factor and rms amplitude of the kHz QPOs are, at least in part, driven by the properties of the accretion flow. This questions the suggestion that the drop of the quality factor of the lower kHz QPO at high QPO frequencies provided evidence of the ISCO in these systems. This, on the other hand, offers an avenue to develop models to explain the radiative properties of the QPO. 

\subsection{The energy-dependent lags and coherence of the kHz QPOs}
\label{sec:lags}

As we described in \S\ref{sec:qpo101}, one can study the phase/time lags of the QPOs using light curves in two different energy bands. Besides the lags, one can also study the coherence function between the two light curves \citep{Vaughan-1997, Nowak-1999, Ingram-2019}. The coherence function \citep{Bendat-2010}, $\gamma^2(\nu)$, measures the degree of linear correlation between two noiseless signals (light curves) as a function of the Fourier frequency. More precisely, $\gamma^2(\nu)$ should be called $\gamma^2(\nu;E_1,E_2)$ to indicate that it is the coherence function between two light curves at energies $E_1$ and $E_2$; when the lags and the coherence function are given as a function of energy, one means that those are the quantities measured at $E=E_2$ with respect to the reference band, in this case $E_1$. However, since the observed light curves are not noiseless, but are affected by Poisson counting noise (see \S\ref{sec:qpo101}), to study the degree of correlation between two X-ray light curves one uses the intrinsic coherence function, usually denoted as $\gamma_{\rm I}^2(\nu)$, that corrects for this \citep[see][for an explanation]{Vaughan-1997}. For simplicity, here we will use the term coherence function, and will write $\gamma^2(\nu)$, to refer to the intrinsic coherence function.

\begin{figure}
\centering
\includegraphics[width=0.6\textwidth, trim=0cm 4cm 0 4cm, clip, angle=0]{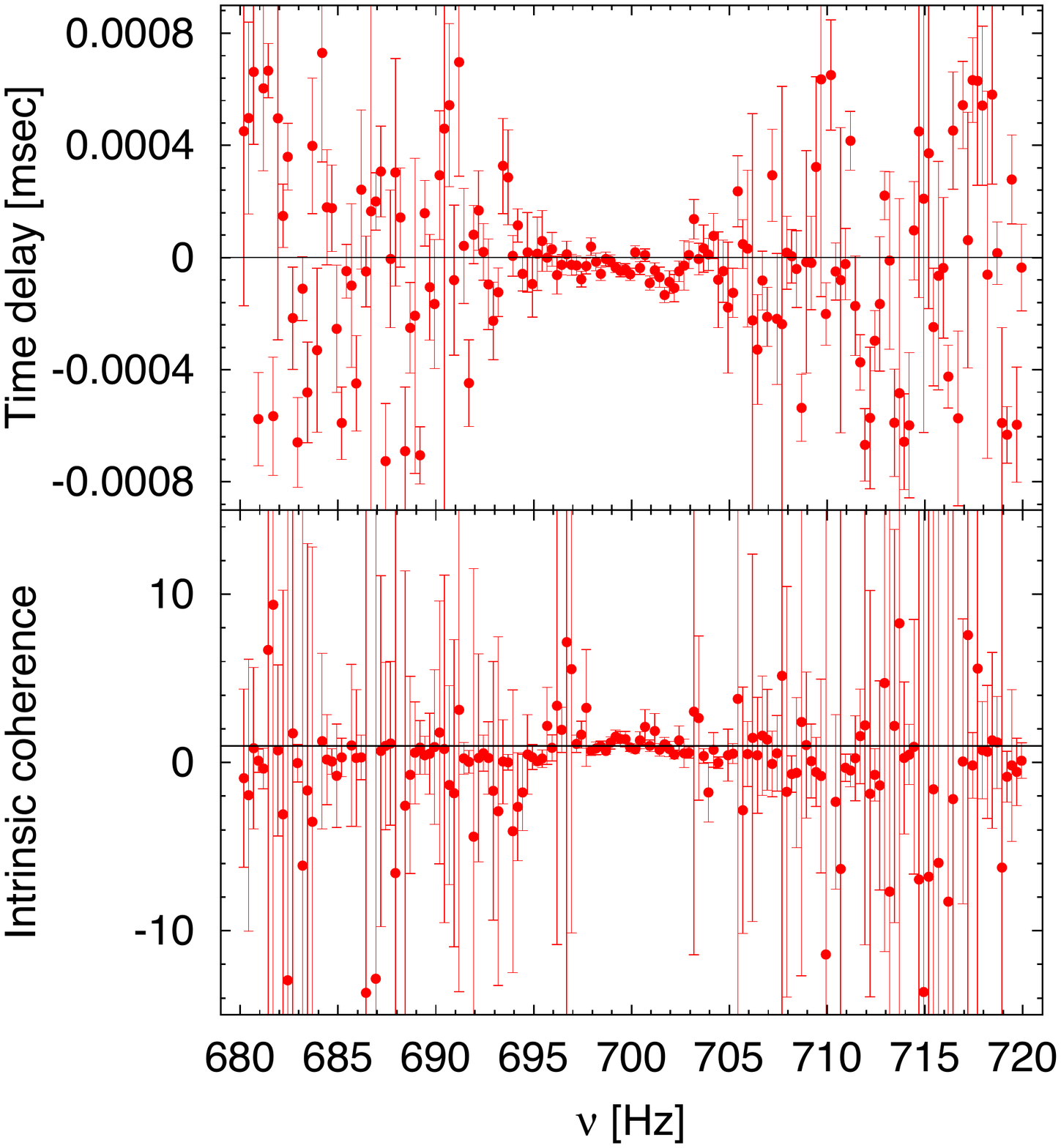}
\caption{Time lags (top panel) and coherence function (bottom panel) as a function of Fourier frequency, at frequencies around that of the lower kHz QPO in 4U~1608--52. The horizontal line in the top panel is at the zero time lag, while the horizontal line in the bottom panel shows the perfect coherence of 1 \citep{deAvellar-2013}.}
\label{fig:tdelay-intcoh-Marcio}
\end{figure}

Figure~\ref{fig:tdelay-intcoh-Marcio} shows the time lags (top panel) and coherence function (bottom panel) around the frequency of the lower kHz QPO in 4U~1608--53, using light curves with energies around $3$ keV and $8$ keV \citep{deAvellar-2013}. This Figure shows that the two light curves used to compute the lags and coherence function are perfectly correlated (coherence function equal to 1 with small errors) in the frequency range at which the QPO signal dominates, and uncorrelated (coherence function consistent with 0 and large errors) outside that frequency range. This Figure also shows that the lags of the lower kHz QPO in this source are soft (negative), meaning that the low-energy photons lag the high-energy ones. 

\begin{figure}
\centering
\subfloat[]{\label{fig:lags-low-Peille}
\includegraphics[width=0.47\textwidth, trim=0 0 1cm 2cm, angle=0]{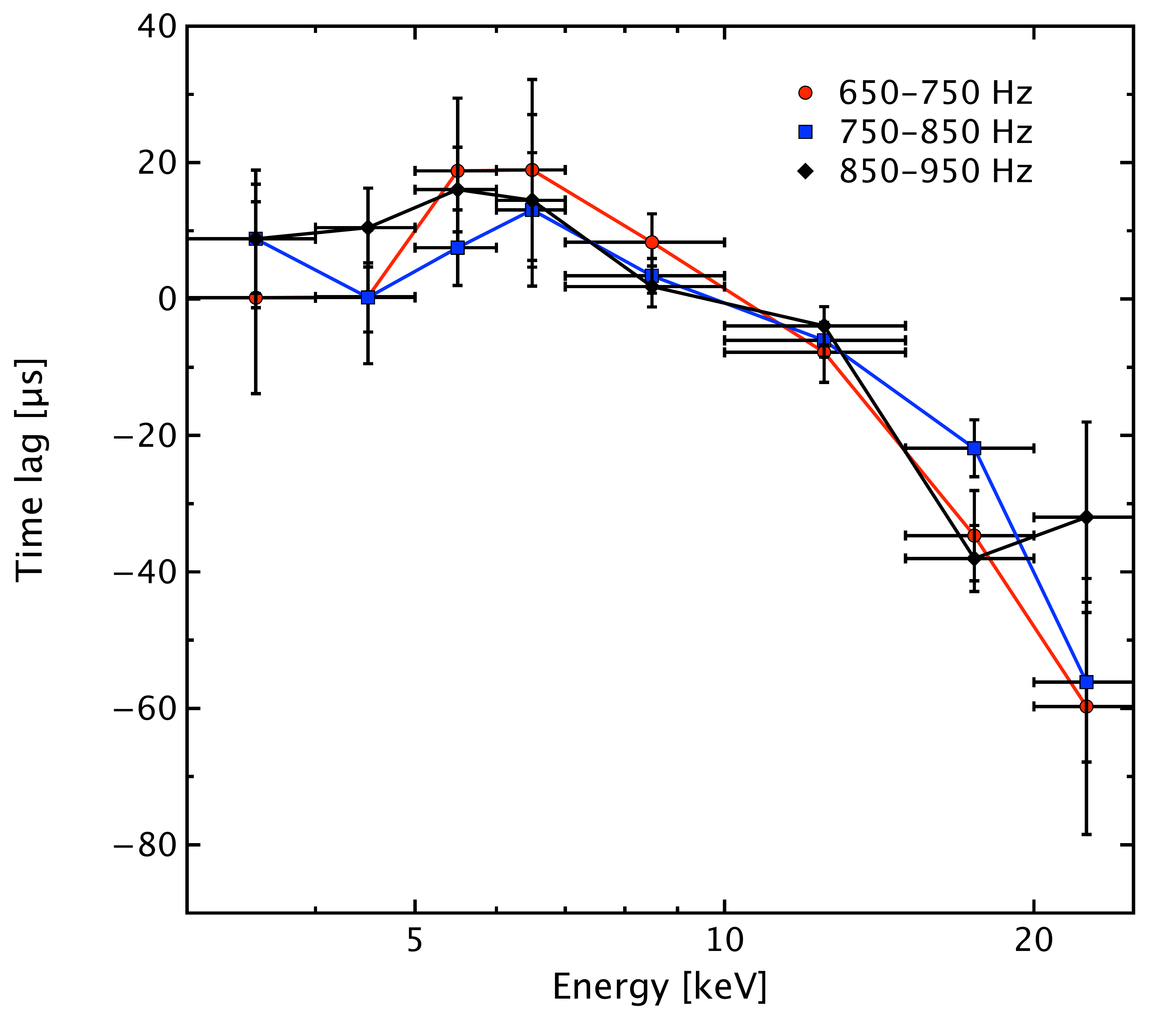}}
\subfloat[]{\label{fig:lags-upp-Peille}
\includegraphics[width=0.47\textwidth, trim=0 0 1cm 2cm, angle=0]{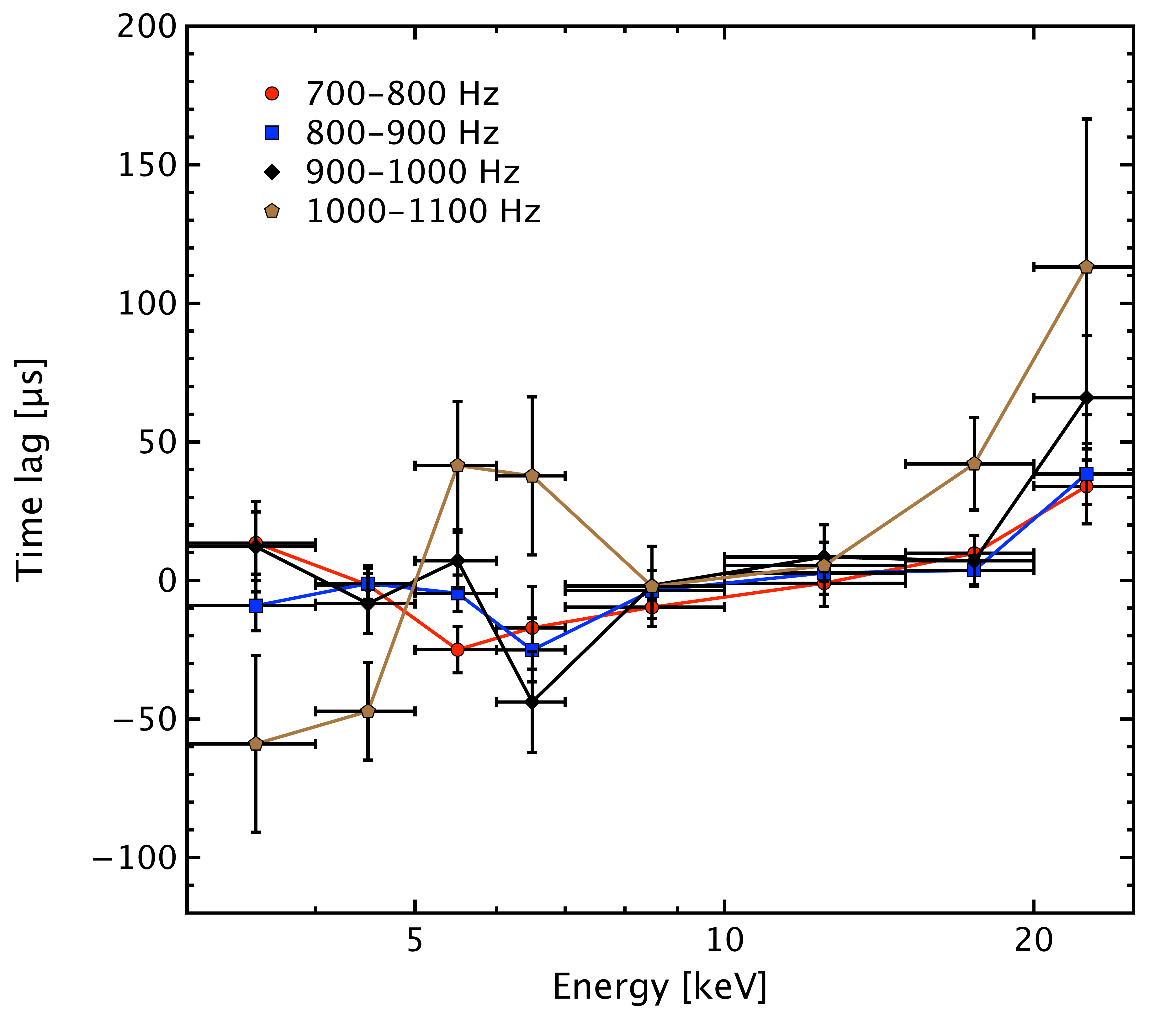}}
\caption{Time lags as a function of energy for the lower (left) and upper (right) kHz QPO in 4U~1728--34 for different QPO frequencies \citep[originally published as Figure 3 in][]{Peille-2015}.}
\label{fig:lags-Peille}
\end{figure}

In \S\ref{sec:qpo101} we showed that the lags of the lower kHz QPO are significantly different from those of the upper kHz QPO \citep{deAvellar-2013}. Specifically, the lags of the lower kHz QPO in 4U~1608--52, 4U~1636--53, Aql~X-1 and 4U~1728--34 are soft and become softer as the energy increases \citep{deAvellar-2013, Barret-2013, Peille-2015, deAvellar-2016, Troyer-2017}. On the contrary, the lags of the upper kHz QPO in 4U~1608--52, 4U~1636--53 and 4U~1728--34 are either consistent with zero or hard and, if anything, they become harder as the energy increases \citep{deAvellar-2013, Peille-2015}. This can be clearly seen in Figure~\ref{fig:lags-Peille} (see also Fig.~\ref{fig:lags-avellar} in \S\ref{sec:qpo101}). The left panel shows the lags of the lower kHz QPO in 4U~1728--34 as a function of energy for different QPO frequencies. The right panel shows the same for the upper kHz QPO in this source. This difference in the lags of the lower and upper kHz QPOs indicates that either the radiative mechanisms that generate each of the kHz QPOs are different, or the mechanism is the same but each of the two signals that are modulated at the frequency of the lower and the upper kHz QPO travel through different parts of the accretion flow before reaching the observer. 

\begin{figure}
\centering
\includegraphics[width=0.7\textwidth, trim=0cm 0cm 0 0cm, clip, angle=90]{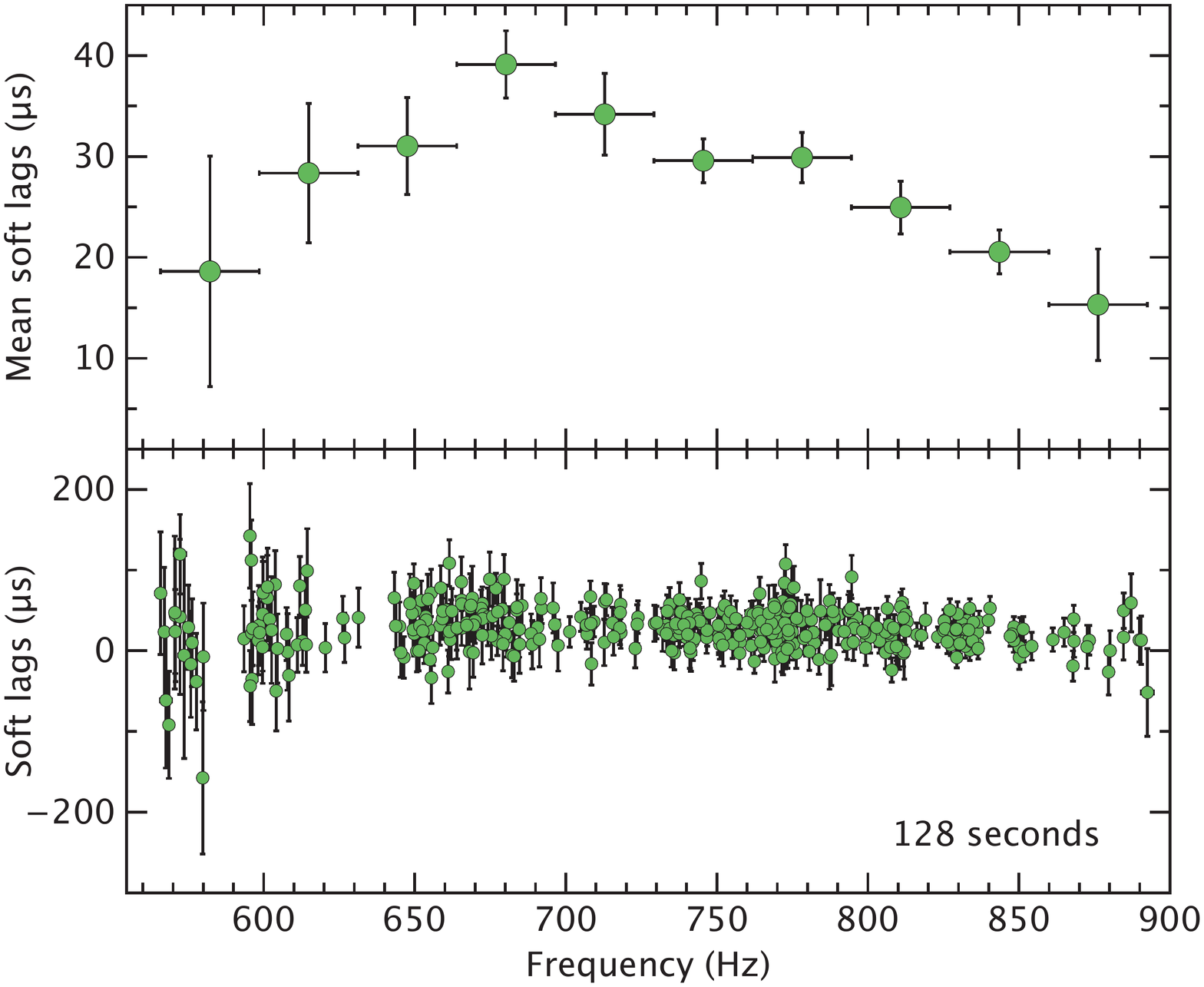}
\caption{Time lags as a function of frequency for the lower kHz QPO in 4U~1608--52 \citep[originally published as Figure 2 in][]{Barret-2013}. The lower panel shows the lags measured over 128-s time intervals, while the upper panel shows the same data binned into ten adjacent QPO-frequency intervals. Notice that by convention, in this Figure negative values correspond to soft lags, which is the opposite convention to the one used in Figures~\ref{fig:lags-avellar} and \ref{fig:tdelay-intcoh-Marcio}.}
\label{fig:lags-nu-Barret}
\end{figure}

In \S\ref{sec:qpo101} (see Figure~\ref{fig:lags-nu-avellar}) we showed that the magnitude of the soft lags of the lower kHz QPO in 4U~1636--53 first increases and then decreases as the frequency of the QPO increases. Since the QPO frequency is a function of $S_a$ (Fig.~\ref{fig:nu-vs-Sa-Zhang}), the dependence of the lags on $S_a$ is similar to that on $\nu_{\rm low}$ \citep[see Fig. 7 in][]{deAvellar-2016}. Figure~\ref{fig:lags-nu-Barret} shows the time lags of the lower kHz QPO in 4U~1608--52 as a function of QPO frequency \citep{Barret-2013}. The magnitude of the lags shows a significant increase as the QPO frequency increases from $\nu_{\rm low}$$\approx$$550$ Hz to $\nu_{\rm low}$$\approx$$700$, and a significant decrease as the frequency increases further. (Notice that in this Figure the convention is that soft lags are positive, therefore the trend appears be the opposite of what is shown in Figure~\ref{fig:lags-nu-avellar}.) In other sources, e.g., 4U~1728--34 \citep{Peille-2015}, Aql~X-1 \citep[][see also \citep{Troyer-2018}]{Troyer-2017}, the trend is not as clear as in 4U~1608--52 and 4U~1636--53. 

The magnitude of the lags of the lower kHz QPO in 4U~1608--52 is between 15$\mu$s and 40$\mu$s. These values translate into light-travel distances between 4.5 km and 12 km, commensurate with the expected distance between the inner edge of the accretion disc and the neutron-star surface \citep[e.g.][]{Miller-1998}. It is therefore tempting to identify the lags with the light travel time in that environment. However, from the trend of the lags of the lower kHz QPO in 4U~1608--52 with QPO frequency in Figure~\ref{fig:lags-nu-Barret}, and that of the frequency of the lower kHz QPO vs. the inner radius of the accretion disc for the same source in Figure~\ref{fig:R_in-nu_Barret}, it is apparent that the relation between the lags of the lower kHz QPO and the inner radius of the accretion disc is not monotonic. If the model used to fit the energy spectra of 4U~1606--52 \citep{Barret-2013} is correct (remember, it is just a model), this non-monotonic relation between the lags and the inner-disc radius implies that the lags cannot be just a delay due only to the light travel-time of the photons from the disc to the neutron star, or vice versa. A similar conclusion can be drawn about the lags of the upper kHz QPO. In this case the lags do not appear to depend upon QPO frequency (but the errors of the lags of the upper kHz QPO are larger than those of the lags of the lower kHz QPO); at the same time, the frequency of the upper kHz QPO changes by a factor of $\sim$$2$ (see Figs.~\ref{fig:lags-nu-avellar} and \ref{fig:lags-Peille}) and, comparing the range of frequencies spanned by the lower and the upper kHz QPOs and the range of inner-disc radii in Figure~\ref{fig:R_in-nu_Barret}, the inner-disc radius changes also by a factor of at least $\sim$$2$. Therefore, the lags of the upper kHz QPO cannot be due only to a light travel-time delay between the disc and the neutron star either.

Since the lags of the lower kHz QPO are soft, whereas inverse Compton scattering in a corona should produce hard lags (\S\ref{sec:qpo101}; but see below), a different mechanism was required to model the lags of the lower kHz QPO. Reverberation due to reflection of hard photons from the corona off the accretion disc had been used to explain the soft lags of the broad-band noise component in the power spectrum of active galactic nuclei \citep{Fabian-2009, Zoghbi-2011, Zoghbi-2012} and galactic black-hole candidates \citep[][see also \cite{Kara-2019}]{Uttley-2011, deMarco-2015}. Reverberation was therefore a promising mechanism to explain the lag spectrum of the lower kHz QPOs. 

\begin{figure}
\centering
\subfloat[]{\label{fig:lag-nu-Cacket}
\includegraphics[width=0.495\textwidth, trim=0.5cm 6.4cm 0 6cm, clip, angle=0]{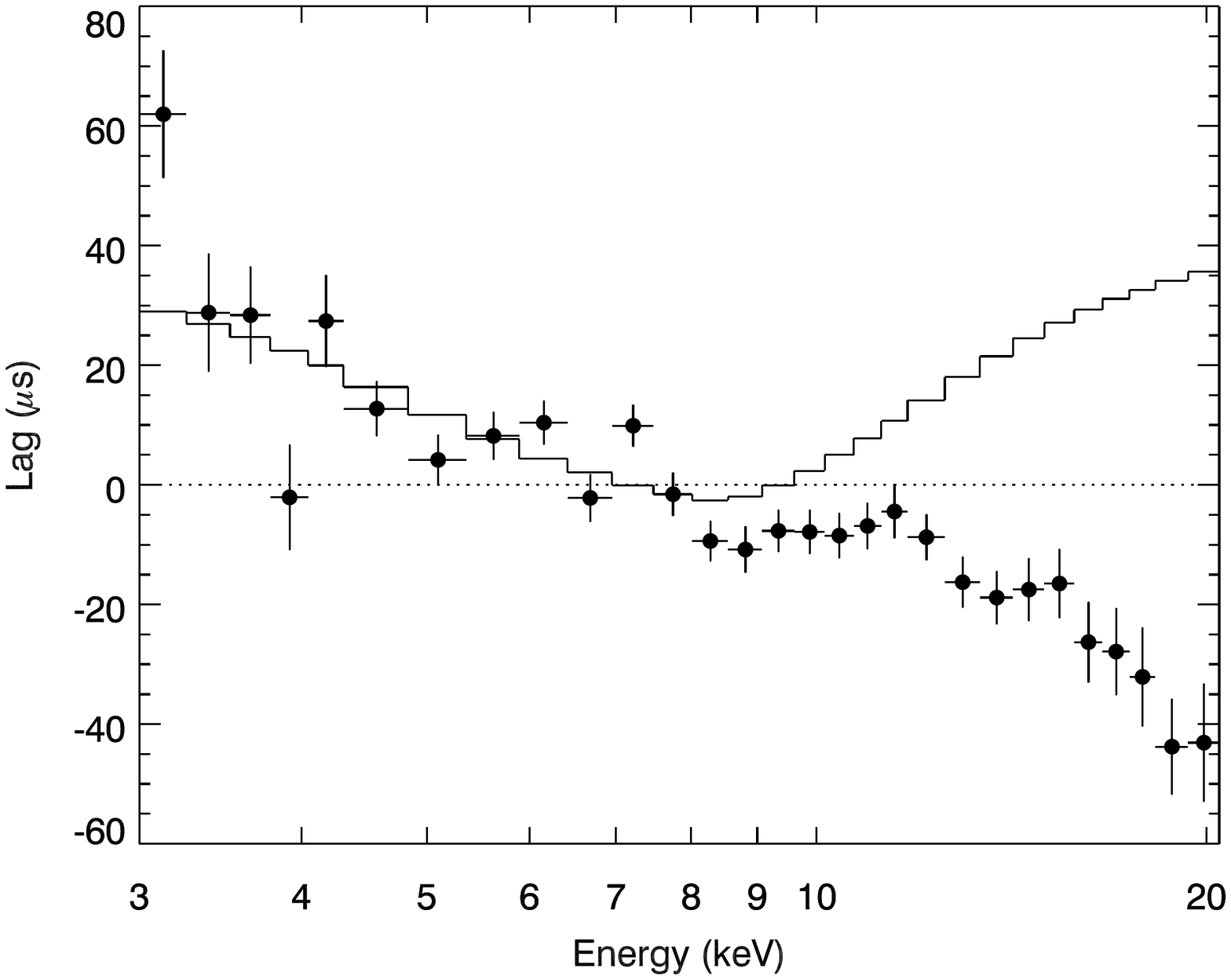}}
\subfloat[]{\label{fig:lag-nu-Coughenour}
\includegraphics[width=0.48\textwidth, trim=5.8cm 4cm 6cm 5.1cm, clip, angle=0]{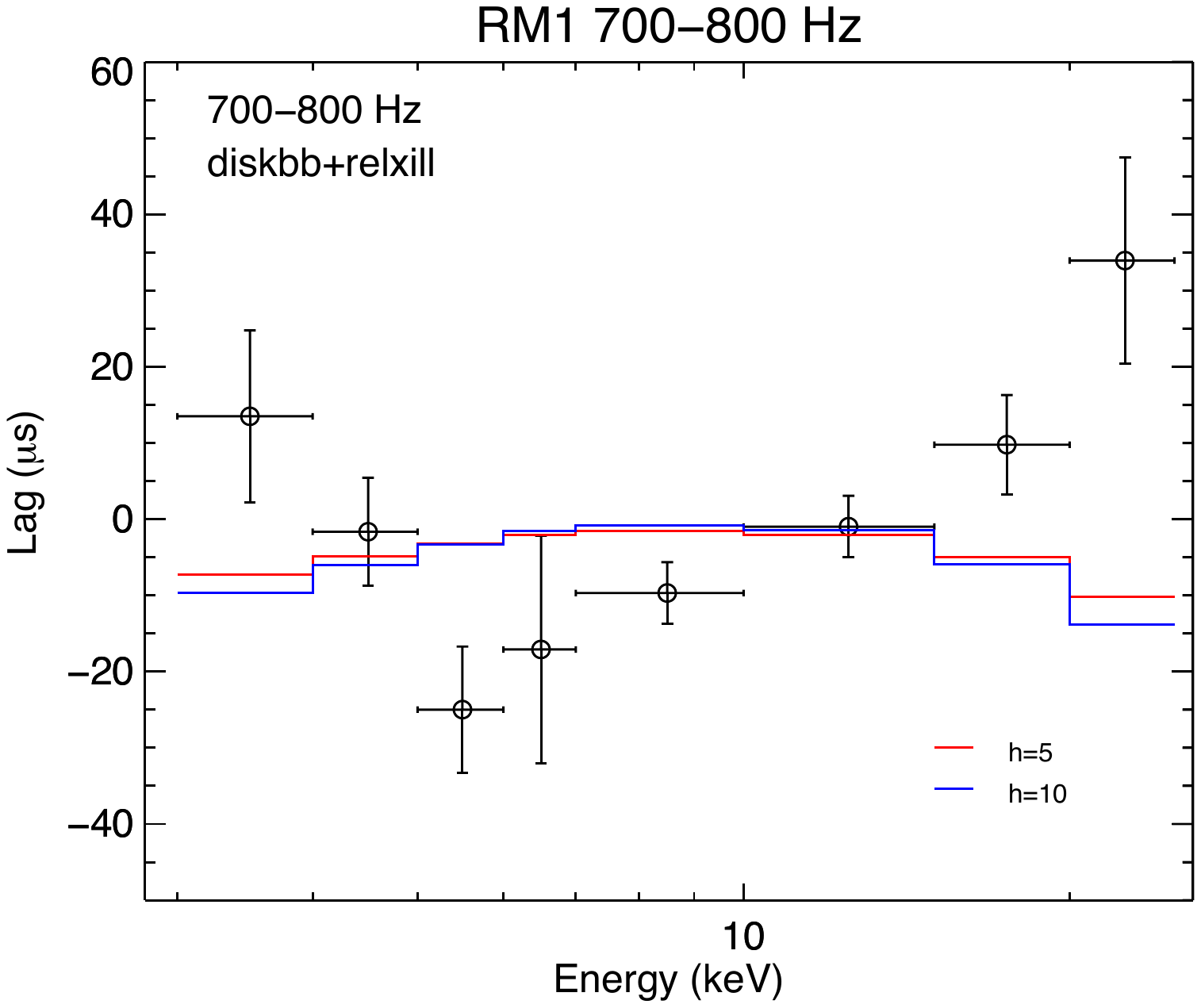}}
\caption{Time lags of the lower kHz QPO in 4U~1608--52 \citep[left; originally published as Figure 6 in][]{Cackett-2016} and the upper kHz QPO in 4U~1728--34 \citep[right; originally published as Figure 6 in][]{Coughenour-2020} as a function of energy, with the best-fitting reverberation model.}
\end{figure}

Figure~\ref{fig:lag-nu-Cacket} shows the lag spectrum of the lower kHz QPO in 4U~1608--52 \citep{Cackett-2016} with the best-fitting reverberation model. While the lags of the QPO decrease more or less monotonically as the energy increases, the best-fitting model predicts that above $E$$\sim$$8$ keV the lags should increase with energy, contrary to the observations. This led to the conclusion \citep{Cackett-2016} that the lags of the lower kHz QPO in 4U~1608--52 cannot be due only to reverberation. Subsequently, the same idea was tested on the upper kHz QPO in 4U~1728--34 \citep{Coughenour-2020}; the result of the fits of a reverberation model to the lags of this QPO is shown in Figure~\ref{fig:lag-nu-Coughenour}. From this Figure it is apparent that, while the lags of the QPO first decrease with energy up to $E$$\sim$$5-6$ keV, and increase again above that energy, the model predicts that the lags should follow the opposite behaviour. This led to the conclusion that, as with the case of the lags of the lower kHz QPO in 4U~1608--52, the lags of the upper kHz QPO in 4U~1728--34 cannot be solely due to reverberation. Notice that, with few exceptions \citep{Ingram-2017, Mastroserio-2018}, so far reverberation models have been used only to fit and explain the time/phase lags, and not the amplitude, of the observed light curves of accreting supermassive black holes in active galactic nuclei and stellar-mass black holes in galactic X-ray binaries, whereas a consistent model of the radiative mechanism that fits the data should be able to explain both the rms and lag spectrum of the variability in these sources.

\begin{figure}
\centering
\subfloat[]{\label{fig:rms-E-low-Troyer}
\includegraphics[width=0.49\textwidth, trim=0 0 0 0, clip, angle=0]{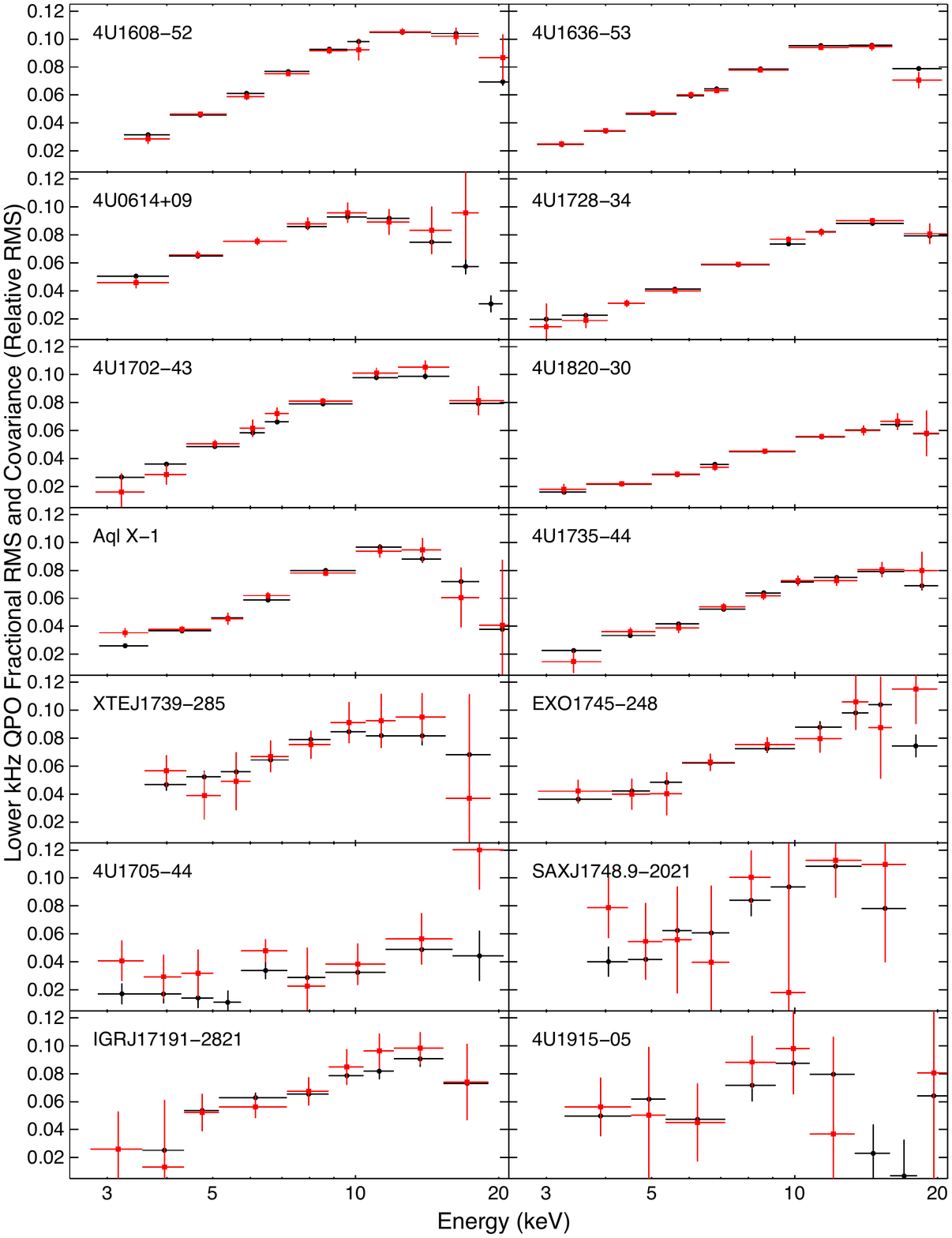}}
\subfloat[]{\label{fig:lags-E-lowTroyer}
\includegraphics[width=0.49\textwidth, trim=0 0 0 0, clip, angle=0]{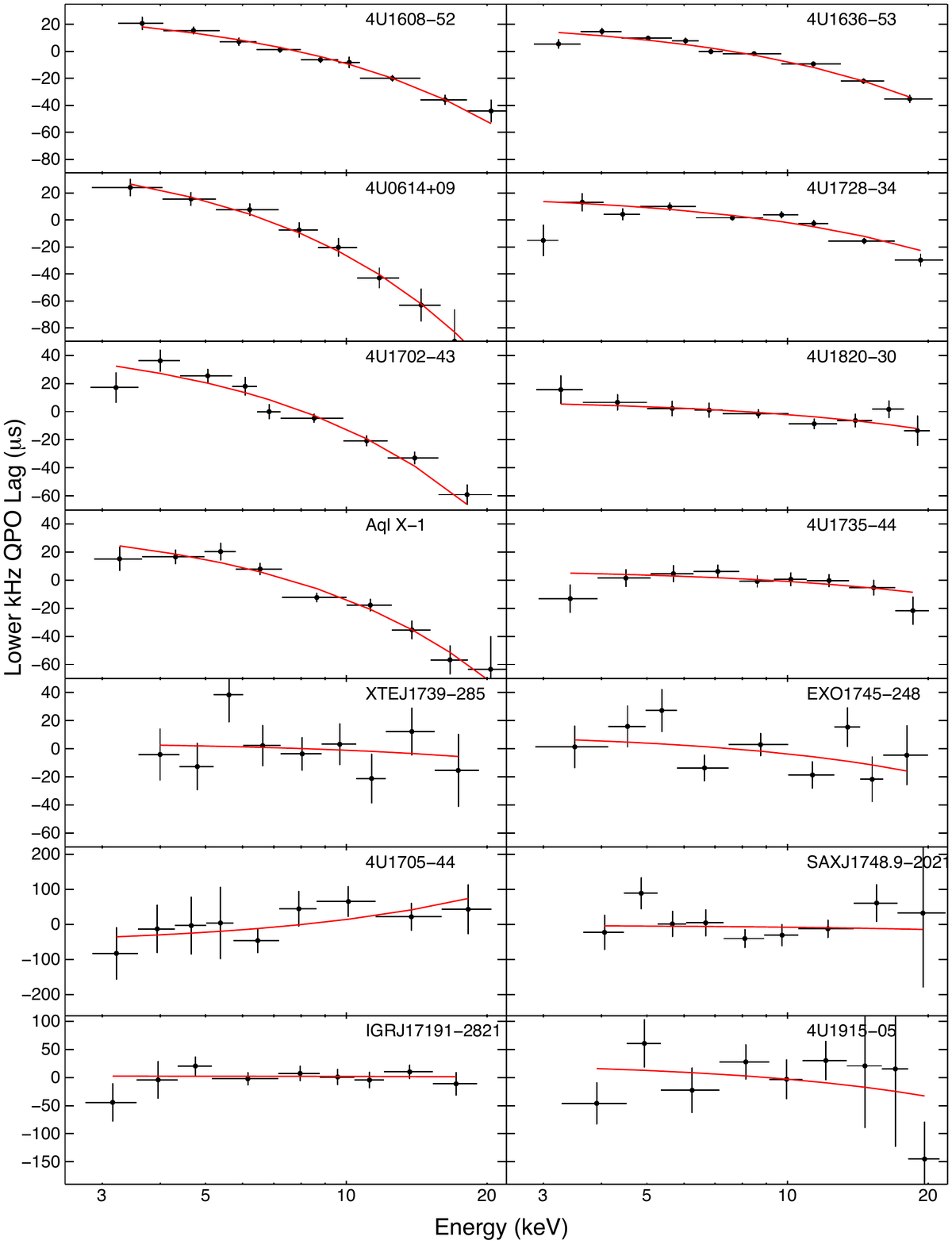}}
\caption{Fractional rms (left) and time lags (right) as a function of energy for the lower kHz QPO in 14 atoll sources \citep[originally published as Figures 4 and 8 in][]{Troyer-2018}.}
\label{fig:Troyer}
\end{figure}

Figure~\ref{fig:Troyer} shows some of the results of the most extensive study yet of the lags and rms spectra of the kHz QPOs in LMXBs \citep{Troyer-2018}. That work presents the energy-dependent rms amplitude, time lags and intrinsic coherence function, plus the time lags as a function of frequency of the lower kHz QPOs in 14 sources, and the same information for the upper kHz QPO in six out of those 14 sources. In Figure~\ref{fig:Troyer} we show only the rms and lag spectra of the lower kHz QPO in those 14 sources (we already showed the rms spectra of the upper kHz QPO in six of those 14 sources in Fig.~\ref{fig:rms-E-upp-Troyer}). The results shown in \citep{Troyer-2018} reinforce the importance of considering both the rms and lag spectra to try and model the radiative properties of the kHz QPOs.

Models that involve inverse Compton scattering did not seem appropriate to explain the lags of the lower kHz QPO, given that those lags are soft, whereas Comptonisation would only produce hard lags (see \S\ref{sec:qpo101}). Contrary to those expectations, inverse Compton scattering could work, and produce soft lags, if there is feedback from the corona to the disc \citep[][see also \citep{Lee-1998}]{Lee-2001}. In this scenario, soft photons from the disc are up-scattered in the corona; part of those up-scattered photons go to the observer, and produce the power-law like component in the spectrum, but a fraction of those corona photons will illuminate back the disc, increasing the disc temperature \citep{Gierlinski-2008}. If the photon flux that is originally produced in the disc is modulated (e.g., at the QPO frequency), the flux that returns to the disc after being scattered in the corona will also be modulated and, as those photons reheat the disc, the disc temperature will vary at the frequency of that modulation. In this feedback loop, the photons that reach the observer from the corona through the disc will be delayed with respect to those photons from the corona that are directly emitted towards the observer. Since the disc temperature is lower than that of the corona, the process results in a delay of the soft with respect to the hard photons, as observed. 

\begin{figure}
\centering
\subfloat[]{\label{fig:rms-Kumar.pdf}
\includegraphics[width=0.49\textwidth, trim=0.1cm 0 0 0, clip, angle=0]{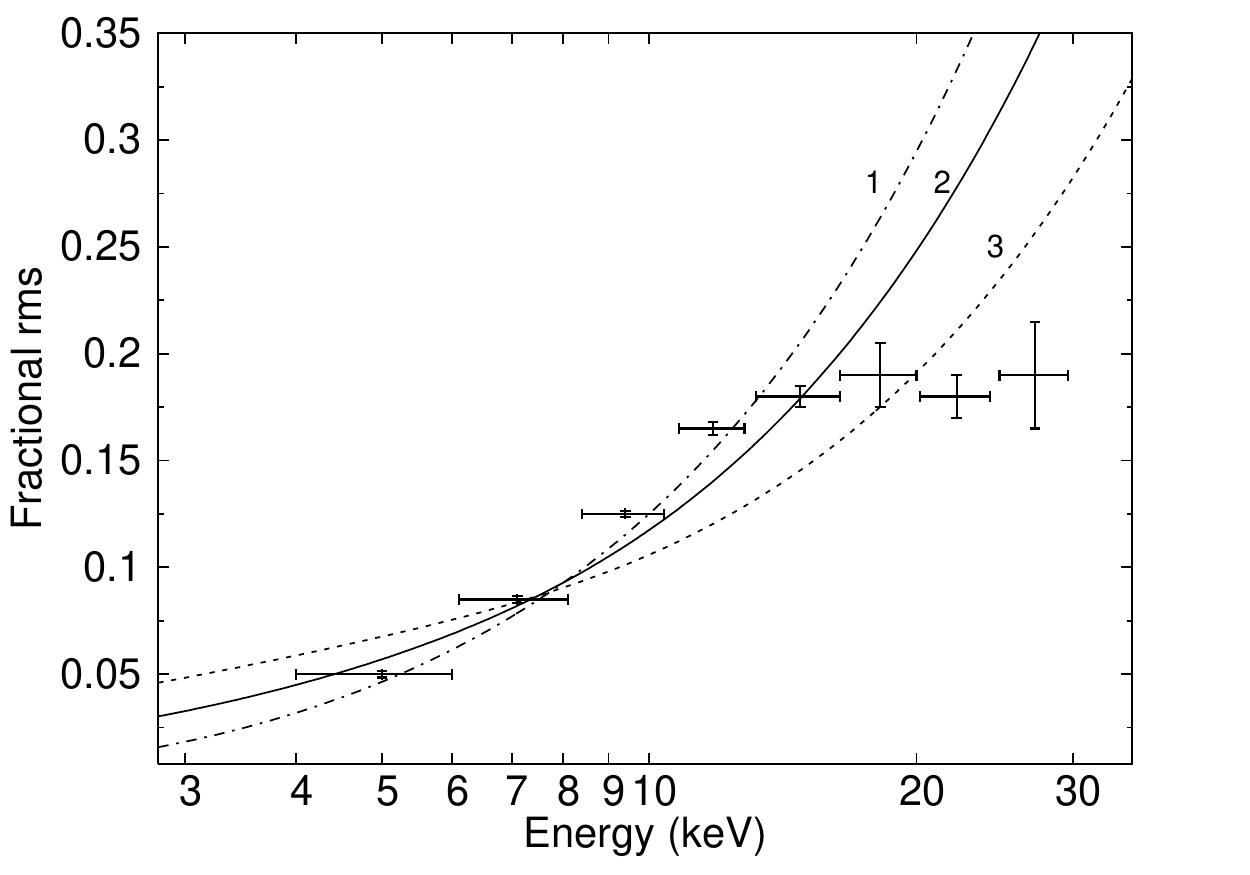}}
\subfloat[]{\label{fig:lag-Kumar.pdf}
\includegraphics[width=0.49\textwidth, trim=0.1cm 0 0 0, clip, angle=0]{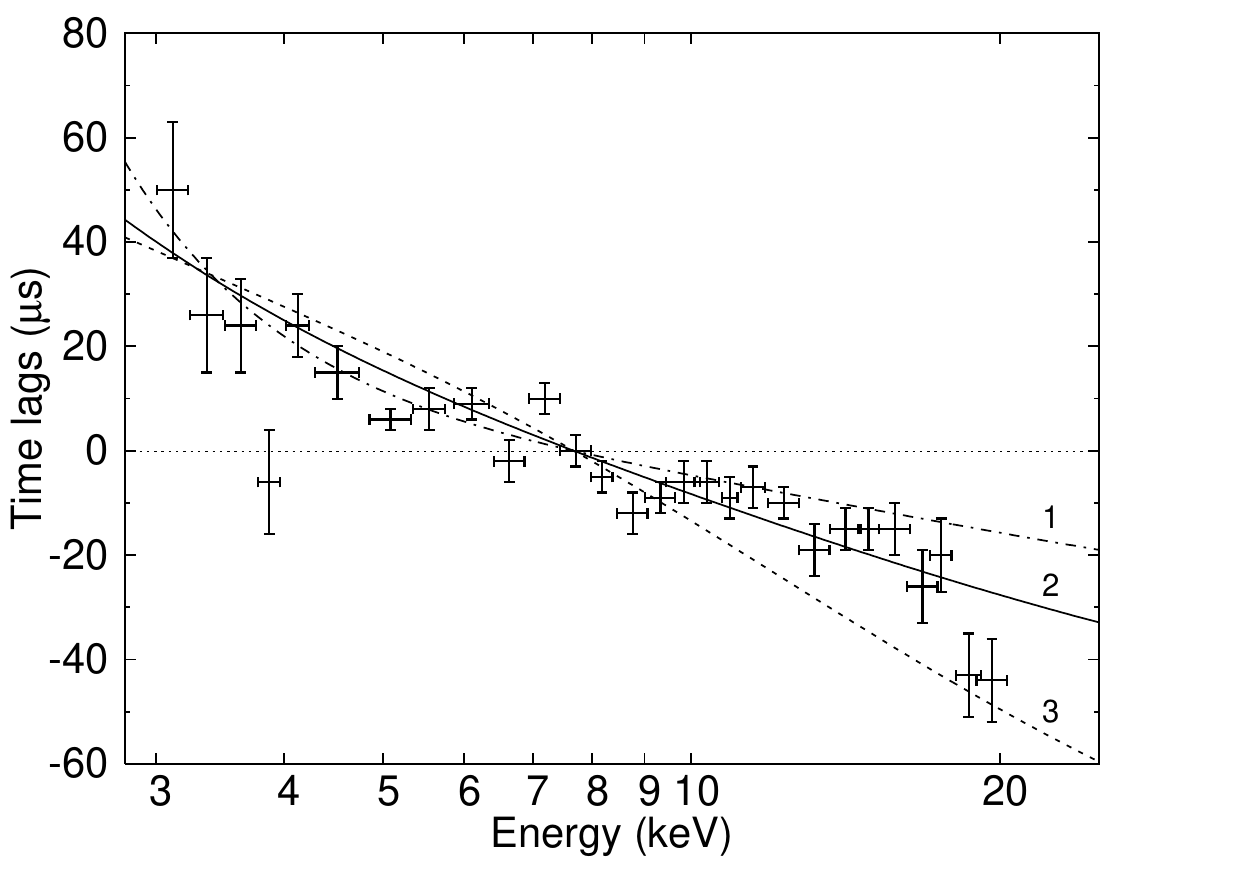}}
\caption{Fractional rms amplitude (left) and time lags (right) of the lower kHz QPO in 4U~1608--52 as a function of energy. The three lines lines labeled 1, 2 and 3 correspond to calculations of the variability produced by a feedback loop between the accretion disc and a 1-km thick corona \citep{Kumar-2014} with values of $\eta$, the fraction of photons of the corona that impinge back onto the disc, of 0.3, 0.4 and 0.5, respectively.}
\label{fig:Kumar}
\end{figure}

\begin{figure}
\centering
\includegraphics[width=0.8\textwidth, trim=0cm 3cm 0 4cm, clip, angle=0]{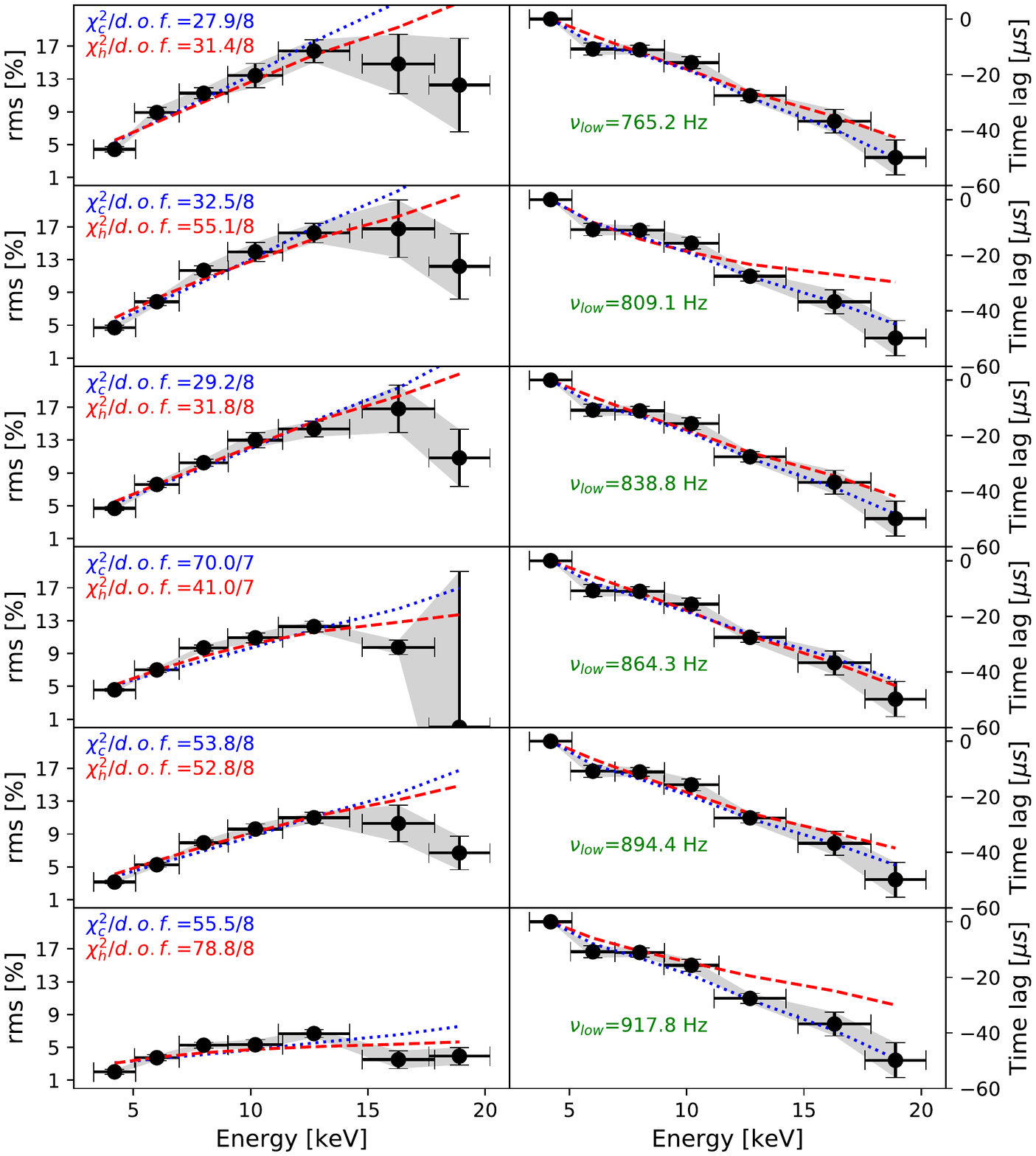}
\caption{Fractional rms (left) and phase lag (right) spectra of the lower kHz QPO in 4U~1636--53 for six values of the QPO frequency as indicated. \citep{Karpouzas-2020}. The red and blue lines correspond to the best-fitting model of the variability model that includes feedback between the accretion disc and the corona for, respectively, a hot and a cold accretion disc \citep[see][for details]{Karpouzas-2020}.}
\label{fig:lags-rms-Karpouzas}
\end{figure}

This process can be modelled by solving the time-dependent version of the Kompaneets equation \citep{Kompaneets-1957}, assuming a geometry of the corona \citep{Lee-1998, Lee-2001, Kumar-2014, Kumar-2016, Karpouzas-2020}. Since the inverse Compton scattering process cools down the electrons in the corona, there has to be an external source of heating \citep{Merloni-2001, Malzac-2004} to explain long-lived coronas in accreting LMXBs. If the photon flux from the disc that cools the corona is variable, the power provided by this external heating source must also be variable to maintain the equilibrium in the system. An interesting aspect of this approach is that the solution of the equation that describes the variability of the flux received by the observer provides both the energy-dependent amplitude and lags of the signal, and hence one can fit, with the same model, both the rms and lag spectrum of the QPOs. 

The two panels in Figure~\ref{fig:Kumar} show the rms and lag spectrum of the lower kHz QPO in 4U~1608--52 together with the results of the model that describes, simultaneously, the energy-dependent amplitude and lag of the variability produced by a feedback loop between the accretion disc and the corona \citep{Kumar-2014}. The models reproduce the rms and lag spectrum of the lower kHz QPO in 4U~1608--52 fairly well when the corona is a few km thick and for feedback fractions between $\sim$$0.1$ and $\sim$$0.5$. 

The plots in Figure~\ref{fig:Kumar} correspond to data obtained \citep{Berger-1996, Barret-2013} when the frequency of the lower kHz QPO in 4U~1608--52 was $\sim$$800$ Hz. We saw, however, that the rms spectrum of the lower kHz QPO changes significantly as the frequency of the QPO changes \citep[see \S\ref{sec:rms-amplitude} and][]{Ribeiro-2019}; while it is likely that the lag spectrum of the lower kHz QPO also changes with QPO frequency, the error bars of the current measurements  \citep{deAvellar-2013, Peille-2015} are too large to tell.

\begin{figure}
\centering
\includegraphics[width=0.8\textwidth, trim=0cm 3cm 0 4cm, clip, angle=0]{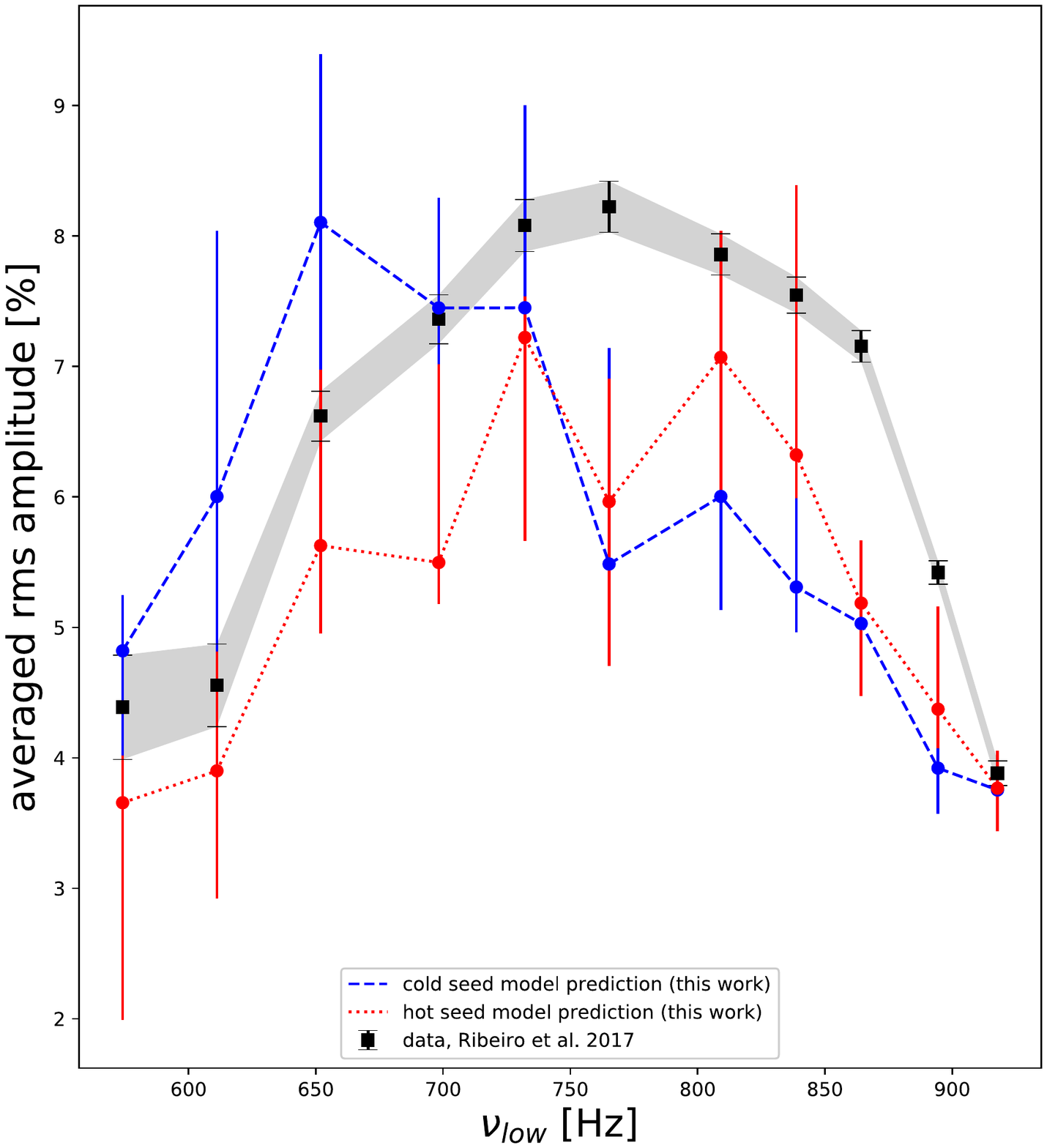}
\caption{Full-band fractional rms amplitude of the lower kHz QPO in 4U~1636--53 as a function of QPO frequency. The grey shaded area represents the same measurements \citep{Ribeiro-2019} of the rms amplitude of the QPO shown in Figure~\ref{fig:rms-nu-Ribeiro}, while the red and blue lines and symbols correspond to the best-fitting model of the variability model for, respectively, a hot and a cold accretion disc \citep{Karpouzas-2020}.}
\label{fig:avg_rms-Karpouzas}
\end{figure}

Figure~\ref{fig:lags-rms-Karpouzas} shows the fits to the rms and lag spectra of the lower kHz QPO in 4U~1636--53 for six different values of the QPO frequency, while Figure~\ref{fig:avg_rms-Karpouzas} shows the full-band fractional rms amplitude vs. QPO frequency that results from these fits  \citep{Karpouzas-2020}. These fits provide \citep[under the assumptions made in the model; see][for details]{Karpouzas-2020} the evolution of the parameters of the corona, size, $kT_e$ and $\tau$, as a function of the QPO frequency that matches reasonably well the observations \citep[e.g., Fig.~\ref{fig:spectral-pars_QPO-freq-Ribeiro}; see][for details]{Karpouzas-2020}.

While the model appears to work reasonably well, it does not reproduce the data completely. As Figures~\ref{fig:Kumar} and \ref{fig:lags-rms-Karpouzas} show, the models fit the lags over the full energy range and the rms spectrum below $\sim$$12$ keV quite accurately, but fail to reproduce the rms spectrum at energies above $\sim$$12-13$ keV. Above those energies the rms spectrum stops increasing and levels off (see also Figs.~\ref{fig:rms-E-Berger} and \ref{fig:rms-nu-E-Ribeiro}), whereas the second derivative of the model is always positive (Fig.~\ref{fig:rms-Kumar.pdf}). This is the energy range at which emission from the Compton hump due to reflection of corona photons off the accretion disc is expected \citep[e.g.,][]{Ross-1996, Ross-2007, Fabian-2009}. If the reflected signal is not (strongly) modulated, that extra emission component at those energies will reduce the rms variability without affecting the lags. This needs to be explored further using time-dependent reflection models. 

\begin{figure}
\centering
\includegraphics[width=0.8\textwidth, trim=1cm 1cm 1cm 1.5cm, clip, angle=0]{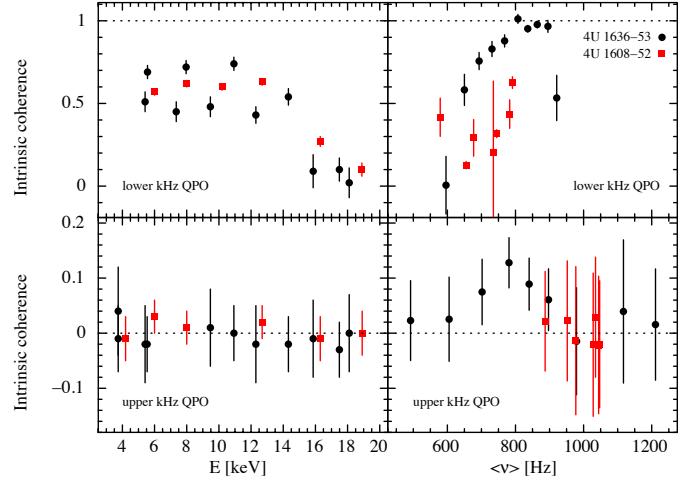}
\caption{Coherence function of the lower (top) and upper (bottom) kHz QPOs in 4U~1636--53 and 4U1608--52 as a function of energy (left) and QPO frequency (right). The plots on the left show the coherence between the photons with energies given by the value on the $x$ axis relative those in the reference bands, for all observations with QPOs combined. The plots on the right show the coherence between photons in the $4$$-$$12$ keV band relative to photons in the $12$$-$$20$ keV band at different QPO frequencies \citep[adapted from][]{deAvellar-2013} }
\label{fig:intCoh-vs-E-vs-freq_deAvellar}
\end{figure}

To end this part, we will now discuss briefly some results of the coherence function of the kHz QPOs. In Figure~\ref{fig:intCoh-vs-E-vs-freq_deAvellar} we show the coherence function of the lower and upper kHz QPOs in 4U~1636--53 and 4U1608--52 as a function of energy and QPO frequency \citep[the plots are from][see that reference for details of the way the coherence function was calculated using all the QPO data for those two sources]{deAvellar-2013}. There are four things that we would like to mention about the plots shown in this Figure: (i) The coherence function of the lower kHz QPO in both sources is independent of energy below $E$$\approx$$12$ keV, and drops as the energy increases beyond that value. This is the same energy at which the rate of increase of the rms spectrum of the lower kHz QPO starts to flatten (Figs.~\ref{fig:rms-E-Berger}, \ref{fig:rms-nu-E-Ribeiro}, \ref{fig:rms-E-low-Troyer} and \ref{fig:lags-rms-Karpouzas}). This suggests that at those energies another signal, independent of the one that produces the variability of the signal at the frequency of the lower kHz QPO, starts to become important in the light curves . 

As mentioned previously, this extra component could be the Compton hump, which is part of the reflection component coming from the accretion disc and dominates the reflected spectrum at those energies \citep{Ross-1996}. (ii) In 4U~1636--53, the degree of linear correlation at the frequency of the lower kHz QPO between the light curves in the two energy bands used in this study (see caption in Fig.~\ref{fig:intCoh-vs-E-vs-freq_deAvellar}) increases from $\gamma^2$$\approx$$0$ to  $\gamma^2$$\approx$$1$ as the QPO frequency increases from $\nu_{\rm low}$$\approx$$600$ Hz to $\nu_{\rm low}$$\approx$$800$ Hz, and then drops to  $\gamma^2$$\approx$$0.5$ as the QPO frequency increases further to $\nu_{\rm low}$$\approx$$900$ Hz. The behaviour of the coherence function of the lower kHz QPO in 4U~1636--53 with QPO frequency is very similar to that of the quality factor and the rms amplitude of the lower kHz QPO in this source (Figs.~\ref{fig:rms-nu-Ribeiro} and \ref{fig:Q-model-Barret}). These similarities provide a strong indication that these three phenomena are related, and suggest that the amplitude and width (or the quality factor) of the lower kHz QPO in 4U~1636--53 (and, by extension, of the lower kHz QPO in all the other sources\footnote{This is possibly also the case for the upper kHz QPO, and all other QPO signals in these sources.}) reflect (at least in part) the degree of linear correlation of the signals in different energies bands at the QPO frequency. In the model of the rms amplitude and the lags in which the corona and the disc are connected through a feedback loop \citep{Lee-2001, Kumar-2014, Karpouzas-2020}, the temperatures of the corona and the accretion disc oscillate coherently; therefore, the soft and hard light curves produced in this scenario are linearly correlated and the intrinsic coherence is high. (iii) The degree of linear correlation between the light curves of the source in the two energy bands used to calculate the coherence function is much lower at the frequency of the upper than at the frequency of the lower kHz QPO. From the discussion in the previous point, this is probably the reason why the upper kHz QPO is much broader (lower quality factor) than the upper kHz QPO \citep{Barret-2006}. (iv) The coherence of the upper kHz QPO shows a small increase at $\nu_{\rm upp}$$\approx$$700-800$ Hz. This is the same frequency at which the rms amplitude of the upper kHz QPO (Fig.~\ref{fig:rms-nu-Ribeiro}) and the slope of the rms spectrum of the QPO (Fig.~\ref{fig:rms-slopes-Ribeiro}) as a function of QPO frequency show a local maximum. In the context of the model in which a feedback mechanism connects the corona and the disc \citep{Lee-2001, Kumar-2014, Karpouzas-2020}, this could be interpreted as the source of soft photons and the Comptonising medium not oscillating coherently for the largest part of the range of QPO frequencies spanned by the upper kHz QPO, and the disc and the corona becoming resonant when the frequency of the upper kHz QPO is at around 800 Hz.

\subsection{Other phenomenology of the kHz QPOs}
\label{sec:other}

In this subsection we will briefly discuss two additional phenomena that are interesting to try and understand the mechanisms that produce the kHz QPOs, but do not fit thematically in the previous subsections: Harmonics of the kHz QPOs and frequency and amplitude modulation of the QPO signal. These topics have not yet been fully explored, partly because the analysis required is not standard, and partly because the data available do not allow us to go beyond what has been done so far. For instance, there is only one paper published exploring the harmonic content of the kHz QPOs, and only four papers about the frequency and amplitude modulation of the QPO signal by other timing phenomena. The few results available, however, suggest that some of these endeavours are worthwhile pursuing further. 

Most of the models that have been proposed to explain the frequencies of the kHz QPOs assume that one of the QPOs reflects the Keplerian orbital motion at some preferred radius in the accretion disc. For instance, in the sonic-point beat-frequency model \citep{Miller-1998}, the upper kHz QPO is produced at the radius where the radial flow velocity in the disc turns from subsonic to supersonic (the sonic radius), and the lower kHz QPO is a beat between the upper kHz QPO and the spin frequency of the neutron star. Besides the main peaks at $\nu_{\rm low}$ and $\nu_{\rm upp}$,  this model predicts some other (weaker) harmonics and sidebands of these QPOs at specific frequencies; for instance, in this model there should be a relatively strong harmonic of the lower QPO at $2\nu_{\rm low}$ \citep[see table 3 of][for a list of other harmonic and sideband peaks predicted by the model]{Miller-1998}.

In the relativistic-precession model \citep{Stella-1999}, the upper kHz QPO is also assumed to be Keplerian, and the lower kHz QPO is the periastron precession frequency of a  slightly non-circular inner  accretion disc. Oscillating disc models \citep[e.g.,][]{Psaltis-2000} in a GR potential yield the same frequencies as those for test particles in the relativistic-precession model, but also predict other frequencies that are linear combinations of the three basic relativistic frequencies in the disc. For instance, in an oscillating disc in which the Keplerian and the periastron precession frequencies are excited, there should be a peak at a frequency equal to $2\nu_{\rm upp} - \nu_{\rm low}$.

Finding one of these other peaks would, on one hand, favour one model over the other, opening up a path to understand the dynamics of the disc and the production of the QPOs and, on the other hand, would confirm that the upper kHz QPO is due to Keplerian orbital motion in the disc. Unfortunately, the only attempt to find any of these peaks using data of Sco~X-1 \citep{Mendez-2000} gave negative results. The 95\% confidence upper limits of a peak at $2\nu_{\rm low}$ and  $2\nu_{\rm upp} - \nu_{\rm low}$ are, respectively, $0.12$ and $0.26$ times the amplitude of the upper kHz QPO in this source \citep[the amplitude of the upper kHz QPO in Sco~X-1 is between 0.6\% and 2.5\%; see table 1 in][for the upper limits at other frequencies]{Mendez-2000}. The signal of these other peaks could be attenuated in the corona \citep{Brainerd-1987, Kylafis-1987, Miller-1998, Miller-2000} depending on the frequency of the peak, and the radius and optical depth of the corona. Considering this effect, the upper limits of an unattenuated signal at those two frequencies would be a factor between $0.15$ and $0.30$ of the unattenuated amplitude of the upper kHz QPO in Sco~X-1 \citep[see table 2 in][for the unattenuated upper limits at other frequencies]{Mendez-2000}. In conclusion, in Sco~X-1 none of the secondary QPO peaks predicted by the two main classes of models of the kHz QPOs are detected, with upper limits that imply that these secondary peaks are one to two orders of magnitude weaker in power than the upper kHz QPO.

\begin{figure}
\centering
\includegraphics[width=0.6\textwidth, trim=0cm 6cm 3cm 6cm, clip, angle=0]{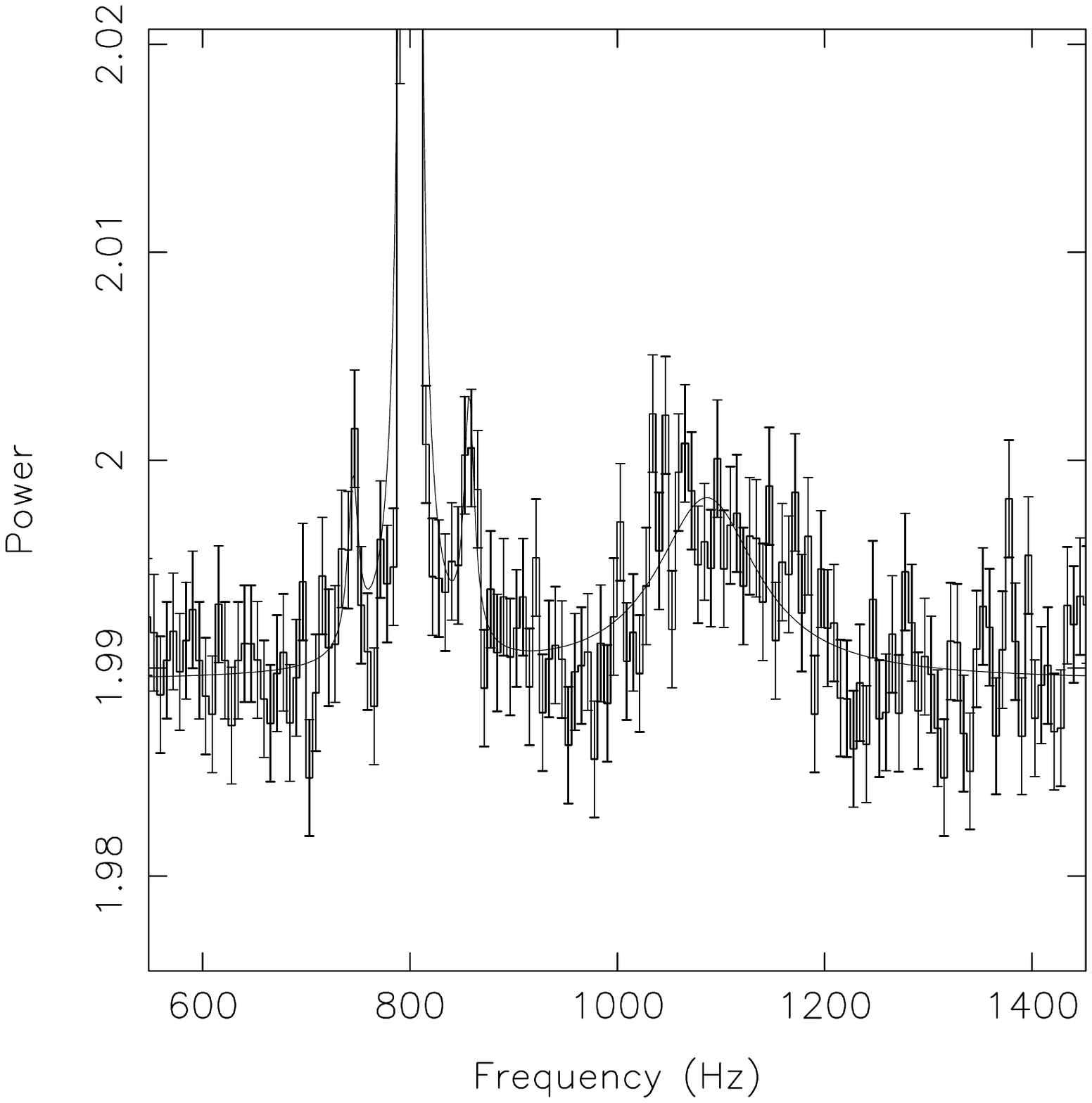}
\caption{Sideband peaks of the lower kHz QPO in 4U~1636--53. The sideband peaks are visible at a frequency that is $\pm55$ Hz of the frequency of the strong and narrow lower kHz QPOs, which is at $\sim$$800$ Hz in this observation. The broad peak at $\sim$$1100$ Hz in the Figure is the upper kHz QPO \citep{Jonker-2005}.}
\label{fig:sidebands-Jonker}
\end{figure}

If the amplitude of the signal of the QPOs is modulated, one would expect to see sidebands to the main oscillation, at frequencies that are equal to the frequency of the QPOs plus or minus the frequency of the mechanism that modulates the QPO amplitude. This could happen if, for instance, on very short scales the amplitude of the kHz QPOs depends upon mass accretion rate. Although there is no guarantee that this is the case, given that on long time scales the amplitude of the QPOs depends upon QPO frequency (e.g., Fig.~\ref{fig:rms-nu-Ribeiro} in \S\ref{sec:qpo101}) and, in turn, QPO frequency depends on inferred mass accretion rate (e.g., Fig.~\ref{fig:nu-vs-Sa-Zhang} in \S\ref{sec:qpo101}), it is not unthinkable that this relation could also hold on very short time scales.

Figure~\ref{fig:sidebands-Jonker} shows the power spectrum of 4U~1636--53 in the range of frequencies of the kHz QPOs \citep{Jonker-2005}. The strong and narrow peak (off the scale of the plot) at $\sim$$800$ Hz and the weak and broad peak at $\sim$$1100$ Hz are, respectively, the lower and the upper kHz QPOs. The scale of the $y$ axis in the plot has been chosen to highlight the weak QPO peaks that appear at frequencies that are $\sim$$50$ Hz below and above the frequency of the lower kHz QPO. These sidebands peaks are very significantly detected in this observation of 4U~1636--53, and had been observed before in this source and in 4U~1608--52 and 4U~1728--34  \citep{Jonker-2000b}. One possible explanation for these phenomena, given in \citep{Jonker-2005}, is that the sideband peaks reflect a modulation in the radiation pattern that produces the upper kHz QPO at the Lense-Thirring precession frequency at the inner edge of the accretion rate, and that this modulation would in turn modulate the formation of the lower kilohertz QPO.

\begin{figure}
\centering
\subfloat[]{\label{fig:freq-modulation-1-Yu}
\includegraphics[width=0.54\textwidth, trim=0cm 4cm 0cm 12cm, clip, angle=0]{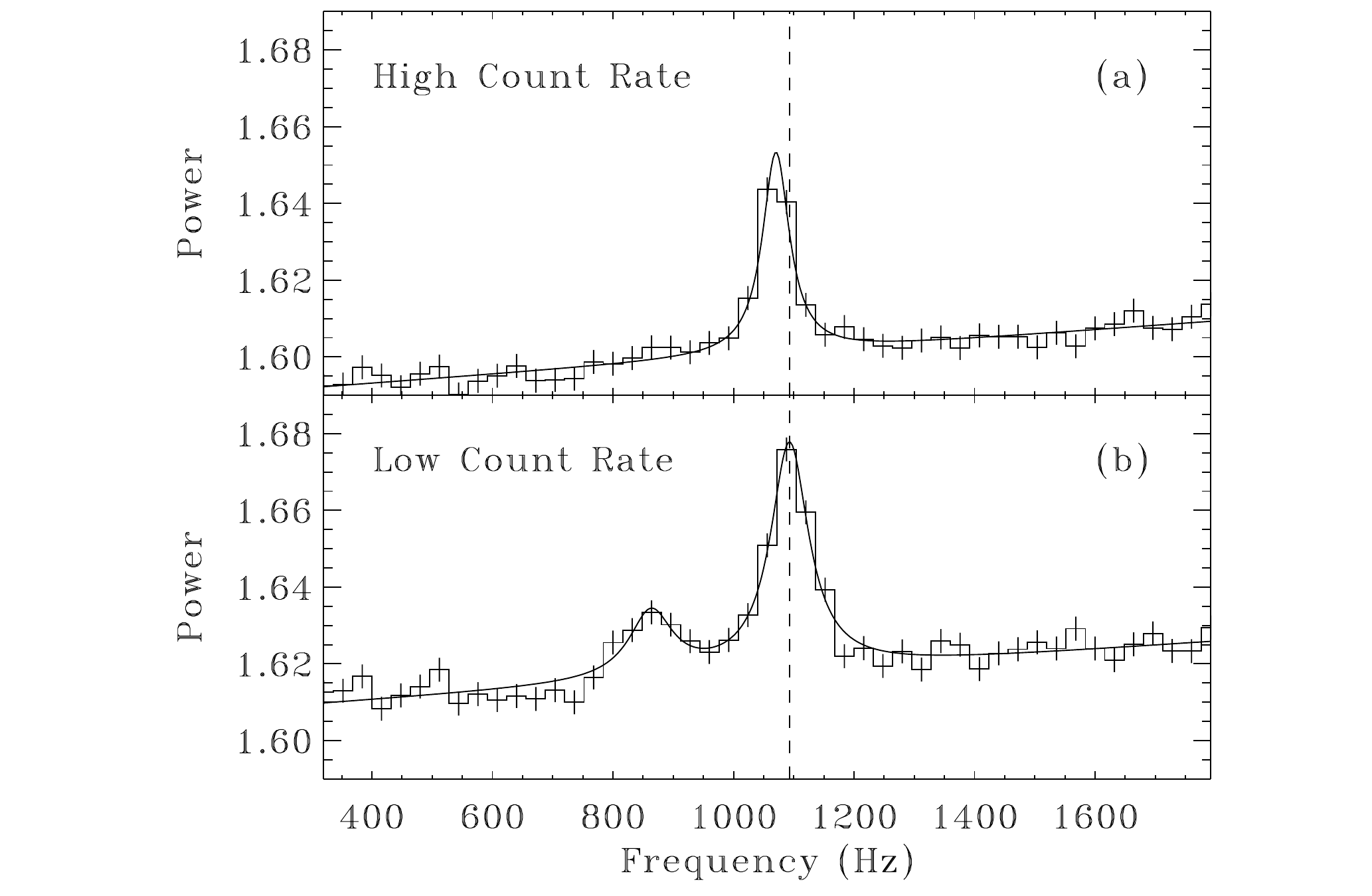}}
\subfloat[]{\label{fig:freq-modulation-2-Yu}
\includegraphics[width=0.42\textwidth, trim=2.7cm 4cm 2.cm 12cm, clip, angle=0]{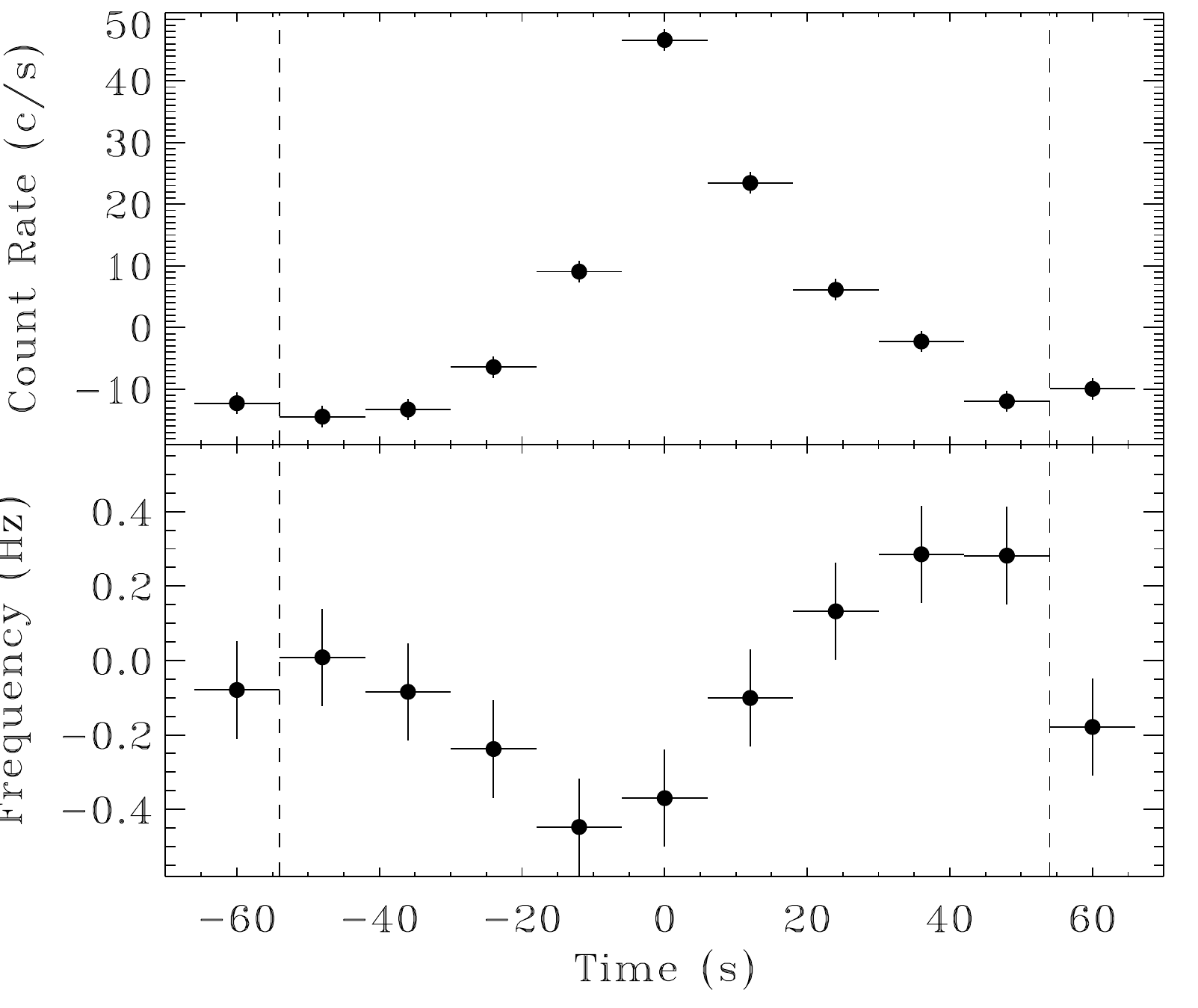}}
\caption{Left: Power spectra in the range of frequencies of the kHz QPOs in Sco~X-1. The power spectrum in the top panel was calculated during intervals of high count rates on the time scale of the normal branch QPO (NBO) at $6-8$ Hz in this source. The power spectrum in the bottom panel corresponds to the intervals of low count rates on the time scale of the NBO \citep[originally published as Figure 2 in][]{Yu-2001}. Right: For the NS LMXB 4U~1608--52, count rate of the source (top) and frequency of the lower kHz QPO (bottom) over the cycle of the millihertz QPO in this source \citep[originally published as Figure 2 in][]{Yu-2002}.}
\label{fig:freq-modulation-Yu}
\end{figure}

Similar to what happens to the amplitude, the frequency of the oscillations that produce the QPOs can also be modulated on the time scale of another variability component in the light curve. Figure~\ref{fig:freq-modulation-1-Yu} shows two power spectra of Sco~X-1 calculated from very short time intervals to sample the maxima and minima of the light curve of the source on time scales of $0.15$ s, corresponding to the frequency range,  $6-8$-Hz, of the so-called Normal Branch Oscillation (NBO) in this source. The top and bottom panels of this Figure show the power spectrum of, respectively, the maxima and minima of the light curve on those time scales. The frequency of the upper kHz QPO changes significantly, by $\sim$$20$ Hz, with the change being anti correlated with the source count rate. This result shows that the frequency of the upper kHz QPO in Sco~X-1 is driven by changes that happen on the time scale of the NBO. 

Figure~\ref{fig:freq-modulation-2-Yu} shows the result of a similar analysis for the lower kHz QPO in 4U~1608--52 on the time scale of the millihertz QPO in this source \citep{Revnivtsev-2001}. In this case, again, the frequency of the QPO changes over the cycle of the millihertz QPO and is anti correlated with the source count rate in that cycle. This result, and the previous one, could be explained if, for instance, the upper kHz QPO in Sco~X-1 and the lower kHz QPO in 4U~1608--52 are produced at the inner inner edge of the accretion disc, and the value of the inner radius of the disc is modulated by radiation coming from the neutron-star surface on the time scale of the NBO signal in the case of Sco~X-1, and the time scale of the millihertz QPO in 4U~1608--52. 


\section{Probing neutron-star interiors and GR with kHz QPO}
\label{sec:probing}

The Keplerian orbital frequency at a radial distance $r$ from the centre of a slowly rotating neutron star (to first order in the specific angular momentum, $j= cJ/GM^2$) is:

\begin{equation}
\displaystyle \nu_{\rm K} = \frac{1}{2\pi} \left(1-j\left(\frac{GM}{rc^2}\right)^{3/2}\right)\sqrt{\frac{G M}{r^3}},
\label{eq:Kepler}
\end{equation}

where $M$ and $J$ are, respectively, the gravitational mass and angular momentum of the star, and $G$ and $c$ are the gravitational constant and the speed of light, respectively \citep{Kluzniak-1990, Miller-1999}. Because the disc must be outside the neutron star, if the frequency of one of the kHz QPOs reflects Keplerian orbital motion at this radial distance in the disc  \citep{Miller-1998, Stella-1999}, it must hold that $r \geq R_{\rm NS}$, where $R_{\rm NS}$ is the neutron-star radius. The existence of a radius, $R_{\rm ISCO}$, inside which no stable circular orbit is possible \citep{Bardeen-1972} also implies that $r \geq R_{\rm ISCO}$. Depending on the equation of state, the radius of the ISCO can be inside or outside the neutron star \citep{Miller-1998b}. Under these conditions, for an observed frequency of a QPO, $\nu_{\rm QPO}$, the maximum allowed mass and radius of the neutron star are, respectively,  $M_{\rm max}=2.2~M_\odot (1+0.75j)(1000~{\rm Hz}/\nu_{\rm QPO})$, and $R_{\rm max}=19.5~{\rm km}(1+0.2j)(1000~{\rm Hz}/\nu_{\rm QPO})$. The highest the QPO frequency, the smallest the maximum allowed mass and radius of a neutron star. 

The maximum frequency of any of the kHz QPO observed with RXTE from all LMXBs is $\nu_{\rm upp}$$=$$1288$$\pm$$8$ Hz in 4U 0614+09, with a $3$-$\sigma$ lower limit of $1267$ Hz \citep{vanDoesburgh-2018}. This measurement puts an upper limit to the mass of the neutron star in this system of $M < 2.1 \Msun$ that is fairly independent of the assumed neutron-star equation of state. Assuming that this QPO reflects the Keplerian orbital frequency at, or just outside, the ISCO, this frequency leads to a mass of the neutron star in 4U 0614+09 of $2.0$$\pm$$0.1\Msun$ \citep{vanDoesburgh-2018}.

One should keep in mind that the identification of the upper kHz QPO with the Keplerian frequency at the inner edge of the accretion disc is model dependent. For instance, in the model proposed in \citep{Osherovich-1999} the lower kHz QPO is the one identified with the orbital frequency at the inner disc radius. On the other hand, as shown for instance in \citep{Alpar-2008, Erkut-2008, Alpar-2016}, it is possible to have frequencies above the Keplerian frequency at the inner edge of the accretion disc if there is a transition region between the disc and the neutron-star surface where matter in the disc has to slow down to match the rotation speed of the neutron star. 

Measurements of QPO frequencies and spectral properties of the source offer another possibility to constrain the mass and radius of a neutron star. A broad iron emission line at $6.5$$-$$7$ keV has been observed in about a dozen accreting neutron stars \citep{DiSalvo-2005, Bhattacharyya-2007, Cackett-2008, Pandel-2008, Cackett-2009, DiSalvo-2009, Iaria-2009, Reis-2009, Shaposhnikov-2009, Cackett-2010, Piraino-2012, Egron-2013, JMiller-2013, Sanna-2013, DiSalvo-2015, Pintore-2015, Iaria-2016, Wang-2017, Ludlam-2018, Wang-2019, Ludlam-2019, Mazzola-2019}. It has been proposed that, like in accreting black-hole systems \citep{Fabian-1989, Ross-1996}, the iron line is due to reflection of corona photons off the accretion disc, and that the line profile is driven by special and GR effects. If this is the case, the shape of the line profile would depend upon the inner disc radius. Therefore, detecting kHz QPOs in the power spectrum and a broad iron line in the energy spectrum of the same object would provide separate ways to constrain the neutron-star parameters in that object \citep{Bhattacharyya-2011}. 

There is only one source, 4U 1636--53, in which both a broad iron line in the energy spectrum and kHz QPOs in the power spectrum were observed simultaneously in four separate occasions \citep{Sanna-2014}. Because these four observations sampled different spectral states of the source, and presumably the inner radius of the disc changed between observations \citep{Gierlinski-2002}, this dataset offered the opportunity to test whether the inner disc radius deduced independently from the line profile and from the kHz QPOs changed in a consistent way. The main result of this analysis \citep{Sanna-2014} was that the inner radius of the accretion disc deduced from the frequency of the upper kHz QPO was correlated with the spectral state of the source, whereas the radius deduced from the profile of the iron line was not. Because of this, the combined results from the kHz QPOs and the iron line do not lead to a consistent value of the neutron-star mass. Since those were the only observations available for this test, and no new observations of kHz QPOs are possible with current missions, we have to wait until the eXTP \citep{Watts-2019} and Athena \citep{Nandra-2013} missions fly to address this question again. For the moment the jury is still out.

We refer the reader to the chapter in this book by Cole Miller for a more detailed discussion on methods of constraining neutron-star masses, radii and the equation of state of the cold dense matter that constitutes these stars.


\section{Conclusions and outlook}

The kHz QPOs in neutron stars unlocked a new door to study the dynamics of matter, and the time-dependent interplay between matter and radiation, on very short time scales in the violent and turbulent environment that surrounds a neutron star, and under the most extreme conditions in GR. These QPOs also offer a chance to constrain the mass, radius and internal constitution of these extremely compact objects, with the potential to unveil the properties of matter under conditions that are unattainable in laboratories on Earth. The amount and quality of the data collected in almost fifteen years with the RXTE satellite are unique and, for the moment, unrivalled by data that have or can be collected by any other satellite in the past or currently in operation. The rich phenomenology of the kHz QPOs discussed in this chapter triggered scores of new ideas about accretion, and will guide this area of research for years to come. The next generation of X-ray satellites with enough time resolution to observe the millisecond time scales in accreting LMXBs with enough collecting area, and the capability to deal with very bright objects, will soon be a reality. The complementary technical capabilities of Athena and eXTP will give a new impulse to this topic in the coming decade. If we were asked to anticipate now what the most fruitful areas of research will be when those missions start providing data, we would dare to say that the most exciting results will come from the correlated study of the properties of the kHz QPOs and the broad iron line (\S\ref{sec:probing}) over the shortest possible time scales, and the prospect of finding signatures of the innermost stable circular orbit around a neutron star (\S\ref{sec:qpo101}). The analysis tools that are needed to achieve these goals have to address {\em simultaneously} the spectral (physical parameters that describe the emission) and timing (rms amplitudes, lags and coherence function) properties of variable signals over the shortest time scales allowed by the data. As we showed in this Chapter, initiatives are already being taken in that direction. This, and the efforts to try and understand the data, will keep us busy for at least two more decades.

\begin{acknowledgement}

This work is part of the research programme Athena with project number 184.034.002, which is (partly) financed by the Dutch Research Council (NWO). TMB acknowledges financial contribution from the agreement ASI-INAF n.2017-14-H.0. This research has made use of data and/or software provided by the High Energy Astrophysics Science Archive Research Center (HEASARC), which is a service of the Astrophysics Science Division at NASA/GSFC. This research made use of NASA's Astrophysics Data System.

\end{acknowledgement}


\bibliography{timing}

\end{document}